\documentclass{jfm}
\usepackage{graphicx,epsf,subfigure}
\usepackage{natbib}
\usepackage{epsfig} 
\usepackage{epstopdf}
\usepackage{amsmath}
\usepackage{wasysym}
\usepackage{booktabs}
\usepackage{mathrsfs}
\usepackage{color}

\let\realverbatim=\verbatim
\let\realendverbatim=\endverbatim
\renewcommand\verbatim{\par\addvspace{6pt plus 2pt minus 1pt}\realverbatim}
\renewcommand\endverbatim{\realendverbatim\addvspace{6pt plus 2pt minus 1pt}}
\makeatletter
\newcommand\verbsize{\@setfontsize\verbsize{10}\@xiipt}
\renewcommand\verbatim@font{\verbsize\normalfont\ttfamily}
\makeatother

\ifCUPmtlplainloaded \else
  \checkfont{eurm10}
  \iffontfound
    \IfFileExists{upmath.sty}
      {\typeout{^^JFound AMS Euler Roman fonts on the system,
                   using the 'upmath' package.^^J}%
       \usepackage{upmath}}
      {\typeout{^^JFound AMS Euler Roman fonts on the system, but you
                   don't seem to have the}%
       \typeout{'upmath' package installed. jfm.cls can take advantage
                 of these fonts,^^Jif you use 'upmath' package.^^J}%
      }
  \else
  \fi
\fi

\ifCUPmtlplainloaded \else
  \checkfont{msam10}
  \iffontfound
    \IfFileExists{amssymb.sty}
      {\typeout{^^JFound AMS Symbol fonts on the system, using the
                'amssymb' package.^^J}%
       \usepackage{amssymb}%
       \let\le=\leqslant  
         
      }{}
  \fi
\fi

\ifCUPmtlplainloaded \else
  \IfFileExists{amsbsy.sty}
    {\typeout{^^JFound the 'amsbsy' package on the system, using it.^^J}%
     \usepackage{amsbsy}}
    {\providecommand\boldsymbol[1]{\mbox{\boldmath $##1$}}}
\fi

\newsavebox{\astrutbox}
\sbox{\astrutbox}{\rule[-5pt]{0pt}{20pt}}

\title[Latitudinal Libration driven flows in triaxial ellipsoids]{Latitudinal libration driven flows in triaxial ellipsoids}
\author[ S. Vantieghem, D. C\'ebron, J. Noir]{S.\ns V\ls A\ls N\ls T\ls I\ls E\ls G\ls H\ls E\ls M$^1$ \footnote{Email adress for correspondance: stijn.vantieghem@erdw.ethz.ch}, \ns 
D.\ns C\ls \'E\ls B\ls R\ls O\ls N$^{1,2}$  \ns  \and  \ns  J.\ns N\ls O\ls I\ls R$^1$}

\affiliation{$^1$Institut f\"ur Geophysik, Sonneggstrasse 5, ETH Z\"urich, Z\"urich, CH-8092, Switzerland. \\ $^2$Universit\'e Grenoble Alpes, CNRS, ISTerre, Grenoble, France.}

\date{2014}

\pagerange{\pageref{firstpage}--\pageref{lastpage}}

\begin{document}

\label{firstpage}
\maketitle

\begin{abstract}
Motivated by understanding the liquid core dynamics of tidally deformed planets and moons, we present a study of incompressible flow driven by latitudinal libration within rigid triaxial ellipsoids. We first derive a laminar solution for the inviscid equations of motion under the assumption of uniform vorticity flow. This solution exhibits a resonance if the libration frequency matches the frequency of the spin-over inertial mode. Furthermore, we extend our model by introducing a reduced model of the effect of viscous Ekman layers in the limit of low Ekman number \cite[]{noir2013precession}. This theoretical approach is consistent with the results of \cite{chan2011simulations-2} and \cite{zhang2012} for spheroidal geometries. Our results are validated against systematic three-dimensional numerical simulations. In the second part of the paper, we present the first linear stability analysis of this uniform vorticity flow. To this end, we adopt different methods \cite[]{lifschitz1991local,gledzer1978finite} that allow to deduce upper and lower bounds for the growth rate of an instability. Our analysis shows that the uniform vorticity base flow is prone to inertial instabilities caused by a parametric resonance mechanism. This is confirmed by a set of direct numerical simulations. Applying our results to planetary settings, we find that neither a spin-over resonance nor an inertial instability can exist within the liquid core of the Moon, Io and Mercury.
\end{abstract}

\begin{keywords}
Rotating flows -- Waves in rotating fluids -- Parametric Instability
\end{keywords}
\section{Introduction}
\begin{figure}  
\begin{center}               
\includegraphics[width=0.75\textwidth]{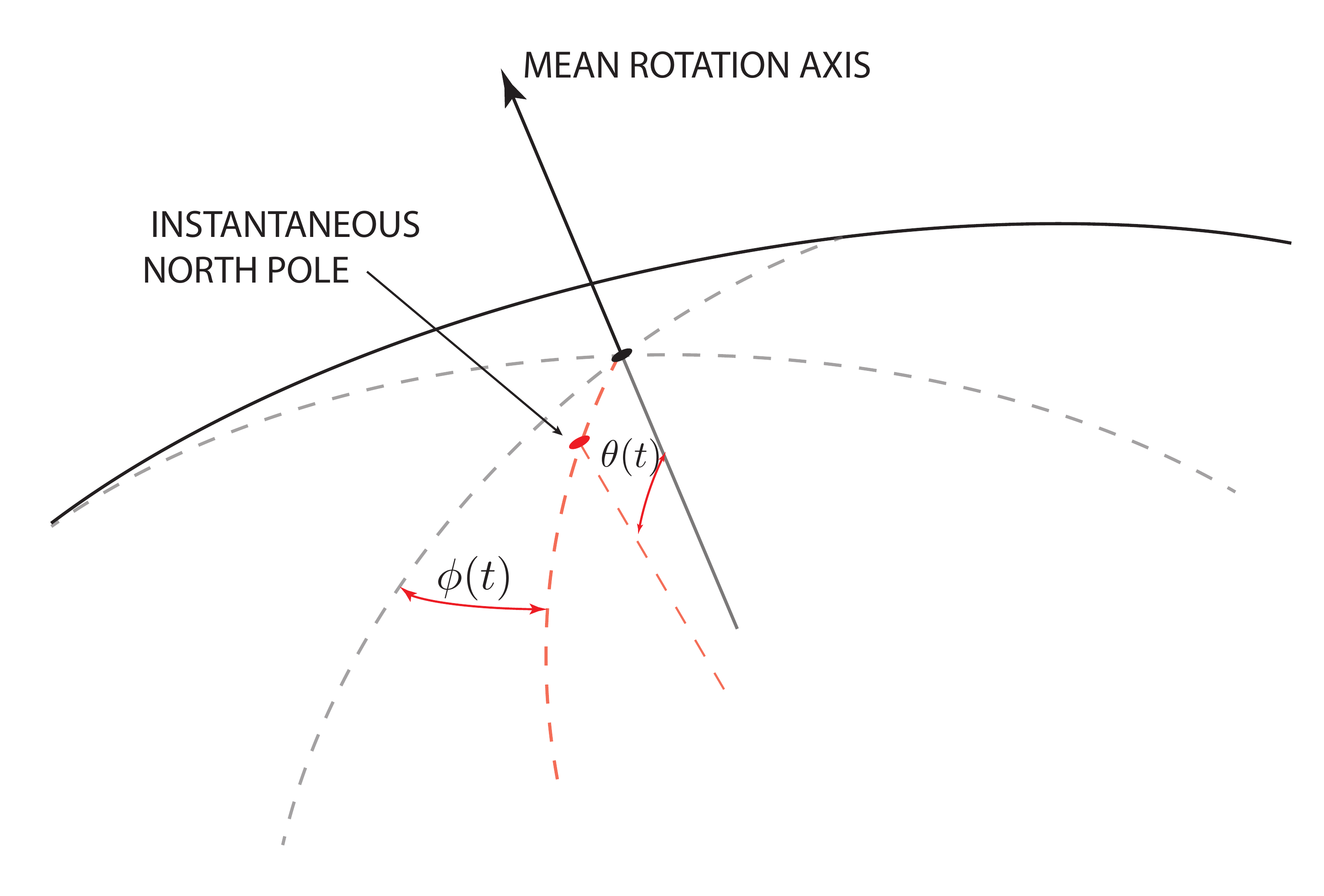}
\label{fig:librations}
    \caption{Illustrative sketch of latitudinal and longitudinal librations, denoted respectively by angles $\theta(t)$ and $\phi(t)$.}                
  \end{center}
\end{figure}

\begin{figure}  
\begin{center}               
\includegraphics[width=\textwidth]{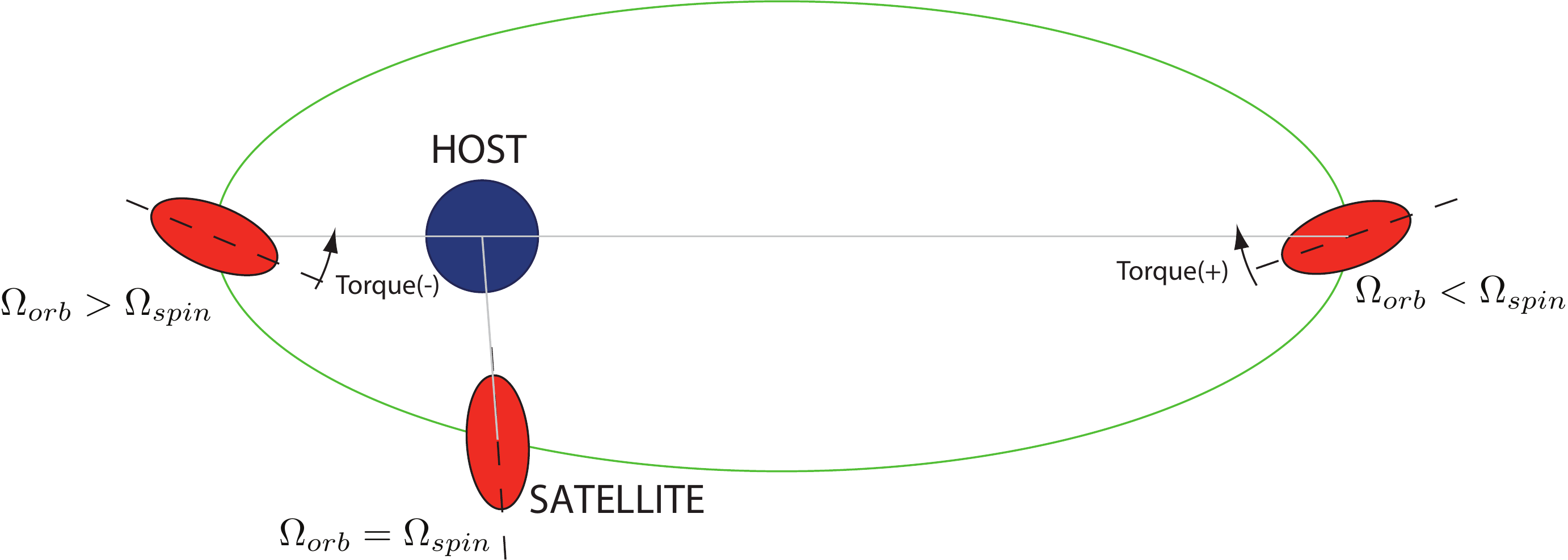}
\includegraphics[width=\textwidth]{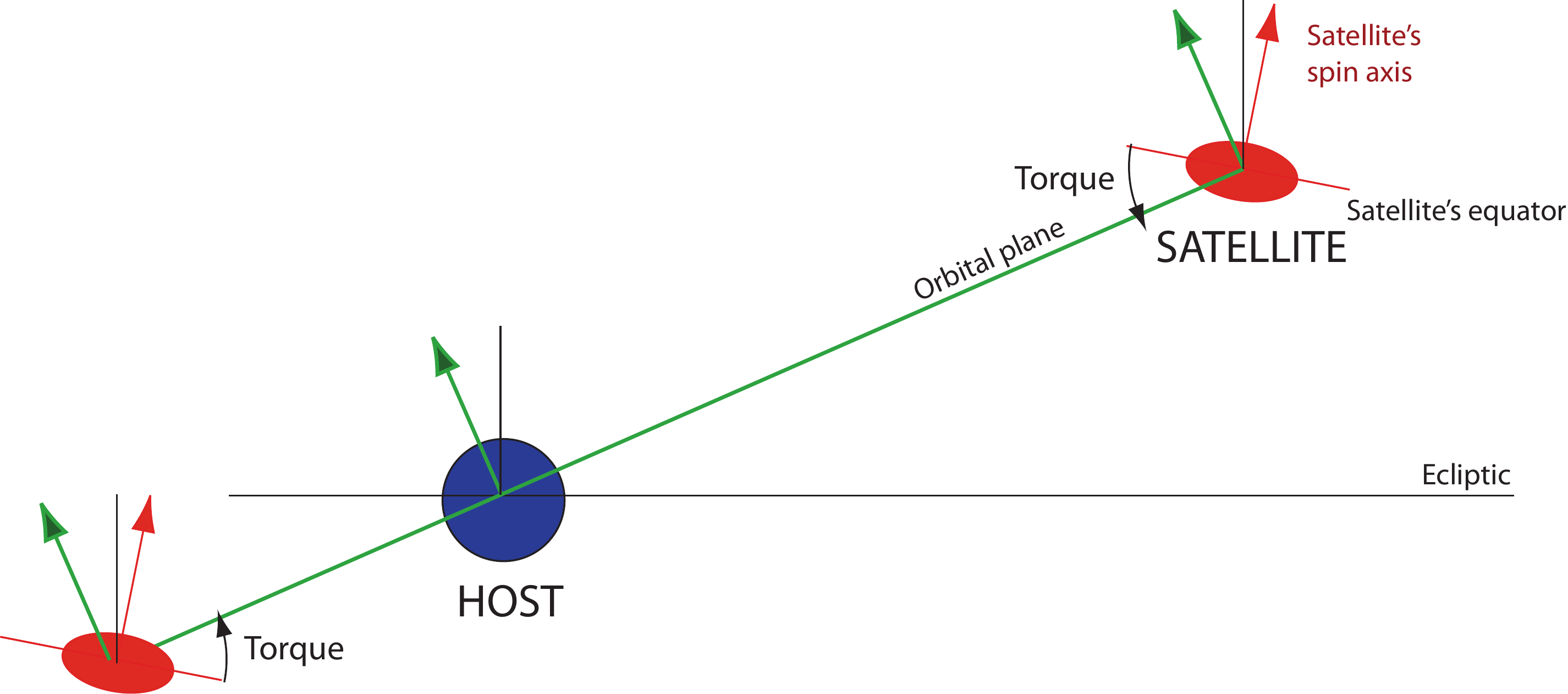}

    \caption{Mechanisms giving rise to librations in longitude (top) and latitude (bottom). Shown are the host body, its satellite at different positions along its orbital trajectory, and the gravitational torques resulting from variations in the orbital speed $\Omega_{orb}$ (top) and a non-alignment between the ecliptic and the orbital plane (bottom).}                
  \label{fig:orbit}
  \end{center}
\end{figure}

As a consequence of gravitational coupling with their orbital partners, the rotation of planets and moons is not constant in time but undergoes periodic variations. One can identify different classes of modulation of the rotational dynamics. Precession and nutation refer to the effect whereby the rotation axis of a body undergoes gyroscopic-like motions. Libration refers to an oscillation of the figure axes of an object with respect to a fixed, mean rotation axis (see figure \ref{fig:librations}). We usually distinguish longitudinal and latitudinal librations, which are east-west and north-south oscillations, respectively. Due to the combination of long-period gravitational interaction and rotation, the shape of synchronous satellites in hydrostatic equilibrium can be represented by a triaxial ellipsoid. In the present paper, we will consider the case of a rigid mantle with a permanent tidal deformation. Figure \ref{fig:orbit} illustrates the physical mechanisms that give rise to librations. Longitudinal librations originate from the fact that, according to Kepler's laws, the orbital speed of a satellite is not constant along its (elliptical) orbit. Together with the body's tidal deformation, this results in time-dependent gravitational torques. Latitudinal librations on the other hand are related to gravitational torques due to the non-alignment of the orbital plane of a satellite and the ecliptic of the satellite-host system.

\par   
As early as the end of the nineteenth century, scientists proposed that observations of these phenomena could be used to infer the internal structure of planets and moons \cite[]{hopkins1839,kelvin1895,Poincare_precession}. More recently, observations of the precessional motion of the Earth's moon have revealed large dissipation that could be associated with a turbulent liquid core whose size is approximately 350 km \cite[]{yoder1981free,williams2001lunar}. Furthermore, Earth-based measurements of the longitudinal librations of Mercury strongly suggest that it possesses a liquid core \cite[]{margot2007}. Since the original suggestions of \cite{larmor1919}, it is believed that planetary magnetic fields originate from a dynamo process in electrically conductive liquid internal layers. For the Earth, it is well accepted that the dynamo is driven by thermochemical convection. However, most terrestrial-like planets that had or still have a self-sustained magnetic field, such as Mercury, the Moon in its early stage, and Jupiter's moon Ganymede, are unlikely to be in a convective state. Flows driven by orbital perturbations may thus provide an alternative dynamo generation mechanism \cite[]{bullard1949electromagnetic,malkus1968precession,kerswell1998tidal,lebarsNature,dwyer2011}.
\par 
Quite a few theoretical, experimental and numerical investigations have addressed the dynamic response of liquid layers resulting from orbital perturbations. Pioneering work was established by \cite{sloudsky1895rotation}, \cite{hough1895oscillations} and \cite{Poincare_precession}, who obtained an elegant uniform vorticity solution for the inviscid flow within a precessing spheroidal cavity. Note that we use the term spheroid to signify an ellipsoid that is axisymmetric with respect to the mean rotation axis. Many decades later, \cite{stewartson1963motion} and \cite{busse1968} extended this work by deriving an expression for the corrective viscous boundary layer. The solution of Busse was later rederived and experimentally verified by \cite{noir2003experimental}. \cite{kerswell1993instability}, on the other hand, investigated the stability of the Poincar\'e solution, and found that it is prone to so-called shear and elliptical instabilities. The underlying instability mechanism can be understood as a parametric resonance between two inertial modes and the strain imposed by the elliptical geometry \cite[]{kerswell2002elliptical}. 
The work of Poincar\'e was extended to triaxial ellipsoids by \cite{roberts2011} and \cite{noir2013precession}. \cite{WR2012} and \cite{Cebron_Pof} studied uniform vorticity solutions for flow driven by longitudinal libration within an ellipsoidal container. Both also addressed the dynamical stability of this solution. As in the case of precession, they found that the strain induced by the ellipsoidal geometry may give rise to elliptical instabilities. \cite{WR2012} also showed that these instabilities may drive a self-sustained dynamo process. 
\par
So far, only a few investigations have been devoted to flows forced by latitudinal libration. \cite{zhang2012} provided an analytical solution for inviscid flow in an oblate spheroidal cavity in the limit of small libration amplitude. This study also included asymptotic corrections induced by the presence of a small but finite value of the viscosity. It was shown that latitudinal libration may drive a flow, analogous to the spin-over mode. Provided that the libration frequency matches the eigenfrequency of this mode, resonance may occur, leading to a divergent flow amplitude in the limit of vanishing viscosity. These theoretical results were confirmed by \cite{chan2011simulations-2} using non-linear numerical simulations. Hence, despite the small amplitude of both orbital perturbations and tidal deformations, the resonant-like dynamics suggest that these phenomena may lead to large-amplitude flows that could contribute to the observed dissipation and magnetic field generation.  
\par
In the present study, we consider the problem of latitudinal libration in a triaxial ellipsoid. The problem is formulated in mathematical terms in section 2. Thereafter, we seek solutions of the governing equations under the assumption of uniform vorticity flow. Furthermore, we carry out extensive three-dimensional non-linear simulations to validate this hypothesis. In section 4, we investigate the linear stability of this solution, both by means of a theoretical apparatus and numerical simulations. Although the stability analysis borrows from results developed in section 3, it is possible to follow the mathematical derivation in section 4 without having read section 3 in full detail or extent. Finally, in section \ref{sec:planeto}, we will discuss our findings at planetary settings.

\section{Governing equations}
We consider motions of an incompressible fluid, enclosed within a rigid container of ellipsoidal shape, representing the mantle, that is undergoing libration in latitude. Because the mantle is considered to be rigid, and the fluid homogeneous and incompressible, the gravitational pull of the host does not induce any fluid motion within the satellite's interior, as it can be absorbed into the pressure gradient term. Hence, the flow in the liquid layer can be forced only through viscous, pressure or electromagnetic coupling with the surrounding shell. The fluid properties, such as the density $\rho$ and kinematic viscosity $\nu$, are assumed to be constant and uniform. The evolution in time $t$ of the flow velocity ${\boldsymbol u}$ and the reduced pressure $p$ are governed by the Navier-Stokes and the mass conservation equations (see equations (\ref{eq:divu})-(\ref{eq:navierstokes}) below). We also note that the orbital trajectory can be represented by a pure translation at all time, i.e. a uniform motion in space. Since the Navier-Stokes equation is invariant with respect to a translation, the acceleration resulting from the orbital  motion of the body has no effect on the dynamics of the fluid. 
\par
We express the governing equation in non-dimensional form with respect to the inverse mean rotation rate $\Omega_{0}^{-1}$ as time scale, and a furthermore unspecified characteristic length scale $R$ (e.g. one of the ellipsoid's semi-axes). At planetary settings, $R$ is typically of order 300-2000 km.  Written in a frame of reference that is attached to the walls of the container, referred to as the mantle frame, these equations read as follows:
\begin{eqnarray}
\nabla \cdot {\boldsymbol u} & = & 0,  \label{eq:divu}\\
\frac{\partial {\boldsymbol u}}{\partial t} + {\boldsymbol u}\cdot \nabla{\boldsymbol u} + 2{\boldsymbol \Omega \times {\boldsymbol u}} & = & - \nabla p + E \nabla^2 {\boldsymbol u} + {\boldsymbol r} \times \dot{\boldsymbol \Omega}. \label{eq:navierstokes} 
\end{eqnarray} 
\begin{figure}  
\begin{center}               
\includegraphics[width=0.75\textwidth]{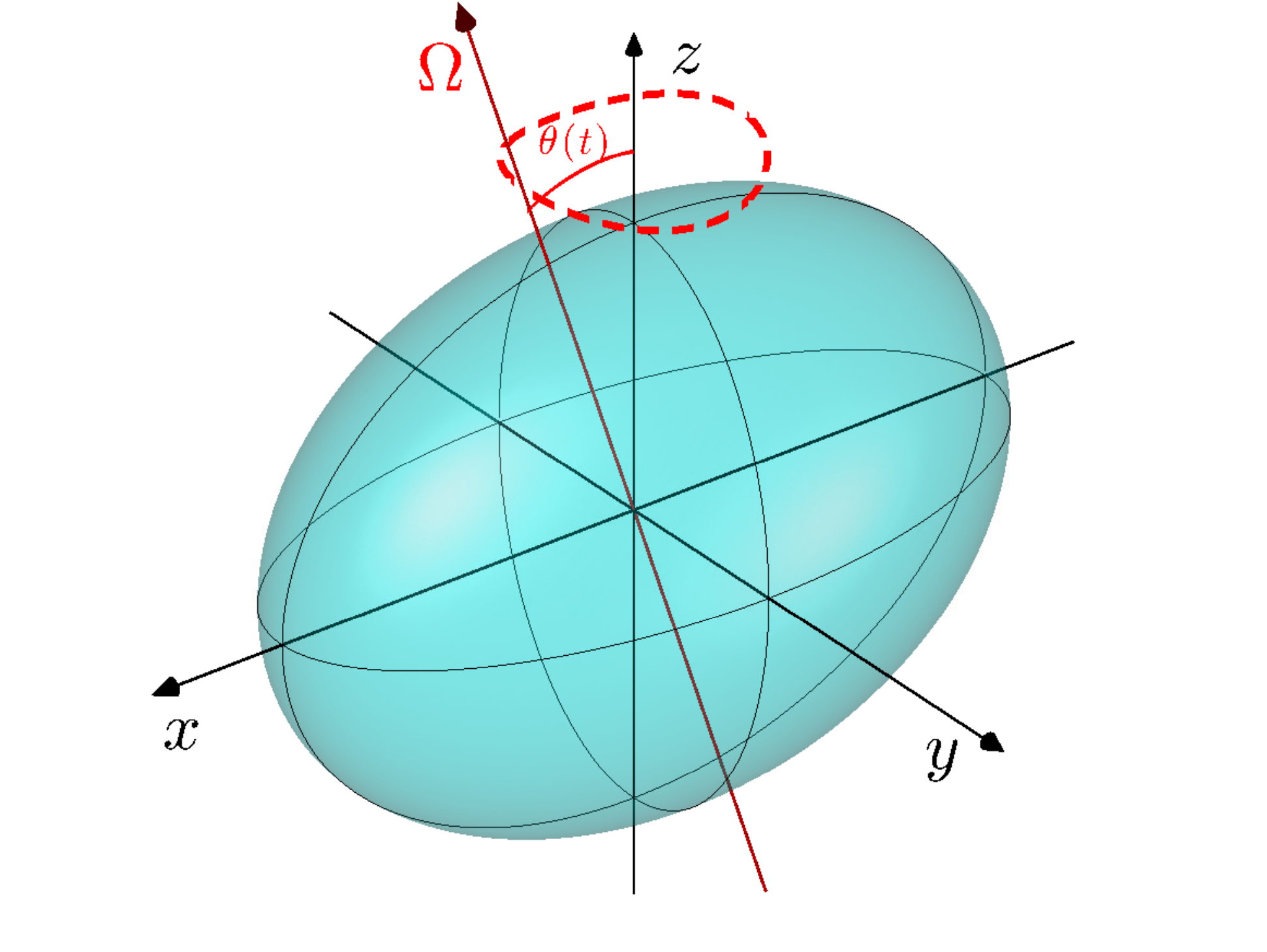}
    \caption{Sketch of the geometrical set-up as seen by an observer in the mantle frame. The dashed red line illustrates the truly calculated trajectory of the total rotation vector ${\boldsymbol \Omega}$ given by (\ref{eq:omega}).}               
    \label{fig:geometry} 
  \end{center}
\end{figure}
Here, ${\boldsymbol u}$ and ${\boldsymbol r}$ denote the flow velocity and the position vector, respectively, and ${\boldsymbol \Omega}$ is the total rotation vector. We choose the mean rotation axis to be aligned with the $z$-direction of an inertial frame of reference. The libration axis on the other hand is fixed in a frame of reference that rotates at the mean angular speed along the inertial $z$-axis, and bears the name `frame of mean rotation'. We adopt the convention that the libration axis is directed along the $x$-direction of the frame of mean rotation.
The container walls appear stationary in the mantle frame, and are given by
\begin{equation}
\frac{x^2}{a^2} + \frac{y^2}{b^2 } + \frac{z^2}{c^2} = 1.
\end{equation}
It will be useful to describe the geometry in terms of two ellipticities $\beta_{ac}$ and $\beta_{bc}$,
\begin{equation}
\beta_{ac} = \frac{a^2-c^2}{a^2+c^2},  \beta_{bc} = \frac{b^2-c^2}{b^2+c^2}. 
\end{equation}
\par
The total rotation vector in the mantle frame can be written as (see Appendix \ref{app:reference_frame})
\begin{equation}
{\boldsymbol \Omega} = (\dot \theta, \sin(\theta), \cos(\theta)), \label{eq:omega}
\end{equation}
where $\theta(t)$ is the tilt between the rotation axis and the figure axes of the container. In the case of latitudinal libration, we have
\begin{equation}
\theta(t) = \Delta {\theta} \,\sin(\omega_L t + \theta_0).  \label{eq:def_theta}
\end{equation}
In figure \ref{fig:geometry}, we have sketched a typical trajectory of ${\boldsymbol \Omega}$ as a function of time.
\par
The vector $\dot{\boldsymbol \Omega}$ has the following expression in the mantle frame:
\begin{equation}
\dot{\boldsymbol \Omega} = (\ddot{\theta}, \,\dot{\theta} \,\cos \theta, - \,\dot{\theta} \sin \theta) \label{eq:omegadot}.
\end{equation}
The momentum equations depends on two non-dimensional parameters. First, the Ekman number 
\begin{equation}
E = \frac{\nu}{\Omega_0 R^2}
\end{equation}
is a measure for the ratio between the viscous and Coriolis forces. We assume that $E^{1/2} \ll \omega_L$, which implies that no spin-up can occur during each libration cycle. 
\par
The second non-dimensional number is the Poincar\'e number
\begin{equation}
\varepsilon = \Delta \theta \,\omega_L,
\end{equation}
which measures the relative angular speed of the libration motion with respect to the mean rotation rate, and appears explicitly in the non-dimensional expression for the rotation vector ${\boldsymbol \Omega}$:
\begin{equation}
{\boldsymbol \Omega} = (\varepsilon \cos(\omega_L t), \sin \theta, \cos \theta).
\end{equation} 
The rationale for using the mantle frame is that the boundary conditions take a particularly simple form. For viscous flows, we will impose the no-slip condition, which expresses that the fluid sticks to the wall, i.e.
\begin{equation}
{\boldsymbol u} = {\boldsymbol 0}.
\label{eq:no-slip}
\end{equation}
For inviscid flows however, we cannot rule out differential motion between the fluid and the container, and can only prescribe the wall-normal component. The impermeability condition
\begin{equation}
{\boldsymbol u} \cdot \hat{ \boldsymbol n}  =  0,
\label{eq:no-penetration}
\end{equation}
with $\hat{ \boldsymbol n}$ being the outward unit normal, expresses that there is no flow across the walls of the container.
\par
The momentum equation (\ref{eq:navierstokes}) may also be expressed in terms of the vorticity ${\boldsymbol \omega} = \nabla \times {\boldsymbol u}$. The corresponding equation for ${\boldsymbol \omega}$ is
\begin{equation}
\frac{\partial {\boldsymbol \omega}}{\partial t} + \nabla \times \large[ \left( 2 \boldsymbol{\Omega} + {\boldsymbol \omega} \right) \times {\boldsymbol u} \large] =E \nabla^2 {\boldsymbol \omega} - 2 \dot{\boldsymbol \Omega}. 
\label{eq:vorticity}
\end{equation}
\par
In planetary science, we are typically concerned with parameters $\beta_{ac}, \beta_{bc}, \varepsilon, E \ll 1$. For most satellites, the main libration frequency $\omega_L$ is close to one as a a consequence of their 1:1 spin-orbit resonance. A notable exception is Mercury which is in a 3:2 resonance and has $\omega_L \approx 2/3$. Henceforth, we will always assume that $E \ll 1$, but  will not yet impose any restrictions on the values of $\beta_{ac}, \beta_{bc}, \omega_L$ and $\varepsilon$. 

\section{Laminar uniform vorticity flow} \label{sec:laminar}
\subsection{The uniform vorticity approach} 
The study of \cite{zhang2012} suggests that the laminar solution in a spheroidal container remains essentially of uniform vorticity, provided that $E \ll 1$. The same was already observed in precessing ellipsoids \cite[]{Poincare_precession,stewartson1963motion,busse1968,noir2013precession}. Along the same line of thought, we propose to seek for solutions of uniform vorticity ${\boldsymbol \omega}$. In the following, however, it will be more convenient to use the rotation rate ${\boldsymbol q}={\boldsymbol \omega}/2$ as primary variable. As such, we may write
\begin{equation}
{\boldsymbol u}= {\boldsymbol q} \times {\boldsymbol r} + \nabla \zeta, 
\label{eq:u_omega_gradphi}
\end{equation}
with $\zeta$ a gauge function that allows the velocity field ${\boldsymbol u}$ to be solenoidal and satisfy the impermeability condition (\ref{eq:no-penetration}). One can show that such a flow is an exact solution of the equations of motion (\ref{eq:divu})-(\ref{eq:navierstokes}) in the bulk of the cavity. This solution satisfies the non-penetration condition (\ref{eq:no-penetration}), but not necessarily the no-slip boundary condition (\ref{eq:no-slip}). As a consequence we expect a boundary layer to develop in the vicinity of the walls, which in turn drives a secondary flow of order $E^{1/2}$ in the interior, the so-called Ekman pumping. In the limit of small Ekman number considered in this study, a classical approach would be to neglect the boundary layer and its associated Ekman pumping. Under this assumption, the vorticity equation (\ref{eq:vorticity}) takes the inviscid form
\begin{equation}
\frac{\partial {\boldsymbol q}}{\partial t} + \nabla \times \large[ \left( \boldsymbol{\Omega} + {\boldsymbol q} \right) \times {\boldsymbol u} \large] =  -\dot{\boldsymbol \Omega}. 
\label{eq:vorticity_inviscid}
\end{equation} 
\par
Hence, the final state of the system always depends on the initial conditions. This can be avoided by reintroducing the viscosity, which requires an exact description of the Ekman layer in a triaxial ellipsoid. Such a calculation is a formidable task, and leads to cumbersome calculations. We thus choose to adopt the simpler but successful approach pioneered by \cite{noir2013precession} in the case of precessing triaxial ellipsoids. It consists of parametrizing the effect of viscous Ekman torques on the bulk flow by  adding a dissipative term in the vorticity equation (\ref{eq:vorticity_inviscid}). These torques act so as to inhibit differential rotation between the motion of the fluid and the container. This term is proportional to $E^{1/2}$ and to the differential rotation between the container and the fluid, i.e. ${\boldsymbol q} -{\boldsymbol q}_M$, where  ${\boldsymbol q}_M$ is the container rotation rate in the considered frame of reference. In the frame attached to the container,  ${\boldsymbol q}_M=\mathbf{0}$, and equation (\ref{eq:vorticity_inviscid}) simply becomes
\begin{equation}
\frac{\partial {\boldsymbol q}}{\partial t} + \nabla \times \large[ \left( \boldsymbol{\Omega} + {\boldsymbol q} \right) \times {\boldsymbol u} \large] = -\dot{\boldsymbol \Omega} -\lambda \sqrt{E}\, \boldsymbol{q}, 
\label{eq:vorticity_modified}
\end{equation}
where $\lambda>0$ parametrizes the rate of dissipation. 
In subsection \ref{subsec:numerics}, we will determine $\lambda$ by fitting the outcome of the uniform vorticity model to three-dimensional (3D) numerical simulations. As noted by \cite{noir2013precession}, the reduced model does not take secondary viscous effects (e.g. internal shear layers) into account.  
\par
Following \cite{hough1895oscillations}, \cite{roberts2011} and \cite{noir2013precession}, the uniform vorticity velocity field satisfying the non-penetration condition (\ref{eq:no-penetration}), is given by
\begin{equation}
{\boldsymbol u} = \left(\frac{2a^2z}{a^2+c^2}q_y - \frac{2a^2y}{a^2+b^2}q_z,\frac{2b^2x}{a^2+b^2}q_z - \frac{2b^2z}{b^2+c^2}q_x, \frac{2c^2y}{b^2+c^2}q_x - \frac{2c^2x}{a^2+c^2}q_y \right) \label{eq:u(q)}.
\end{equation}
Substituting (\ref{eq:u(q)}) into (\ref{eq:vorticity_modified}) yields
\begin{eqnarray}
\dot{q}_x & = & 2a^2\left( \frac{b^2-c^2}{(a^2+b^2)(a^2+c^2)}  q_y q_z +  \frac{\cos \theta}{a^2+c^2}q_y -  \frac{\sin \theta}{a^2+b^2}q_z \right) - \ddot{\theta} -\lambda \sqrt{E} q_x, \label{eq:qx}\\
\dot{q}_y & = &  2b^2 \left(\frac{c^2-a^2}{(a^2+b^2)(b^2+c^2)} q_z q_x +  \frac{\dot{ \theta}}{a^2+b^2}q_z -  \frac{\cos \theta}{b^2+c^2}q_x \right) - \dot{\theta} \cos \theta - \lambda \sqrt{E} q_y, \label{eq:qy}\\
\dot{q}_z & = &  2c^2\left(\frac{a^2-b^2}{(a^2+c^2)(b^2+c^2)} q_x q_y +  \frac{\sin \theta}{b^2+c^2}q_x -  \frac{\dot {\theta}}{a^2+c^2}q_y \right) + \dot{\theta} \sin \theta- \lambda \sqrt{E} q_z. \label{eq:qz}
\end{eqnarray}
In contrast to (\ref{eq:vorticity_modified}), this formulation of the equations of motion takes the form of a set of ordinary differential equations (ODEs), and is computationally thus far less demanding.
\subsection{Analysis in the limit of weak forcing} \label{subsec:theory}
In this section, we seek an analytical solution of  (\ref{eq:qx})-(\ref{eq:qz}) in the limit of small libration amplitude, i.e. $\varepsilon \ll 1$. We assume an asymptotic development of the form
\begin{equation}
{\boldsymbol q} = {\boldsymbol q}^{(0)} + \varepsilon {\boldsymbol q}^{(1)} + \mathcal{O}(\varepsilon^2). 
\end{equation}
We now substitute this ansatz into (\ref{eq:qx})-(\ref{eq:qz}), and successively analyse for increasing orders of $\varepsilon$. Starting at order 0, and invoking that $\cos \theta  = 1 - \mathcal{O}(\varepsilon^2)$, $\sin \theta = \mathcal{O}(\varepsilon)$ and $\dot{\theta} = \varepsilon \cos(\omega_L t) = \mathcal{O}(\varepsilon)$, $\ddot{\theta}=-\omega_L \varepsilon \sin(\omega_L t) = \mathcal{O}(\varepsilon)$, we find
\begin{eqnarray}
\dot{q}^{(0)}_x & = &  2a^2 \frac{b^2-c^2}{(a^2+b^2)(a^2+c^2)}  q^{(0)}_z q^{(0)}_y + \frac{2a^2}{a^2+c^2}q^{(0)}_y  - \lambda \sqrt{E} q^{(0)}_x \label{eq:qx0},\\
\dot{q}^{(0)}_y & = &  2b^2\frac{c^2-a^2}{(a^2+b^2)(b^2+c^2)}  q^{(0)}_x q^{(0)}_z -  \frac{2b^2}{b^2+c^2}q^{(0)}_x - \lambda \sqrt{E} q^{(0)}_y, \label{eq:qy0}\\
\dot{q}^{(0)}_z & = &  2c^2 \frac{a^2-b^2}{(a^2+c^2)(b^2+c^2)}  q^{(0)}_x q^{(0)}_y - \lambda \sqrt{E} q^{(0)}_z \label{eq:qz0}.
\end{eqnarray}
To analyse this system, we introduce  the Lyapunov function $F$ \cite[]{manneville},
 \begin{equation}
F = \left (bc {q}^{(0)}_x \right)^2 + \left (ac {q}^{(0)}_y\right)^2 +  \left (ab {q}^{(0)}_z\right)^2.
 \end{equation}
Using  (\ref{eq:qx0})-(\ref{eq:qz0}) and after some algebra, we obtain
\begin{equation}
\dot{F} =-\lambda \sqrt{E} F, 
\end{equation} 
which is strictly negative definite. From this, it follows immediately that the solution of (\ref{eq:qx0})-(\ref{eq:qz0}) tends to ${\boldsymbol q}^{(0)} = {\boldsymbol 0}$ for any initial condition.
 \par 
We now proceed with our analysis to first order in $\varepsilon$. Using the solution ${\boldsymbol q}^{(0)} = {\boldsymbol 0}$,  we find
\begin{eqnarray}
\dot{q}^{(1)}_x & = & \frac{2a^2}{a^2+c^2}q^{(1)}_y - \lambda \sqrt{E} q^{(1)}_x +\omega_L \sin (\omega_L t) \label{eq:qx1},\\
\dot{q}^{(1)}_y & = & -\frac{2b^2}{b^2+c^2}q^{(1)}_x - \lambda \sqrt{E} q^{(1)}_y- \cos (\omega_L t)  \label{eq:qy1},\\
\dot{q}^{(1)}_z & = & - \lambda \sqrt{E} q^{(1)}_z \label{eq:qz1}.
\end{eqnarray}
As a first step, we consider the inviscid case $E=0$, for which (\ref{eq:qx1})-(\ref{eq:qz1}) has the particular integral
\begin{eqnarray}
q^{(1)}_x & =& \frac{f_0^2 - \omega_L^2}{\omega_L^2 - f^2} \cos (\omega_L t), \label{eq:sol_qx}\\
q^{(1)}_y & = & \frac{\omega_L \beta_{bc} }{\omega_L^2 - f^2 }\sin (\omega_L t), \label{eq:sol_qy} \\
q^{(1)}_z & = & 0 \label{eq:sol_qz},
\end{eqnarray}
with
\begin{eqnarray}
 f   \equiv & \pm \displaystyle \sqrt{\frac{4a^2b^2}{(a^2+c^2)(b^2+c^2)}}  & =   \pm \sqrt{(1+\beta_{ac})(1+\beta_{bc})}, \label{freq:spin_over}  \\
 f_0   \equiv & \displaystyle  \sqrt{\frac{2a^2}{a^2+c^2}}  & =  \pm \sqrt{1 + \beta_{ac}}.
\end{eqnarray}
The most remarkable feature of this solution is the singularity that occurs for $\omega_L = f$, where $f$ is the eigenfrequency of the so-called spin-over mode \cite[]{vantieghem2014inertial}. This suggests that there is a close connection between the present problem and the theory of inertial modes. In order to substantiate this, we first recast (\ref{eq:qx1})-(\ref{eq:qz1}) with $E=0$ in matrix-vector form:
\begin{equation}
\left(\begin{array}{c} \dot{q}^{(1)}_x \\ \dot{q}^{(1)}_y \\ \dot{q}^{(1)}_z \end{array} \right) - 
\left(
\begin{array}{ccc}
0 & \frac{2a^2}{a^2+c^2} & 0 \\
-\frac{2b^2}{b^2+c^2} & 0 & 0 \\
0 & 0 & 0
\end{array}
\right)
\left(
\begin{array}{c}
q^{(1)}_x \\ q^{(1)}_y \\ q^{(1)}_z
\end{array}
\right)
=
\left(
\begin{array}{c}
\omega_L \sin(\omega_L t) \\ -\cos(\omega_L t) \\ 0
\end{array}
\right) \label{eq:oscillator}
\end{equation}
This notation emphasizes that the dynamical system that we study is simply a harmonic oscillator driven by the Poincar\'e force. Resonance, i.e. unbounded growth of the flow, takes place if the driving frequency $\omega_L$ matches one of the natural frequencies $\mu$ of the unforced system. These are solutions of the eigenvalue problem
\begin{equation}
\left(
\begin{array}{ccc}
0 & \frac{2a^2}{a^2+c^2} & 0 \\
-\frac{2b^2}{b^2+c^2} & 0 & 0 \\
0 & 0 & 0
\end{array}
\right)
\left(
\begin{array}{c}
q^{(1)}_x \\ q^{(1)}_y \\ q^{(1)}_z
\end{array}
\right) = \mathrm{i} \mu \left( \begin{array}{c}
q^{(1)}_x \\ q^{(1)}_y \\ q^{(1)}_z
\end{array}
\right), \label{eq:eigenvalue}
\end{equation}
where $\mathrm{i}$ is the imaginary unit. This yields $\mu_{1,2} = \pm f$ and $\mu_{3} = 0$. Following \cite{vantieghem2014inertial} and using (\ref{eq:u(q)}), (\ref{eq:eigenvalue}) is the projection of the inertial mode equation in vorticity formulation, 
\begin{equation}
\nabla \times \left( \hat{\boldsymbol z} \times {\boldsymbol u} \right) = \mathrm{i}\mu {\boldsymbol q} \label{eq:inertialmode},
\end{equation} 
on the subspace of solenoidal uniform vorticity flows that satisfy the impermeability condition (\ref{eq:no-penetration}). The eigenmodes associated with $\mu_{1,2}= \pm f$ are associated with the so-called \textit{spin-over mode}, i.e. an inertial mode whose vorticity is uniform and located in the equatorial plane. Resonant coupling is not possible for the eigenvalue $\mu_3=0,$ since its eigenvector  $(0,0,1)$ is orthogonal to the right-hand side of (\ref{eq:oscillator}).  Summarizing, the solution (\ref{eq:sol_qx})-(\ref{eq:sol_qz}) embodies that latitudinal libration drives a spin-over mode, and that resonance takes place if the libration frequency matches the spin-over frequency. The linearized uniform vorticity theory thus confirms the inviscid theory of \cite{chan2011simulations-2} and \cite{zhang2012}, and extends it to triaxial ellipsoids.
\par 
For the biaxial ellipsoid with $b=c$, i.e. $\beta_{bc}=0$, the solution (\ref{eq:sol_qx})-(\ref{eq:sol_qz}) reduces after some algebra to
\begin{eqnarray}
q^{(1)}_x & =& -\cos(\omega_L t), \\
q^{(1)}_y & = & 0, \\
q^{(1)}_z & = & 0,
\end{eqnarray}
for any value of $\omega_L$ and does not exhibit resonant behaviour.  In the frame of mean rotation (denoted by subscript $R$), this yields
\begin{equation}
{\boldsymbol q}^{(1)}_{R} = {\boldsymbol q^{(1)}} + \varepsilon \cos(\omega_L t)\hat{\boldsymbol x} = {\boldsymbol 0}.
\end{equation}
This shows that the motion of the container cannot be transmitted to the fluid in absence of container topography in planes perpendicular to the libration axis.
\par
A further interesting choice is $\omega_L = f_0$, which cancels $q^{(1)}_x$. This implies that the streamlines of this flow are ellipses located in planes $y=cst$. For  $a=c$, i.e. $\beta_{ac}=0$,  this situation occurs if $\omega_L=1$, and is characterized by circular streamlines.
\par
To fix the non-physical behaviour at resonance, we return to the viscous case $(E \neq 0)$. It is still possible to work out a particular solution for (\ref{eq:qx1})-(\ref{eq:qy1}), although it is more intricate:
\begin{eqnarray}
q^{(1)}_x & =& \frac{(\omega_L^2-f^2)(f_0^2-\omega_L^2)+\mathcal{O}(E)}{(\omega_L^2 -f^2)^2 + 2\lambda^2 E(\omega_L^2+f^2) + \lambda^4 E^2 } \cos(\omega_L t) \nonumber \\
& + & \lambda \sqrt{E} \frac{  \omega_L(\omega_L^2+f^2-2f_0^2)+\mathcal{O}(E)}{(\omega_L^2 -f^2)^2  + 2\lambda^2 E(\omega_L^2+f^2) + \lambda^4E^2}\sin(\omega_L t) , \label{eq:sol_qx_visc} \\
q^{(1)}_y & = & \beta_{bc} \frac{\omega_L(\omega_L^2-f^2)+ \mathcal{O}(E)}{ (\omega_L^2 -f^2)^2 + 2\lambda^2E(\omega_L^2+f^2) + \lambda^4E^2} \sin(\omega_L t) \nonumber \\ & - & \lambda \sqrt{E}\beta_{bc} \frac{\omega_L^2 +f^2 -4\omega_L^2b^2/(b^2+c^2) + \mathcal{O}(E)}{ (\omega_L^2 -f^2)^2 + 2\lambda^2E(\omega_L^2+f^2) +\lambda^4 E^2 } \cos(\omega_L t) \label{eq:sol_qy_visc}, \\
q^{(1)}_z & = & 0. \label{eq:sol_qz_visc}
\end{eqnarray}
We can now distinguish between two cases, depending on whether the first term dominates the second one in the denominator of (\ref{eq:sol_qx_visc})-(\ref{eq:sol_qy_visc}) or vice versa. Assuming that $\omega_L, f, \lambda = \mathcal{O}(1)$, the former occurs if $|\omega_L - f| \gg \sqrt{E}$, i.e. for libration frequencies far away from resonance. The first term in each of (\ref{eq:sol_qx_visc}) and (\ref{eq:sol_qy_visc}) is then of order of magnitude 1, whereas the second ones are $\displaystyle \mathcal{O}(\sqrt{E})$. For $E \rightarrow 0$, we recover the inviscid solution (\ref{eq:sol_qx})-(\ref{eq:sol_qy}). 
\par
For $|\omega_L - f| \ll \sqrt{E}$, the first term in (\ref{eq:sol_qx_visc}) and (\ref{eq:sol_qy_visc}) is of order of magnitude 1, as the leading-order contributions in their numerator and denominator both scale as $\mathcal{O}(E)$. The second term in (\ref{eq:sol_qx_visc}) and (\ref{eq:sol_qy_visc}) now scales as $E^{-1/2}$, and thus dominates the first term. We find thus that for $E \ll 1$, the total solution scales as $E^{-1/2}$ and is established within a frequency window of width $\mathcal{O}(E^{1/2})$. These scalings are consistent with \cite{zhang2012}. Readers whose main interest is in the dynamical stability of the solutions derived above, can now skip the remainder of this section without great loss.
\begin{figure}                 
  \begin{center}
  \setlength{\epsfysize}{5.0cm}

    \begin{tabular}{ccc}
      \setlength{\epsfysize}{5.0cm}
      \subfigure[]{\epsfbox{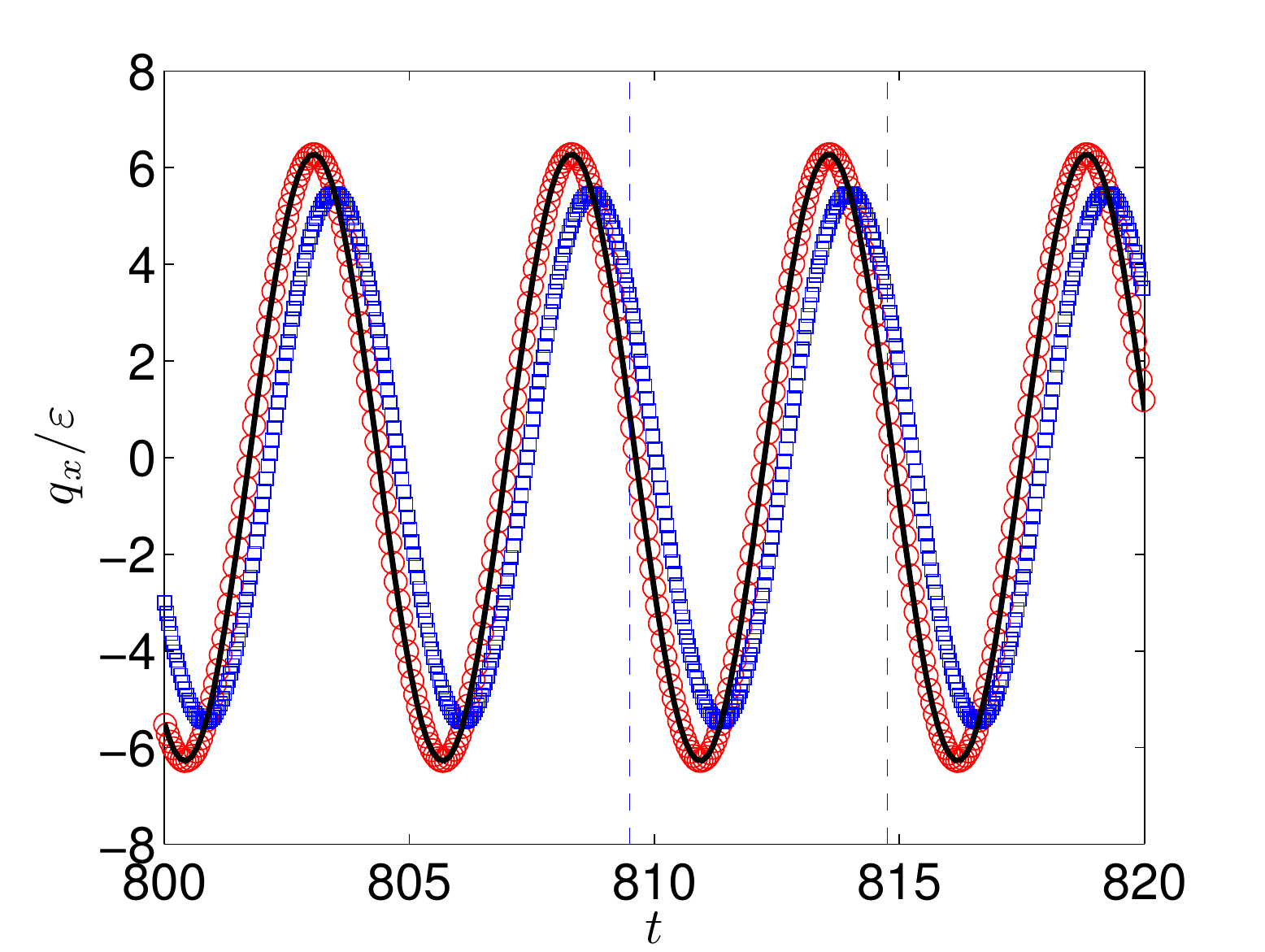}}  &
      \setlength{\epsfysize}{5.0cm}
      \subfigure[]{\epsfbox{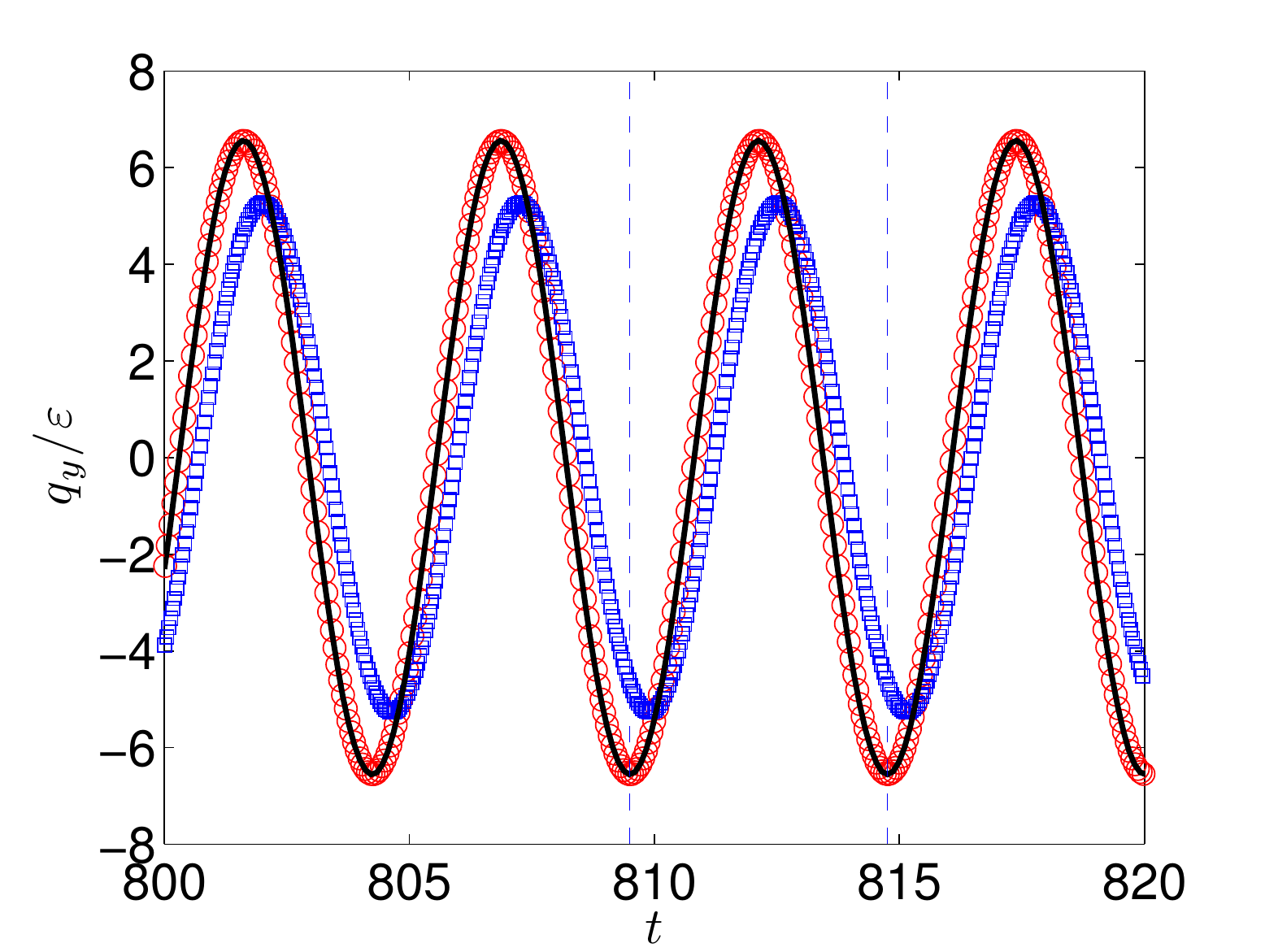}} &    \\  
      \setlength{\epsfysize}{5.0cm}
      \subfigure[]{\epsfbox{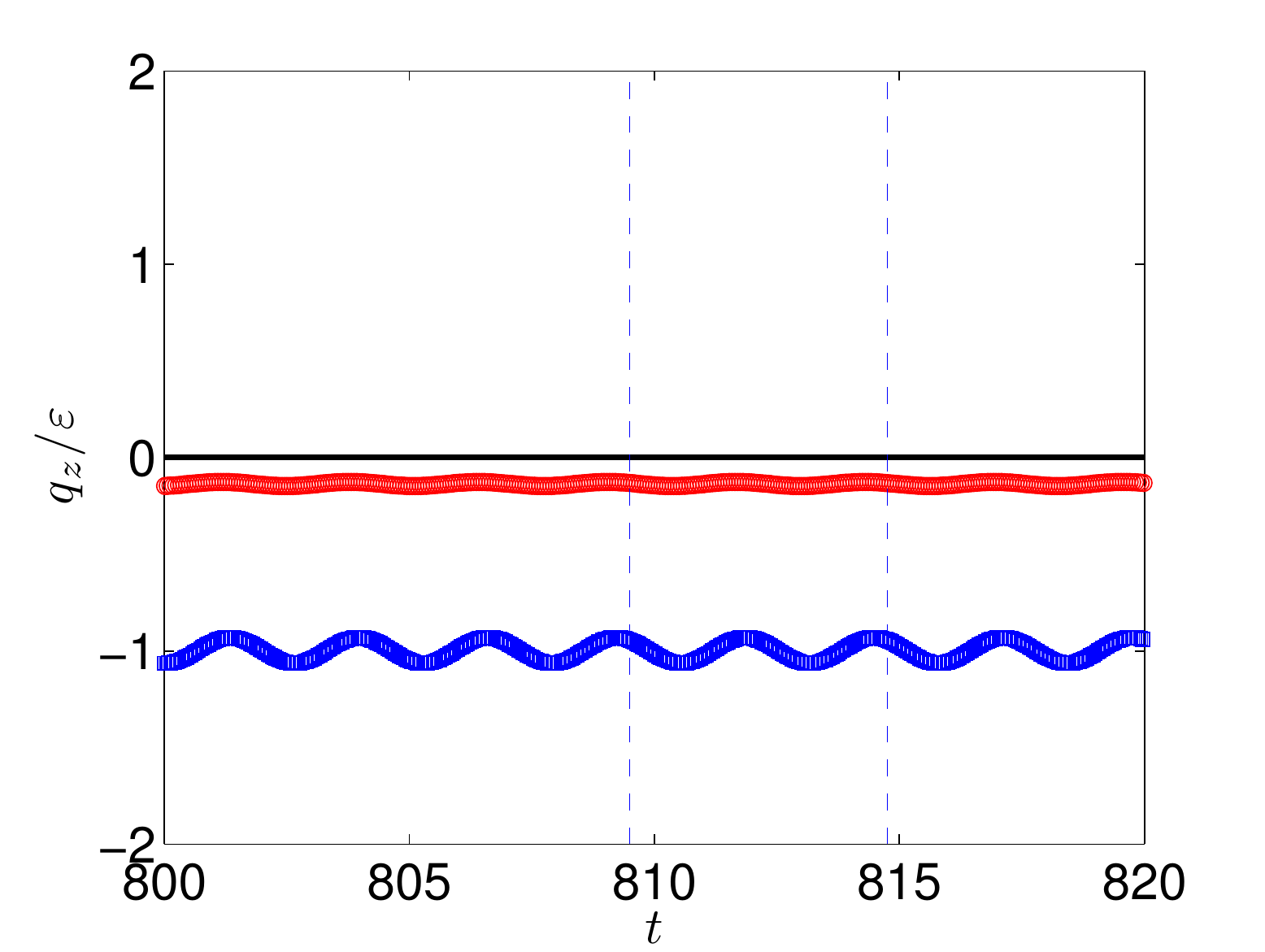}} &
      \setlength{\epsfysize}{5.0cm}
      \subfigure[]{\epsfbox{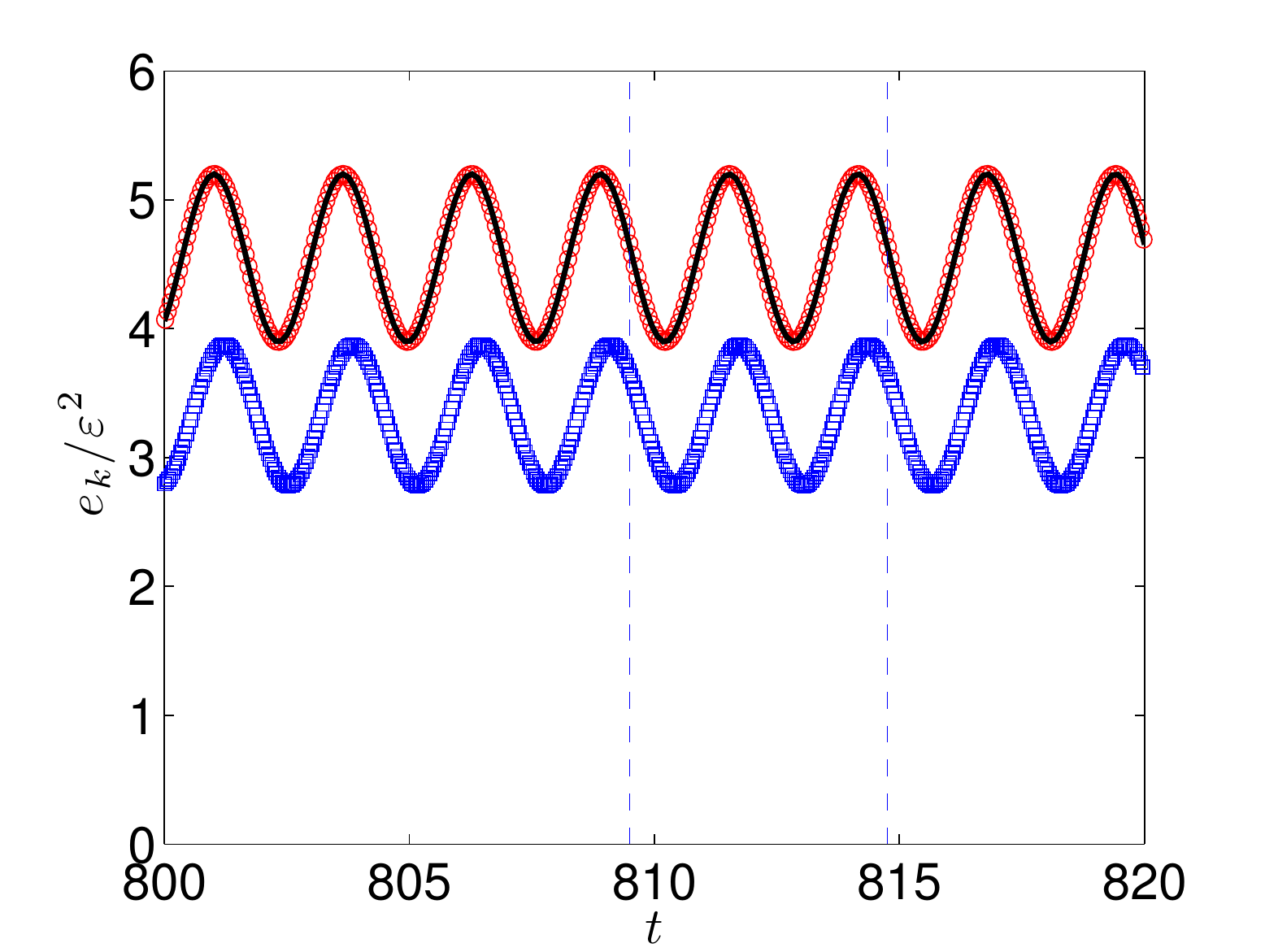}} 
    \end{tabular}
    \caption{Comparison between the exact solution (\ref{eq:sol_qx_visc})-(\ref{eq:sol_qy_visc}) of the linearized ODEs (black, solid line) and direct integration of the non-linear ODEs (\ref{eq:qx})-(\ref{eq:qz}) for $\varepsilon=0.005$ (red circles) and $0.05$ (blue squares) for  $(a,b,c)=(1,\sqrt{5/4},\sqrt{3/4})$ and $\omega_L=f=\sqrt{10/7}$ The distance between the dashed vertical lines spans one libration cycle $T=2\pi/\omega_L$. }
         \label{fig:linear_ode_triax}                
  \end{center}
\end{figure}
\par
Anticipating the numerical validation of our analysis, we introduce $e_k$ and $\overline{e_k}$, the mean kinetic energy density and its time average, which can be readily computed from (\ref{eq:u(q)}), and read
\begin{eqnarray} 
e_k & = & \frac{1}{2V} \iiint_{V} {\boldsymbol u^2}\,\mathrm{d}V, \label{eq:energy_density_def} \\
        & = & \frac{2}{5} \left(\frac{b^2c^2}{b^2+c^2}q_x^2 + \frac{a^2c^2}{a^2+c^2}q_y^2 + \frac{a^2b^2}{a^2+b^2}q_z^2\right),  \label{eq:energy_density}\\
\overline{e_k} & = & \frac{1}{T}\int_{t_0}^{t_0+T} e_k \,\mathrm{d}t, \label{eq:energy_density_time_avg}
\end{eqnarray}
where $V = 4a b c \pi/3$ is the volume of the ellipsoid, and $T = 2 \pi/\omega_L$ the libration period.
Upon substitution of (\ref{eq:sol_qx_visc})-(\ref{eq:sol_qy_visc}) into this expression, we obtain the following result for resonant flows:
\begin{equation}
\overline{e_k}  =  \varepsilon^2 \left( \frac{1}{10\lambda^2 E} \right. \frac{a^2c^2}{(a^2+c^2)^2} \underbrace{\frac{(b^2-c^2)^2}{(b^2+c^2)^2}} _{\beta_{bc}^2}+ \left. \frac{1}{160}\frac{(3b^2+c^2)^2c^2}{(b^2+c^2)b^2}+ \mathcal{O}(E) \right). \label{eq:energy_density_final}
\end{equation}
The second term in $\mathcal{O}(E)$ may dominate if $\beta_{bc} \ll \sqrt{E}$. In particular, if $\beta_{bc}=0$, the expression for $\overline{e_k}$ for resonant and non-resonant flows is
\begin{eqnarray}
\overline{e_k}& = &\varepsilon^2 \left(\frac{c^2}{10} + \mathcal{O}(E)\right), \hspace{1cm} |\omega_L - f| \gg \sqrt{E}, \label{eq:ek_biax1}  \\ 
\overline{e_k}& = & \varepsilon^2 \left( \frac{c^2}{20} + \mathcal{O}(E) \right) , \hspace{1cm} |\omega_L - f| \ll \sqrt{E} . \label{eq:ek_biax2}
\end{eqnarray}
For an observer in the frame of mean rotation, $\overline{e_k}$ for $\beta_{bc}=0$ is then given by
\begin{eqnarray}
\overline{e_k}&=&\mathcal{O}(\varepsilon^2 E), \hspace{2.5cm} |\omega_L - f| \gg \sqrt{E},   \\ 
\overline{e_k}&=&\varepsilon^2 \left( \frac{c^2}{20} + \mathcal{O}(E) \right), \hspace{1cm} |\omega_L - f| \ll \sqrt{E}.
\end{eqnarray}
This indicates that, in the absence of topography (i.e. for $\beta_{bc}=0$), viscosity can transmit the librating motion of the mantle to the fluid. However, the flow becomes vanishingly weak for non-resonant frequencies as $E \rightarrow 0$. At resonance, the amplitude of the flow is independent of $E$, and thus $\mathcal{O}\left(E^{-1/2}\right)$ times stronger than the non-resonant flow.  Motivated by the experimental study of \cite{aldridge1969axisymmetric}, \cite{zhang2013} have argued that similar scalings hold for the case of longitudinal libration of a spherical container. 
\par
We recall that the above analytical solutions have been obtained under the \textit{a priori} assumption that the non-linear terms in the equation of motion remain negligible with respect to the linear ones. \textit{A posteriori}, we find that this is indeed established off-resonance, given that the flow scales as $\varepsilon$. The amplitude of the resonant flow, however, scales as $ \varepsilon \beta_{bc} E^{-1/2}$, and the non-linear terms will therefore be negligible if $\varepsilon \beta_{bc} E^{-1/2} \ll 1$.
\par
In figure \ref{fig:linear_ode_triax}, we compare the solution (\ref{eq:sol_qx_visc})-(\ref{eq:sol_qz_visc}) of the linearized ODEs to numerical solutions of the non-linear system (\ref{eq:qx})-(\ref{eq:qz}) for $(a,b,c)=(1,\sqrt{5/4},\sqrt{3/4})$. The libration frequency $\omega=\sqrt{10/7}$ is chosen such that resonance occurs; the value of $\lambda$ is set to $2.7$ and the Ekman number to $5 \cdot 10^{-5}$. This choice of parameters is representative of the numerical results that will be discussed in subsection \ref{subsec:numerics}. The non-linear solutions for $q_x$, $q_y$ and $e_k$ virtually collapse with their linear counterparts at $\varepsilon=0.005$. For $\varepsilon=0.05$, the amplitudes of $q_x$ and $q_y$ are approximately 10-20 $\%$ lower than predicted by the linear theory. We see furthermore that a non-zero retrograde solution for the vorticity component $q_z$ emerges when $\varepsilon$ increases; this feature is generated by non-linear interactions between the components of the linear solutions ${\boldsymbol q}^{(1)}$. Indeed, in this case the quantity $\varepsilon \beta_{bc} E^{-1/2} \approx 1.77$, and thus nonlinearities are no longer negligible, as argued in the previous paragraph. The $q_z$-component consists of the superposition of a non-zero mean, referred to as a \textit{zonal flow} \cite[]{busse2010mean,sauret2010experimental} and a harmonic modulation of frequency $2 \omega_L$. However, a complete investigation of the zonal flow requires us to take into account higher-order effects, such as non-linear interactions in the viscous Ekman layer. Such a detailed study would involve an exact description of the Ekman boundary layer and will not be pursued here. 
\subsection{Numerical validation and discussion} \label{subsec:numerics}
We now wish to validate our model of uniform vorticity flow. To this end, we will compare the results presented above against 3D non-linear numerical simulations obtained by means of a non-structured finite-volume code \cite[]{Vantieghem_phd}. It uses a collocated arrangement of the variables, and a second-order centred-finite-difference-like discretization stencil for the spatial differential operators. The time advancement algorithm is based on a canonical fractional-step method \cite[]{kim1985}.  More specifically, the procedure to compute the velocity and reduced pressure ${\boldsymbol u}^{j+1},p^{j+1}$ at time step $t^{j+1}=t^{j}+\Delta t$, given the respective variables at time step $t^j$, is as follows.
\begin{enumerate} 
\item We first solve the intermediate velocity ${\boldsymbol u}^{\star}$ from the equation  
\begin{eqnarray}
\frac{{\boldsymbol u}^{\star} - {\boldsymbol u}^{j}}{\Delta t} & = &  - {\boldsymbol u}^{j+1/2}_{AB} \cdot \nabla {\boldsymbol u}^{j+1/2}_{CN} - \nabla p^{j} \\
&& - 2  {\boldsymbol \Omega}^{j+1/2} \times {\boldsymbol u}^{j+1/2} + E \, \nabla^2 {\boldsymbol u}^{j+1/2}_{CN} +\dot{\boldsymbol \Omega}^{j+1/2}\times {\boldsymbol r} \nonumber, 
\end{eqnarray}
with no-slip boundary condition ${\boldsymbol u}^{\star}_{bnd} = {\boldsymbol 0}$. In this expression, ${\boldsymbol u}^{j+1/2}_{AB}$ and ${\boldsymbol u}^{j+1/2}_{CN}$ denote the velocity at time step $j+1/2$ computed using a second-order Adams-Bashforth, respectively Crank-Nicolson approach, i.e.
\begin{eqnarray}
{\boldsymbol u}^{j+1/2}_{AB} & = &\frac{3}{2}{\boldsymbol u}^{j} - \frac{1}{2} {\boldsymbol u}^{j-1}, \\ 
{\boldsymbol u}^{j+1/2}_{CN}& = & \frac{1}{2}({\boldsymbol u}^{j} + {\boldsymbol u}^{\star}). 
\end{eqnarray}
The mixed Adams-Bashforth/Crank-Nicolson formulation for the advective term has the advantage of being kinetic-energy conserving and time-stable for any $\Delta t$, and it does not require the solution of a non-linear system for the unknown ${\boldsymbol u}^{\star}$.
\item The new velocity ${\boldsymbol u}^{j+1}$ is then related to ${\boldsymbol u}^{\star}$ by
\begin{equation}
 {\boldsymbol u}^{j+1}={\boldsymbol u}^{\star} - \Delta t \left( \Delta p^{j+1} \right),
 \end{equation}
with $\Delta p^{j+1} = p^{j+1} - p^{j}$. Imposing the incompressibility constraint on ${\boldsymbol u}^{j+1}$ leads to a Poisson equation for  $\Delta p^{j+1}$,
\begin{equation}
\nabla^2 \left( \Delta p^{j+1} \right) = (\Delta t)^{-1} \nabla \cdot {\boldsymbol u}^{\star},
\end{equation}
with boundary condition $\hat{\boldsymbol n} \cdot \nabla \left( \Delta p^{j+1} \right) = 0$. This Poisson equation is solved with the algebraic multigrid method BoomerAMG \cite[]{boomeramg}. Discussions regarding the performance of this code can be found in \cite{marti2014} and \cite{cebron2014multipolar}.
\end{enumerate}
\par
The computational grid consists of a mixture of tetrahedral, prismatic and hexahedral elements. The total number of control volumes (CV's) at the lowest Ekman number, $E=5\cdot 10^{-5}$, is approximately 2.3 million. This corresponds to a typical CV size $\Delta x$ of about 0.012. In the bulk of the mesh, the CVs are quasi-isotropic. Near the boundaries however, they become more anisotropic as we reduce the wall-normal spacing to $\Delta x \approx 0.0012 \lesssim 0.2 E^{1/2}$ in order to resolve the viscous Ekman layers.  
\par
In the following, we will discuss numerical results for four different geometries, which we label with Roman numerals I-IV, and whose properties are summarized in Table \ref{tab:geometry}. Case I is exactly the same geometry as studied by \cite{zhang2012} so as to allow direct comparison between both theories. In case II, we have a biaxial ellipsoid with $b=c$ for which no resonance can occur. Case III is also a biaxial ellipsoid, but now with $a=c$. Finally, case IV is a triaxial ellipsoid with three different semi-major axes, and will allow us to demonstrate the generality of the present theoretical and numerical approach.
\begin{table}
\begin{center}
\begin{tabular}{ c c c c c c c c c}
\hline \hline
label & $a$ & $b$ & $c$ & $\beta_{ac}$ & $\beta_{bc}$ & $f$ & $\lambda_{asym}$ & $\lambda_{num}$ \\ \hline
I & 1 & 1 & $\sqrt{3/4}$& 1/7 & 1/7 & $8/7$& $2.74 - 0.493\mathrm{i}$ & 2.81 \\ 
II & $\sqrt{3/4}$& 1 & 1 & -1/7  & 0 & $\sqrt{6/7}$ & $2.76 - 0.155\mathrm{i}$ & - \\
III & 1 & $\sqrt{5/4}$ & 1 &0& 1/9 & $\sqrt{10/9}$ &  $2.53 - 0.322 \mathrm{i}$ & 2.61 \\
IV & $1$ & $\sqrt{5/4}$ &$\sqrt{3/4}$& 1/7 & 1/4& $\sqrt{10/7}$ & $2.63 - 0.552\mathrm{i}$ & 2.72 \\
\hline
\hline

\end{tabular} 
\caption{Characteristics of numerically investigated geometries. The theoretical value of the decay rate of the spin-over mode $\lambda_{asym}$ has been computed using a procedure outlined in \cite{vantieghem2014inertial}. No inversion for $\lambda_{num}$ was carried out for geometry II because $\overline{e_k}$ is independent of $\lambda$ according to (\ref{eq:ek_biax1})-(\ref{eq:ek_biax2}).}
\label{tab:geometry}
\end{center}
\end{table}
\par
\begin{figure}                 
  \begin{center}
  \setlength{\epsfysize}{5.0cm}

    \begin{tabular}{ccc}
      \setlength{\epsfysize}{5.0cm}
      \subfigure[]{\epsfbox{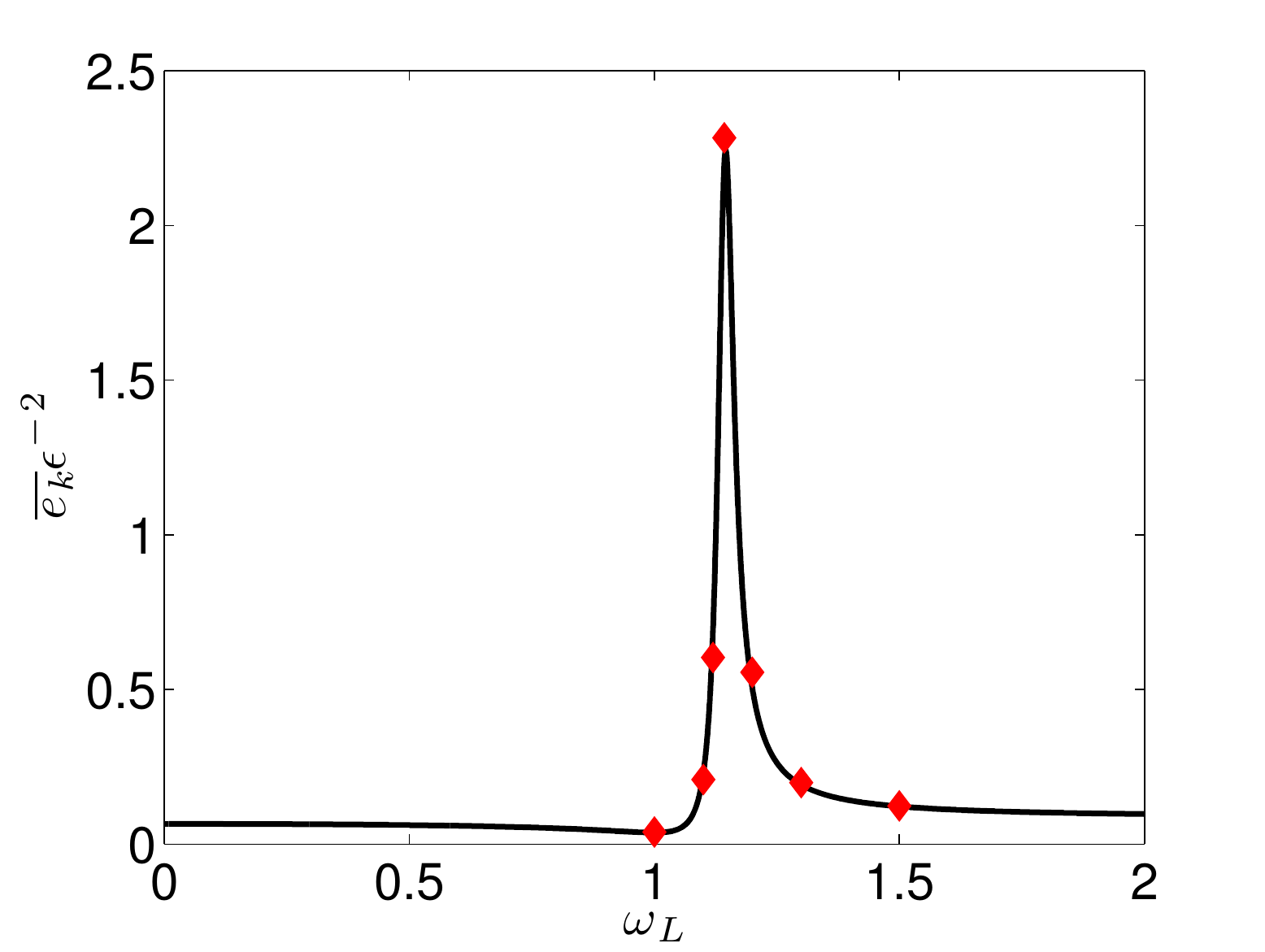}}  &
      \setlength{\epsfysize}{5.0cm}
      \subfigure[]{\epsfbox{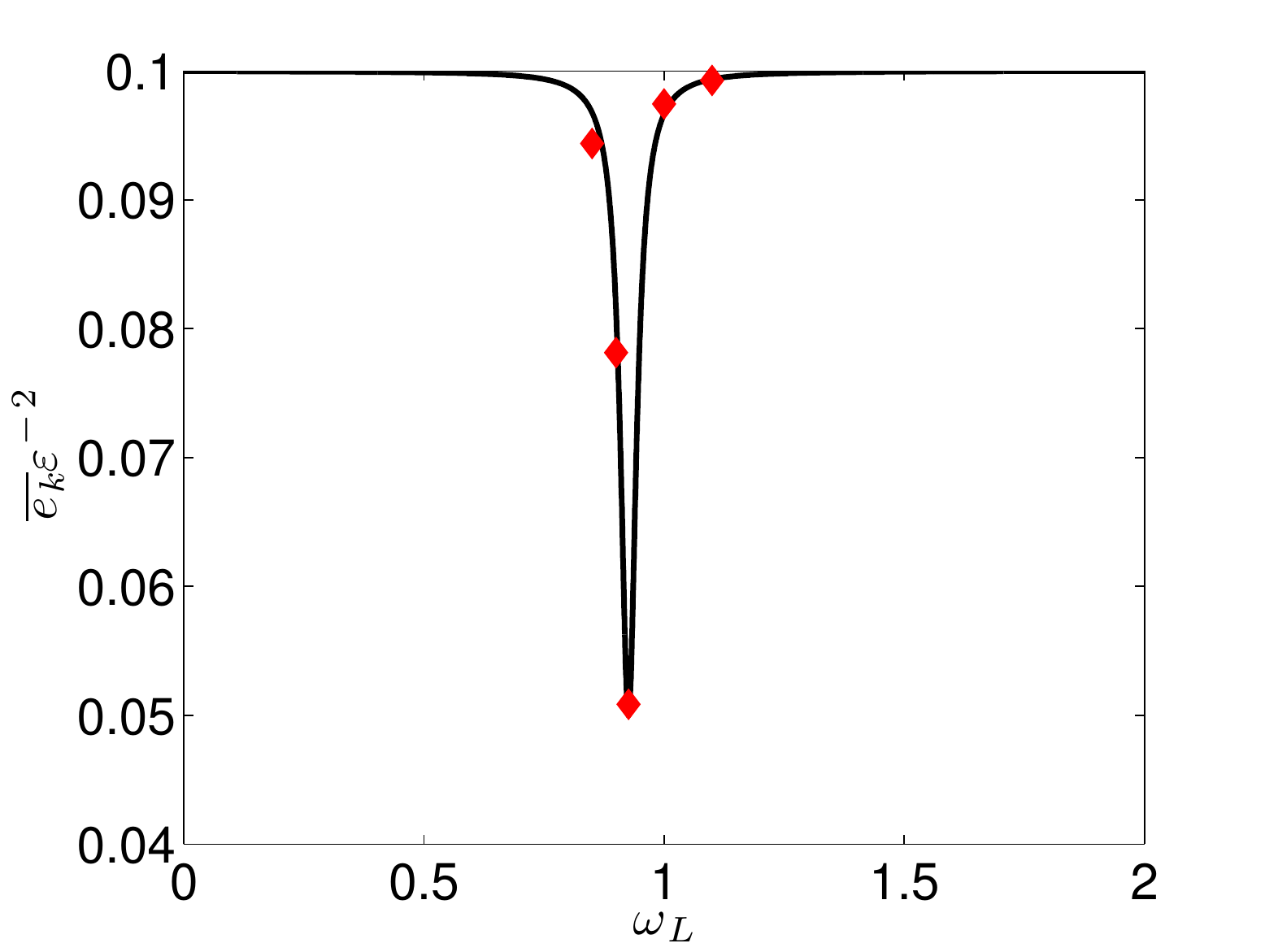}}  \\
      \setlength{\epsfysize}{5.0cm}
      \subfigure[]{\epsfbox{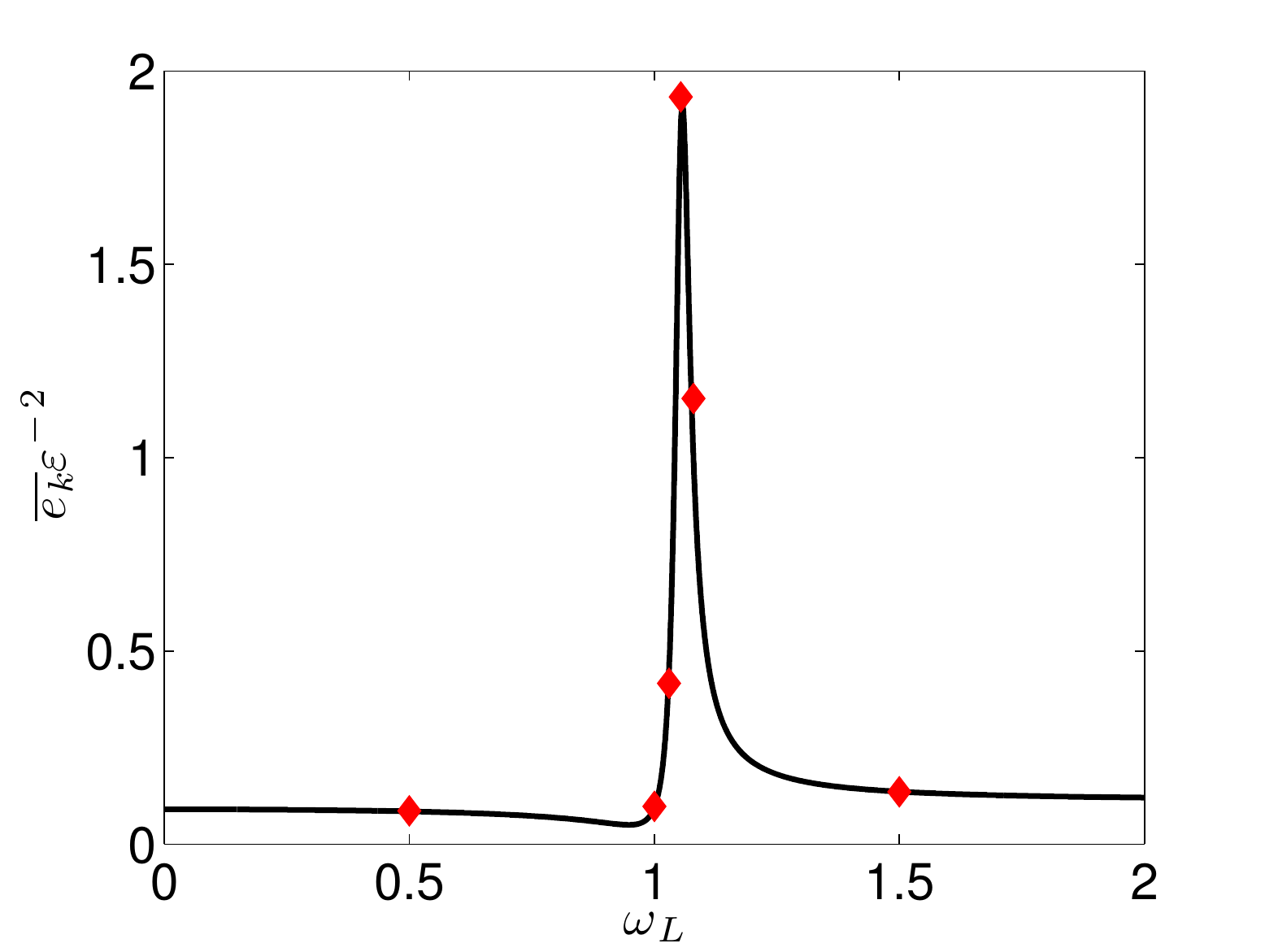}}  &
      \setlength{\epsfysize}{5.0cm}
      \subfigure[]{\epsfbox{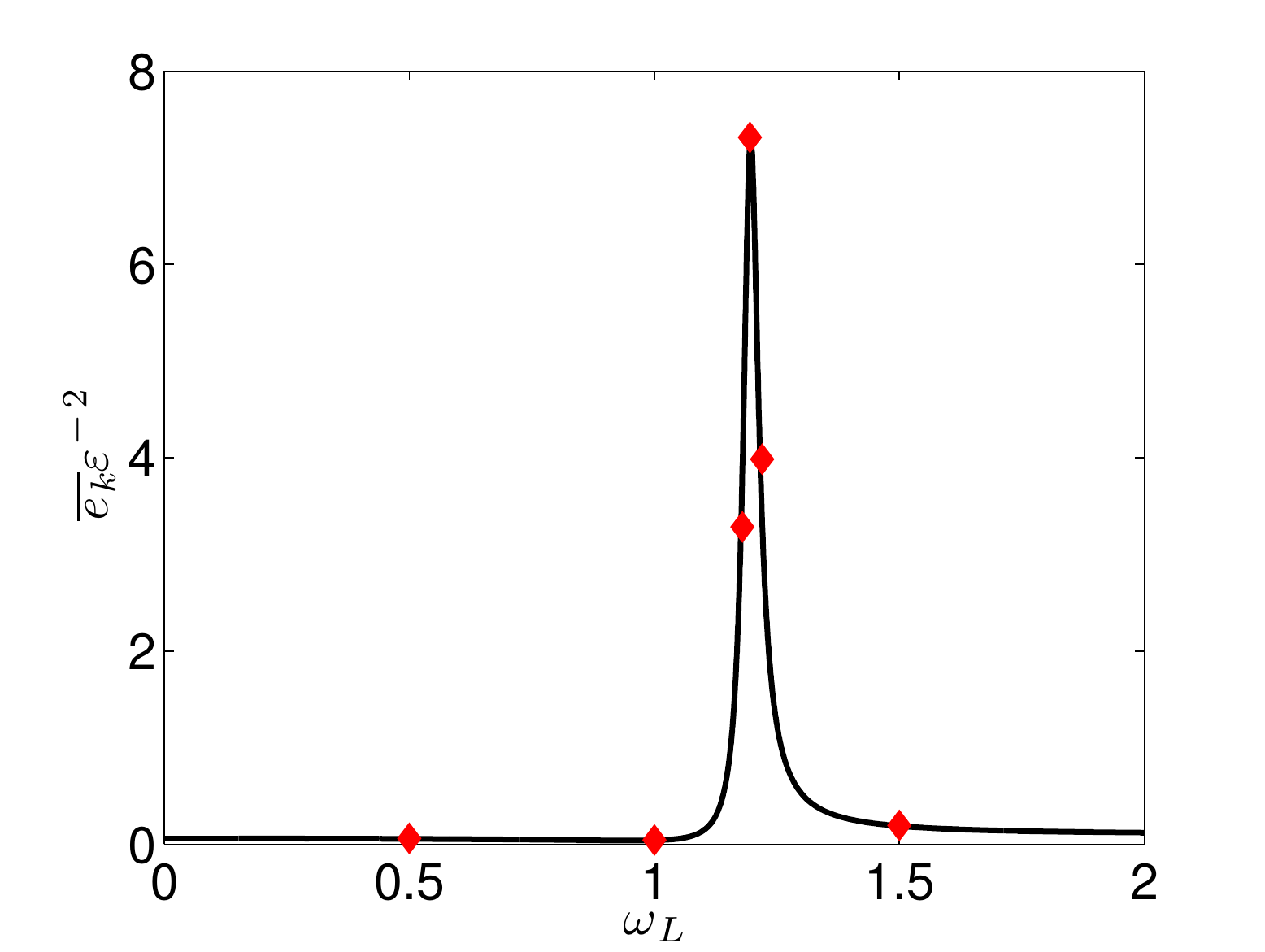}}

    \end{tabular}
    \caption{Mean kinetic energy density $\overline{e_k}$ for $E=5 \cdot 10^{-5}$ and $\varepsilon=0.005$ at resonance. Comparison between the result of the linear theory (\ref{eq:sol_qx_visc})-(\ref{eq:sol_qz_visc}), (\ref{eq:energy_density})-(\ref{eq:energy_density_time_avg}) and the numerical simulations (symbols). Results for geometries I - IV. Note the difference in scale between the different subfigures. }
         \label{fig:numeric_theory_spheroid}                
  \end{center}
\end{figure}

In figure \ref{fig:numeric_theory_spheroid}, we compare the total kinetic energy density (\ref{eq:energy_density_def}) retrieved from our numerical simulations with the linear result (\ref{eq:sol_qx_visc})-(\ref{eq:sol_qz_visc}), (\ref{eq:energy_density_time_avg}). We emphasize that the numerically obtained energy density also contains non-uniform vorticity contributions, as the integral over the ellipsoidal volume also takes into account the viscous boundary layers for example. Nevertheless, there is excellent agreement between the two approaches. The coefficient $\lambda_{num}$ used to trace the theoretical curve is chosen such as to minimize the discrepancy with the numerical data points, and is given in the rightmost column of Table \ref{tab:geometry}. This value can be compared with the the theoretical asymptotic decay rate $\lambda_{asym}$ of the spin-over mode, given in the penultimate column of Table \ref{tab:geometry}. We see that there is a striking quantitative resemblance for the three geometries with topographic coupling (i.e cases I, III and IV). The slight systematic higher value of the dissipation rate is due to the simplified expression of the dissipation used in our reduced model, and the  moderate non-asymptotic values of the Ekman number. Therefore, at first order, one can interpret the dissipation rate in our model as the decay rate of the spin-over mode.
We note finally, that we have not inverted $\lambda$ for the non-resonant case II. According to (\ref{eq:ek_biax1})-(\ref{eq:ek_biax2}), the energy density is independent of $\lambda$ to leading order in $E$ for this geometry. Far away from resonance, we see that $\overline{e_k} \approx 0.1$; near the spin-over frequency, however, $\overline{e_k}$ tends to 0.05.  This agrees well with (\ref{eq:ek_biax1})-(\ref{eq:ek_biax2}). 
\par
\begin{figure}                
  \begin{center}
  \setlength{\epsfysize}{5.0cm}

    \begin{tabular}{ccc}
      \setlength{\epsfysize}{5.0cm}
      \subfigure[]{\epsfbox{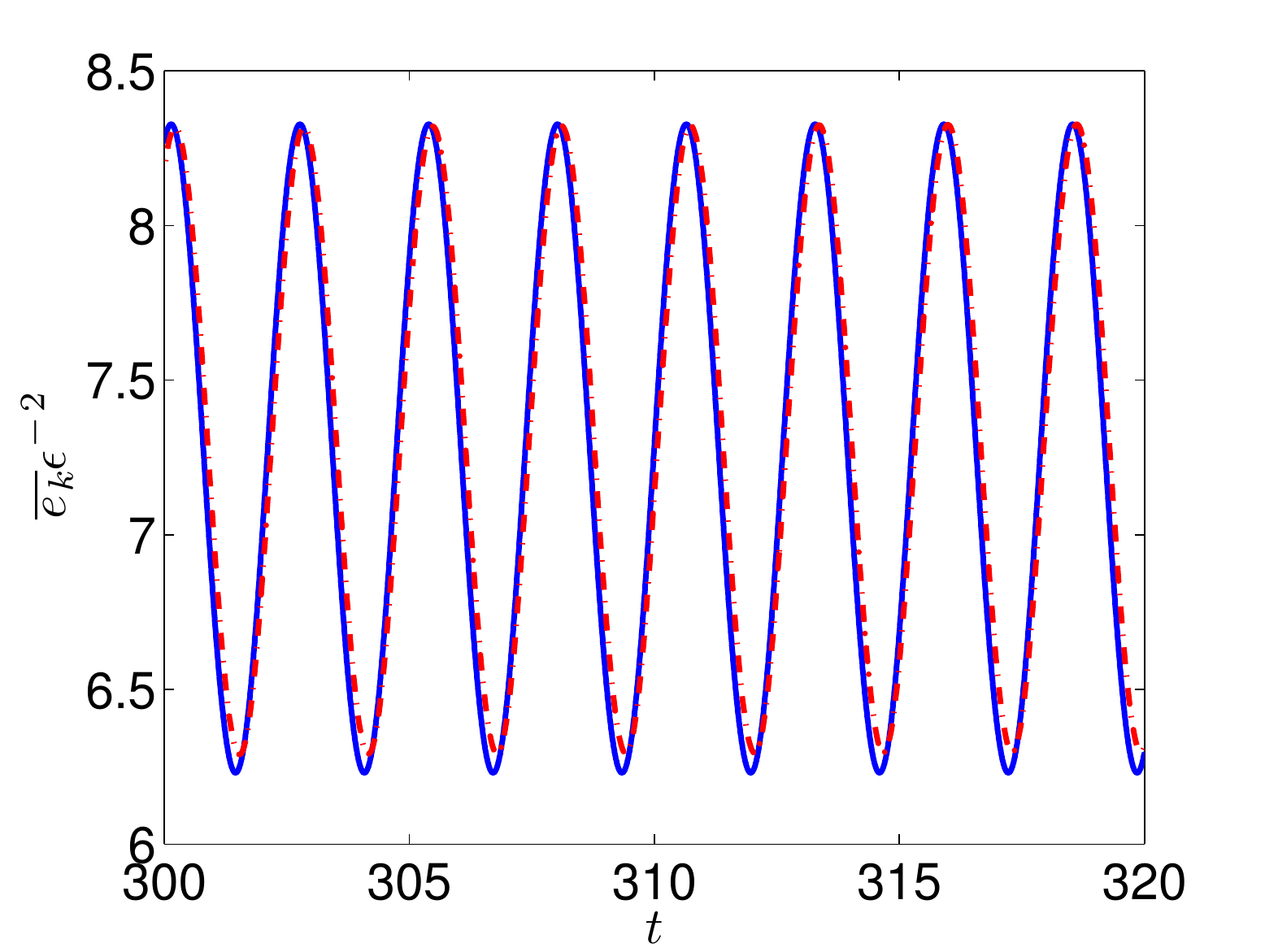}}  &
      \setlength{\epsfysize}{5.0cm}
      \subfigure[]{\epsfbox{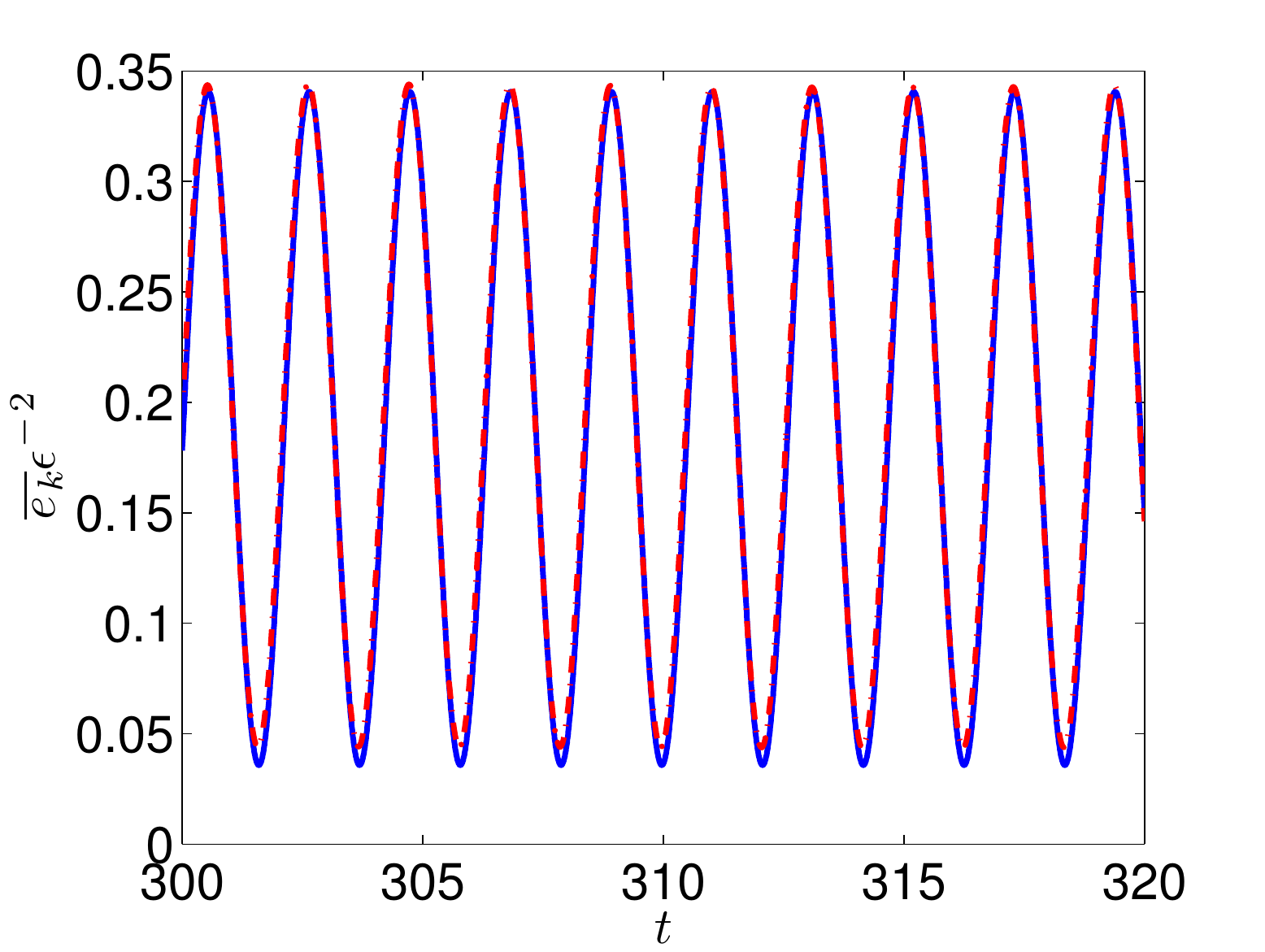}}  \\
      
    \end{tabular}
    \caption{Time series of the mean kinetic energy density $e_k$ for $E=5 \cdot 10^{-5}$, $\varepsilon=0.005$, and geometry IV. Results for $\omega_L=\sqrt{10/7}$ (resonant, subfigure (a)) and $\omega_L=1.5$ (non-resonant, subfigure (b)). Comparison between the result of the linear theory (\ref{eq:energy_density}) (blue, solid line) and the numerical simulations (red, dashed line). }
         \label{fig:numeric_theory_timeseries}                
  \end{center}
\end{figure}

In figure \ref{fig:numeric_theory_timeseries}, we compare the time series of $e_k$ retrieved from the numerical simulations with those from the linear theory. In order to trace the theoretical curve, the coefficient $\lambda$ was set to the value extracted from the frequency scan, i.e. the one provided in the rightmost column of Table \ref{tab:geometry}. Both curves virtually collapse, and this further validates our initial ansatz of uniform vorticity flow and the reduced viscosity model.  

\par
\begin{figure}          
  \begin{center}
 \includegraphics[width=0.7\textwidth]{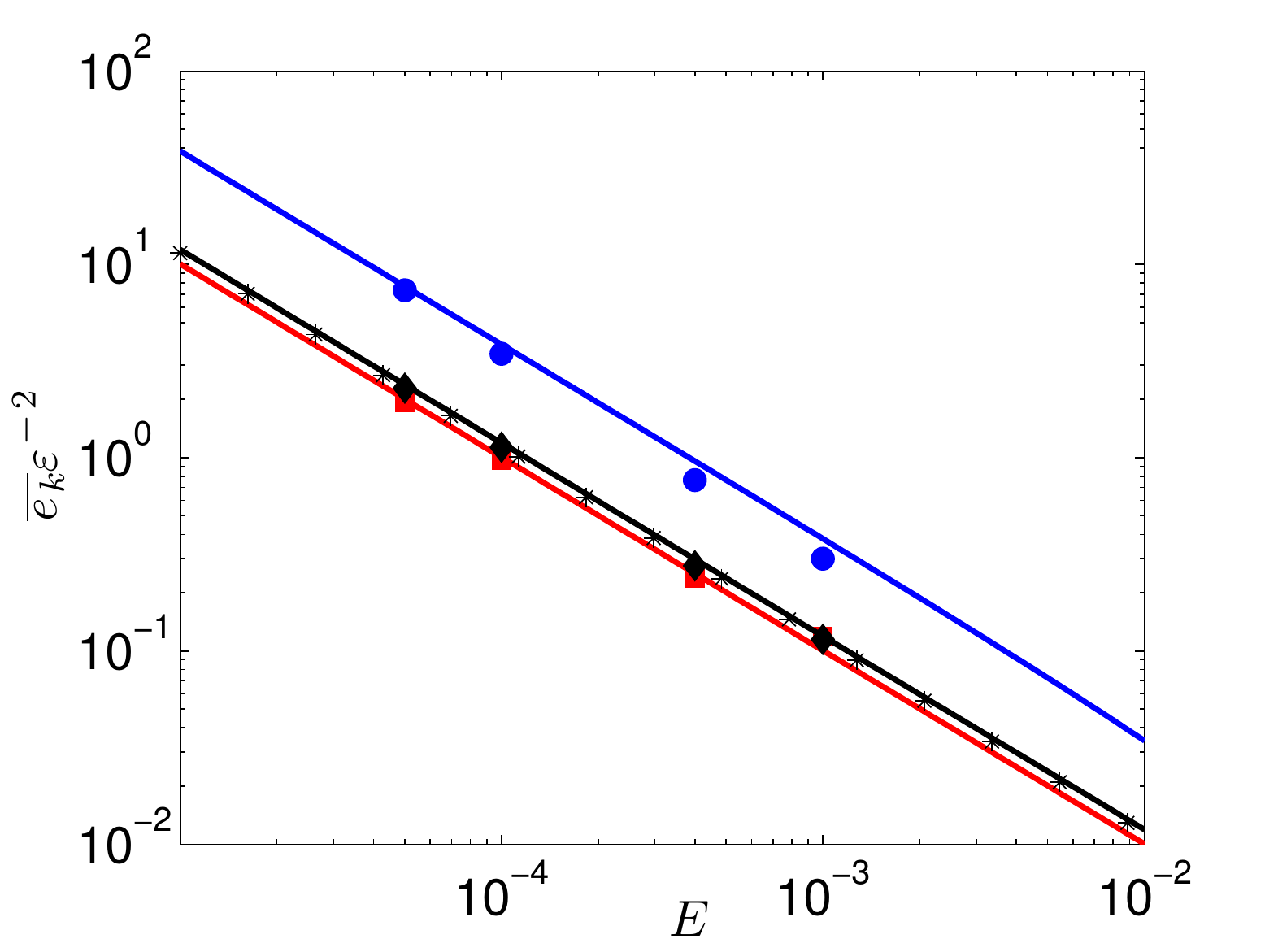}
     \caption{Mean kinetic energy density $\overline{e_k}$ for $E=5 \cdot 10^{-5}$ and $\varepsilon=0.005$. Comparison between the result of the linear theory (\ref{eq:sol_qx_visc})-(\ref{eq:sol_qz_visc}), (\ref{eq:energy_density})-(\ref{eq:energy_density_time_avg}) (solid line) and the numerical simulations (symbols) for the cases with topographic coupling I (black,diamonds), III (red,squares) and IV (blue,circles). Black stars correspond to the asymptotic theory of \cite{zhang2012} for a spheroidal geometry. }
         \label{fig:ekscan}                
  \end{center}
\end{figure}

We now investigate how well the analytical expression (\ref{eq:energy_density}) is established at different values of the Ekman number; our results are summarized in figure \ref{fig:ekscan}. For the spheroidal geometry I, we can also compare our results to the asymptotic expression of \cite{zhang2012}. Generally, we find that the agreement between the theory and the numerics is excellent. Noticeable differences are only observed for $E \gtrsim 4 \cdot 10^{-4}$ for the triaxial case IV. 
\par
\begin{figure}                
  \begin{center}
  \setlength{\epsfysize}{5.0cm}

    \begin{tabular}{ccc}
      \setlength{\epsfysize}{5.0cm}
      \subfigure[]{\epsfbox{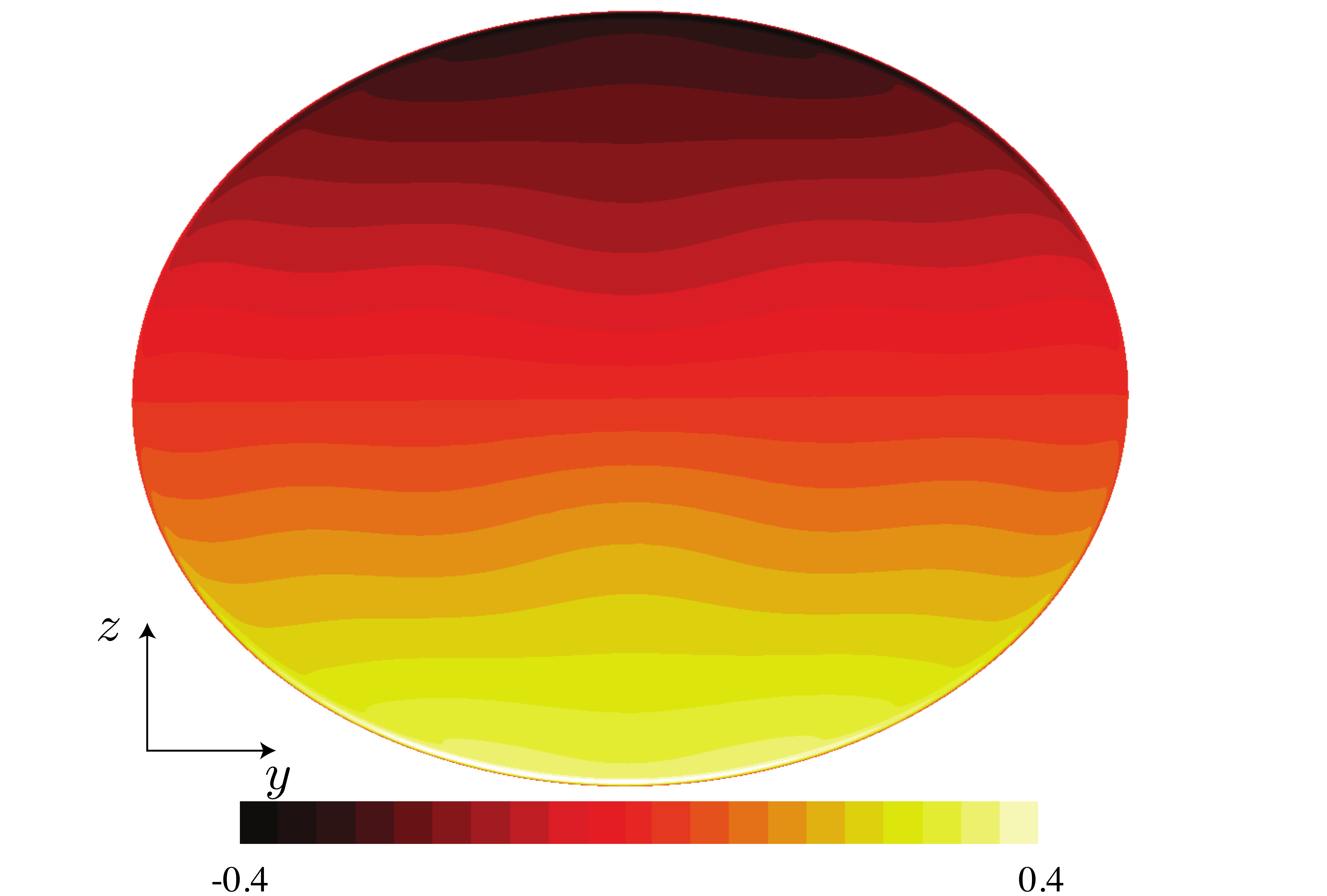}}  &
      \setlength{\epsfysize}{5.0cm}
            \subfigure[]{\epsfbox{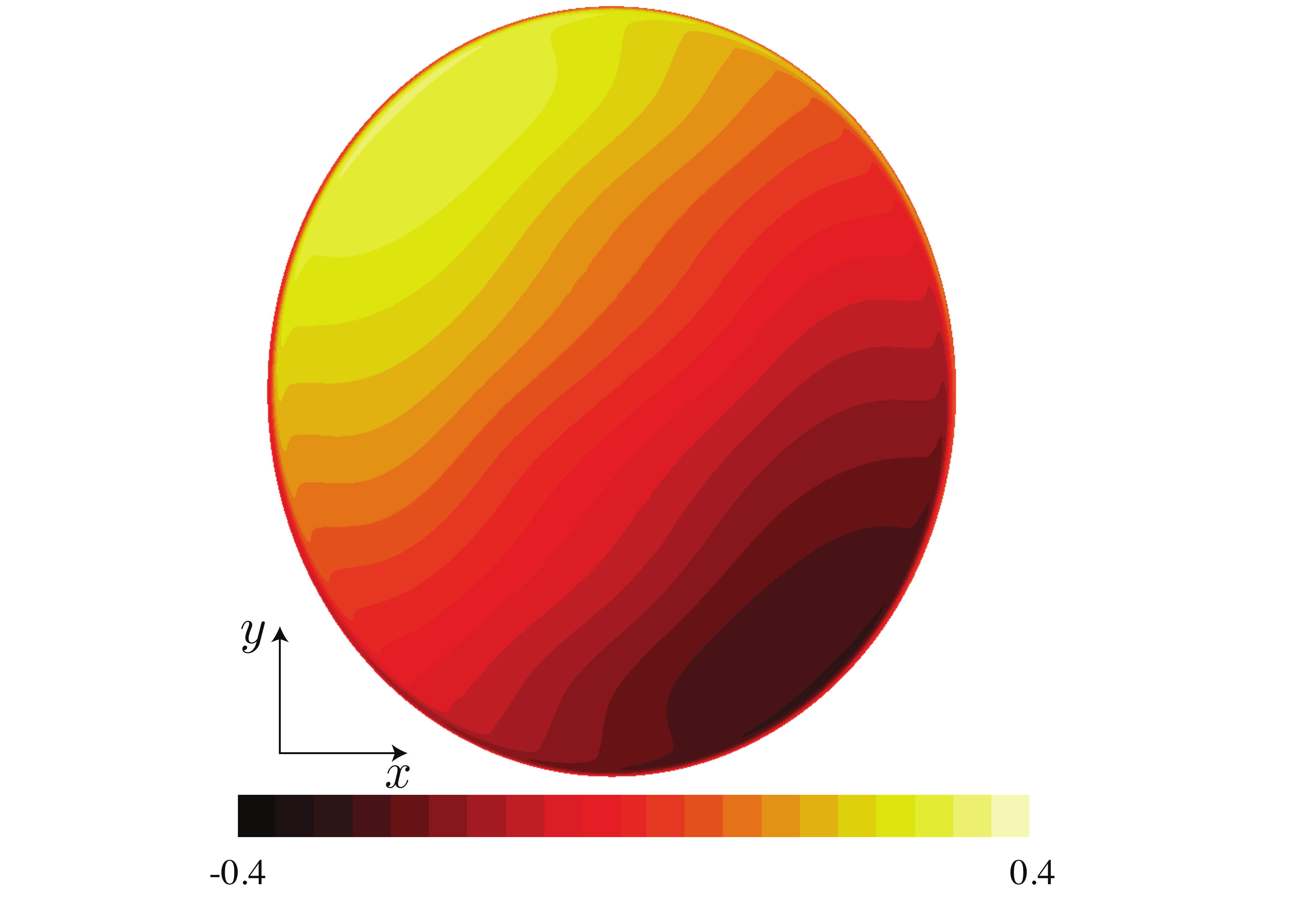}}   \\
      \setlength{\epsfysize}{5.0cm}
         \subfigure[]{\epsfbox{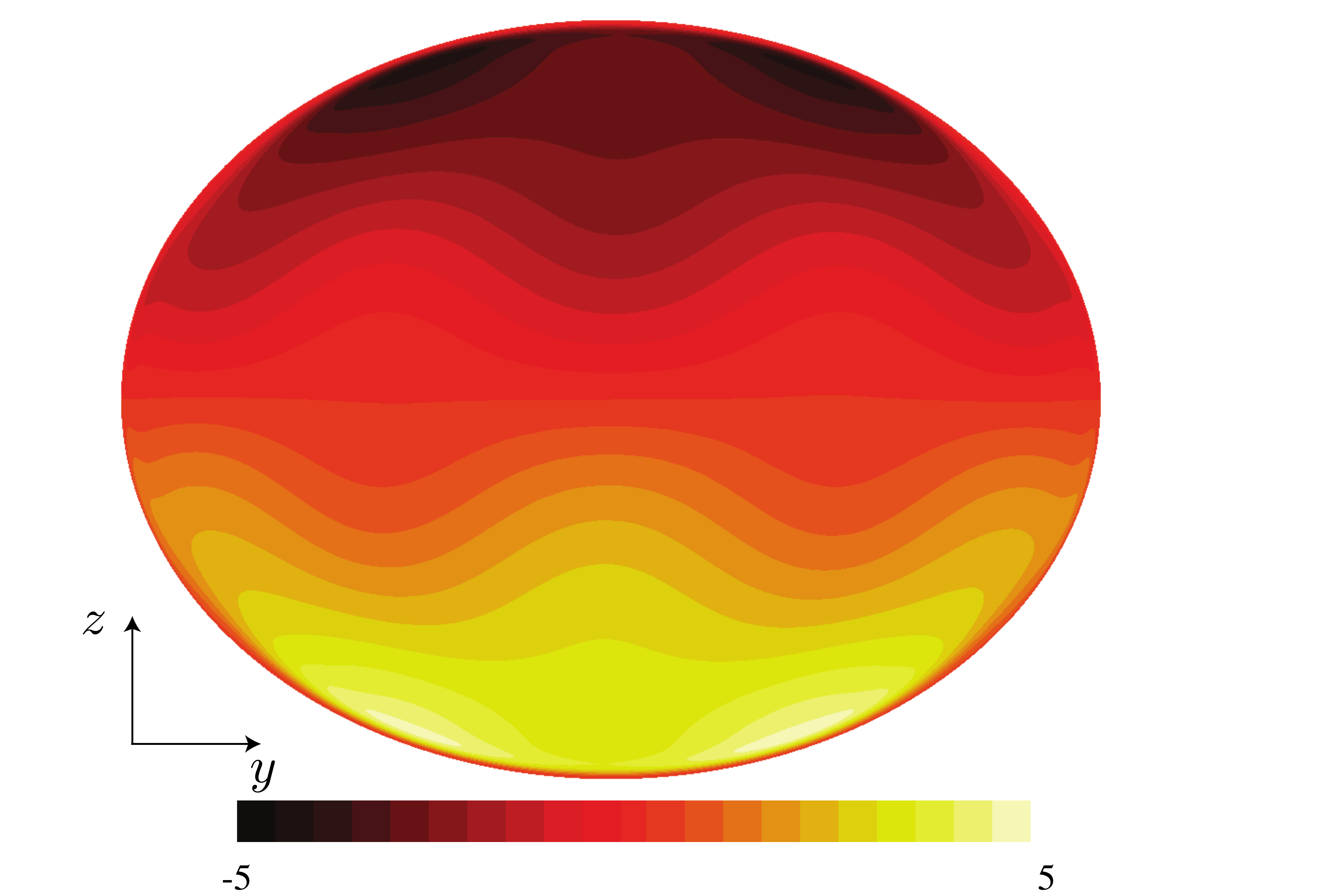}}  &
      \setlength{\epsfysize}{5.0cm}
      \subfigure[]{\epsfbox{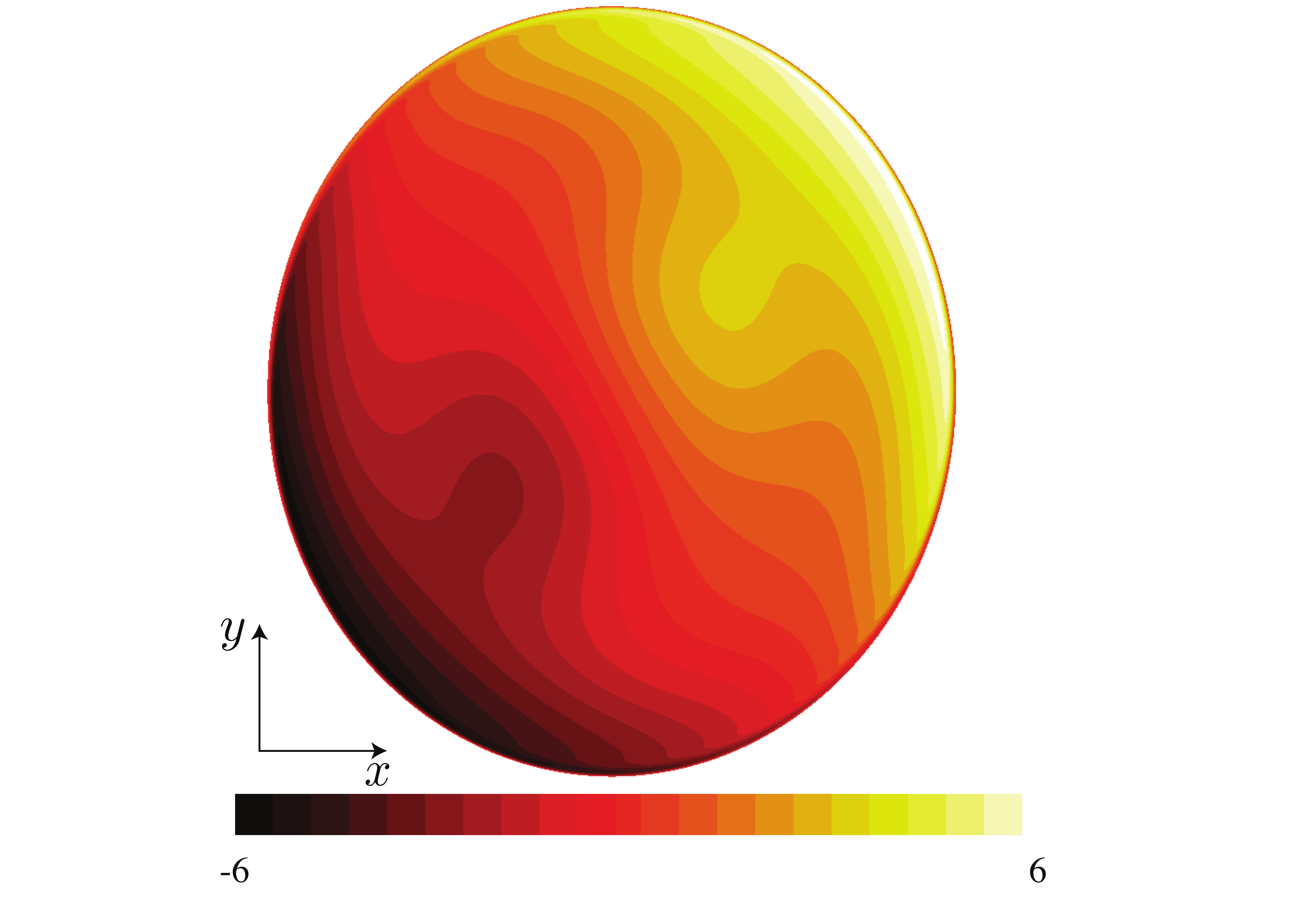}}       
      
    \end{tabular}
    \caption{Isocontours of $u_y$ in the plane $x=0$ (a,c) and $u_z$ in the plane $z=0$ (b,d) for geometry IV and libration frequencies $\omega=1$ (a,b) and $\omega=f=\sqrt{10/7}$ (c,d).}
         \label{fig:spatial}                
  \end{center}
\end{figure}
\begin{figure}                 
  \begin{center}
      \includegraphics[width=0.7\textwidth]{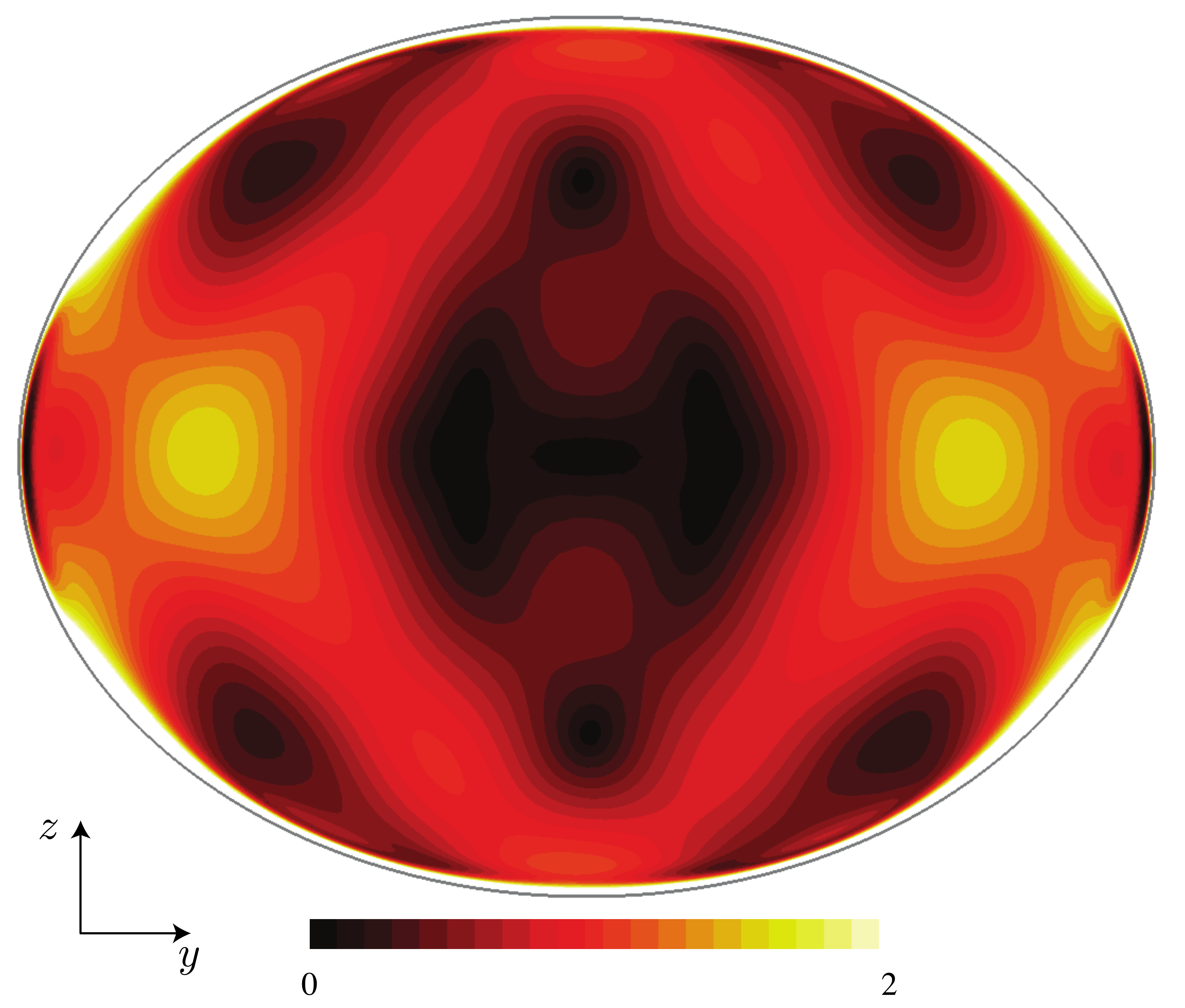}  
    \caption{Instantaneous isocontours of the numerically obtained magnitude of the non-uniform vorticity contribution to the flow in the meridional plane $x=0$ for geometry IV, $E=5\cdot 10^{-5}$ and the resonant driving frequency $\omega_L=f=\sqrt{10/7}$.}
         \label{fig:non_uniform_vorticity}                
  \end{center}
\end{figure}

The mean kinetic energy density is a global measure of the amplitude of the driven flow, but we also wish to assess how well the uniform vorticity solution is established locally. To this end, we show snapshots of $u_y$ in the plane $x=0$ and $u_z$ in the plane $z=0$ in figure \ref{fig:spatial}. According to the uniform vorticity assumption (equation (\ref{eq:u(q)})), we expect these isolines to be straight curves, with  $u_y$ being independent of $y$ in the plane $x=0$.  For non-resonant frequencies, we see that the uniform vorticity solution is reasonably well established in the bulk of the flow, whereas there are noticeable differences for the resonant flow. Finally, in both cases, the theoretical solution breaks down near the container walls, i.e. in the viscous Ekman layers. Similar behaviour was observed by \cite{zhang2012}. The resonant case is further investigated in figure \ref{fig:non_uniform_vorticity}, where we have removed the uniform vorticity component from the flow and show a snapshot of $u_y$ in the plane $x=0$. We observe a conical-like structure that is reminiscent of internal shear layers \cite[]{kerswell1995internal}. As the thickness and amplitude of these layers scale as $E^{1/5}$ and $E^{3/10}$ respectively, this pattern is diffuse and smeared out over a broad area. 

\par
\begin{figure}                 
  \begin{center}
      \includegraphics[width=0.7\textwidth]{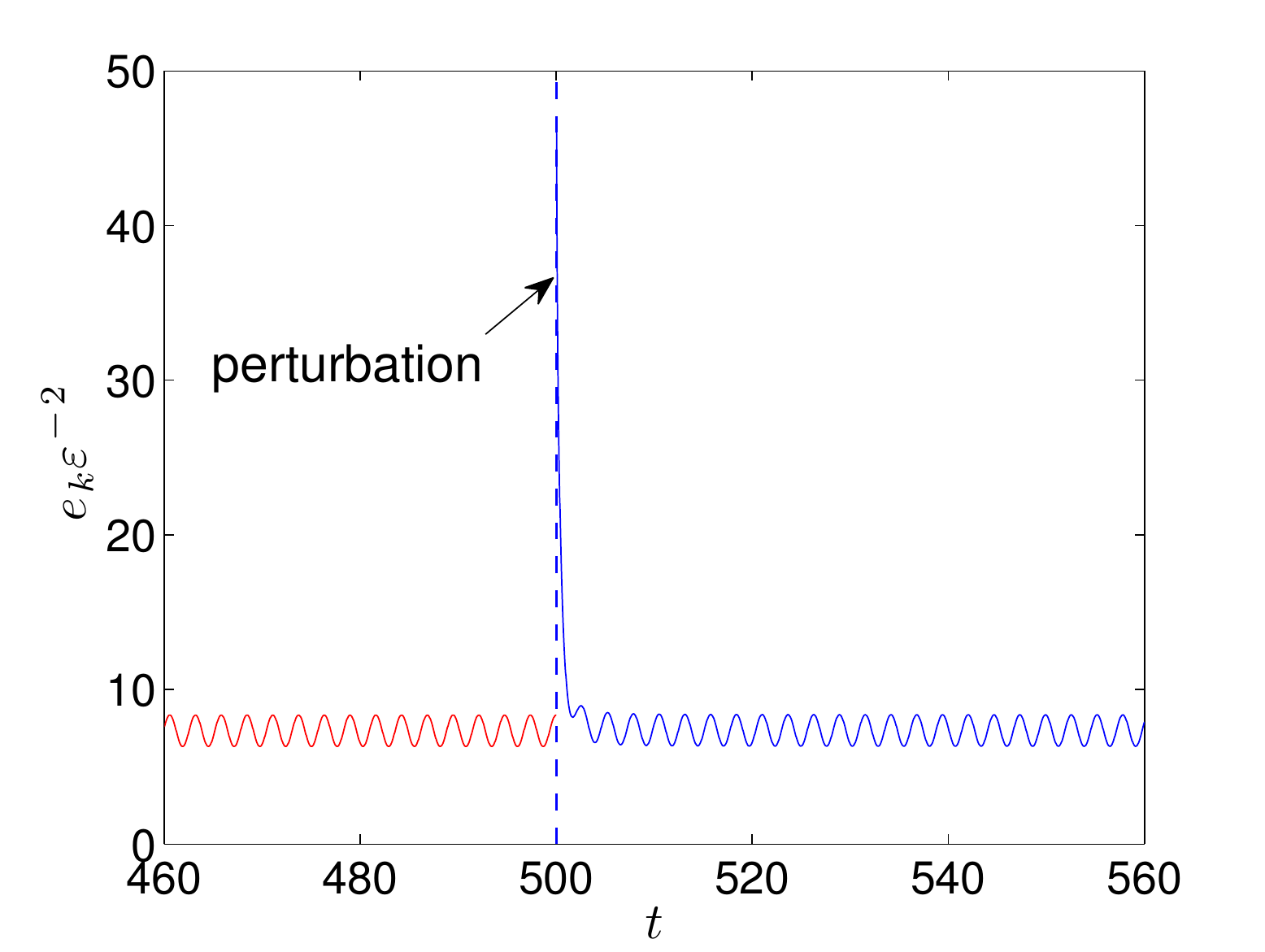}  
    \caption{Kinetic energy density $e_k \varepsilon^{-2}$ before and after perturbation with a random velocity field (at $t=500$) for the triaxial geometry IV, $E=5\cdot 10^{-5}$ and the resonant driving frequency $\omega_L=f=\sqrt{10/7}$.}
         \label{fig:perturbation}                
  \end{center}
\end{figure}

Finally, we also assess the stability of the numerically obtained solutions. To this end, we adopt the following approach. For a given solution, we consider a perturbation that consists of a random velocity field with zero mean and a root-mean-square amplitude exceeding that of the original solution by a factor of approximately two. Starting from this `initial condition', we integrate the Navier-Stokes equation in time. Systematically, we observe that the system quickly returns to its original state, which gives evidence for the stability of the numerically obtained solutions. This is illustrated in figure \ref{fig:perturbation}, where we show time series of the kinetic energy density $e_k$ before and after the perturbation for the triaxial geometry IV, $\omega_L=f$ and $E=5 \cdot 10^{-5}$

\section{Stability analysis} \label{sec:stability}
We now wish to investigate whether the uniform vorticity flow described in the previous section remains stable against small perturbations. The canonical approach to this problem is via linear stability analysis. As a starting point, we write the total flow field as the sum of a prescribed base flow ${\boldsymbol U}$ and a small perturbation ${\boldsymbol v}$ of order of magnitude $\delta \ll 1$,
\begin{equation}
{\boldsymbol u} = {\boldsymbol U} + {\boldsymbol v} \label{eq:base_plus_perturb}.
\end{equation}
We make the choice that ${\boldsymbol U}$ is a uniform vorticity solution of the non-linear problem (\ref{eq:u(q)}),(\ref{eq:qx})-(\ref{eq:qz}). From now on, we will mainly restrict ourselves to non-resonant libration frequencies. Anticipating the following discussion, we recall that we have only pursued an asymptotic development of ${\boldsymbol U}$ up to order $\varepsilon$, i.e. we have
\begin{equation}
{\boldsymbol U} = \underbrace{{\boldsymbol U}_0}_{\mathcal{O}(\varepsilon)} + \underbrace{{\boldsymbol U}_1}_{\mathcal{O}(\varepsilon^2)} ,
\end{equation}
with ${\boldsymbol U}_0$ corresponding to the solution (\ref{eq:sol_qx})-(\ref{eq:sol_qz}), and ${\boldsymbol U}_1$ a uniform vorticity flow for which no analytic expression has been sought. It will also be instructive to consider an asymptotic development of the rotation vector ${\boldsymbol \Omega}$:
\begin{equation}
{\boldsymbol \Omega}  =  \underbrace{(0,0,1)}_{{\boldsymbol \Omega}_0} +  \underbrace{\varepsilon \left( \cos(\omega_L t),\omega_L^{-1}\sin(\omega_L t),0\right)}_{{\boldsymbol \Omega}_1} + \underbrace{\mathcal{O}(\varepsilon^2)}_{\boldsymbol{\Omega}_2}.
\end{equation}
We now substitute (\ref{eq:base_plus_perturb}) into the inviscid equations of motion, and invoking that ${\boldsymbol U}$ is a solution of (\ref{eq:u(q)}),(\ref{eq:qx})-(\ref{eq:qz}), we find
\begin{eqnarray}
\nabla \cdot {\boldsymbol v} & = & 0, \\
\frac{\partial {\boldsymbol v}}{\partial t} + {\boldsymbol U}\cdot \nabla{\boldsymbol v} + {\boldsymbol v}\cdot \nabla{\boldsymbol U}  + {\boldsymbol v}\cdot \nabla{\boldsymbol v} + 2 {\boldsymbol \Omega} \times{\boldsymbol v} & = & - \nabla p . 
\end{eqnarray}
This can be further developed into 
\begin{eqnarray}
\frac{\partial {\boldsymbol v}}{\partial t} + \underbrace{{\boldsymbol U}_0\cdot \nabla{\boldsymbol v}+{\boldsymbol v}\cdot \nabla{\boldsymbol U}_0}_{\mathcal{O}(\delta \varepsilon)} + \underbrace{{\boldsymbol U}_1\cdot \nabla{\boldsymbol v}+{\boldsymbol v}\cdot \nabla{\boldsymbol U}_1}_{\mathcal{O}(\delta \varepsilon^2)} + \underbrace{{\boldsymbol v}\cdot \nabla{\boldsymbol v}}_{\mathcal{O}(\delta^2)} \nonumber & &\\+ \underbrace{2 {\boldsymbol \Omega}_0 \times{\boldsymbol v}}_{\mathcal{O}(\delta)} +  \underbrace{2 {\boldsymbol \Omega}_1 \times{\boldsymbol v}}_{\mathcal{O}(\delta \varepsilon)} +  \underbrace{2 {\boldsymbol \Omega}_2 \times{\boldsymbol v}}_{\mathcal{O}(\delta \varepsilon^2)}    &= &  - \nabla p . 
\end{eqnarray}
Assuming now that $\delta \ll \varepsilon \ll 1$, we can neglect terms of order of magnitude $\delta \varepsilon^2 $ and $\delta^2$ with respect to the other terms. This leaves us with
\begin{equation}
\frac{\partial {\boldsymbol v}}{\partial t} + {\boldsymbol U}_0\cdot \nabla{\boldsymbol v} + {\boldsymbol v}\cdot \nabla{\boldsymbol U}_0 + 2 \left( {\boldsymbol \Omega}_0 + {\boldsymbol \Omega}_1 \right) \times{\boldsymbol v}  =  - \nabla p . \label{eq:disturbance_v2}
\end{equation}
We can anticipate at this point that there will be two separable time scales associated to the Coriolis force, the short time scale ($\tau\sim\Omega_0^{-1}$) and the long one associated with $\Omega_1(\tau \sim \varepsilon^{-1}$).
\par
We now proceed as follows. In a first step, we will invoke energy considerations to identify parameters that govern and bound the growth of ${\boldsymbol v}$. Then, we will deploy two different techniques that solve (\ref{eq:disturbance_v2}). Both of them approach the issue of multiple time scales by considering wave-like solutions whose phase fluctuates on the fast rotation time scale ($ \sim \Omega_0^{-1}$), and whose amplitude varies on the slower time scale $\varepsilon^{-1}$.  The first one is a local method \cite[]{lifschitz1991local}, henceforth abbreviated the LH method, and considers (in)stability along a streamline.  The second one, introduced in \cite{gledzer1978finite,gledzer1992instability} (referred to as GP), is global and considers perturbations that are polynomial in the cartesian coordinates.  The notions global and local refer to whether or not the perturbations satisfy the non-penetration condition (\ref{eq:no-penetration}). As the LH method yields the fastest growing solution of (\ref{eq:disturbance_v2}), regardless of the boundary conditions, it gives us an upper bound for the growth rate. The GP method on the other hand, only considers certain subsets of all possible perturbations, and therefore gives a lower bound on $\sigma$. Finally, in subsection \ref{subsec:instab_simu}, we will compare these approaches against direct numerical simulations. 

\subsection{Energy considerations} \label{subsec:energy}
Prior to seeking direct solutions for equation (\ref{eq:disturbance_v2}), we will first investigate it through the prism of the power statement
\begin{equation}
  \frac{1}{2} \frac{\mathrm{d}}{\mathrm{d}t}\displaystyle \langle{\boldsymbol v}, {\boldsymbol v} \rangle = \displaystyle \langle {\boldsymbol v},{\mathcal H} \cdot {\boldsymbol v}\rangle, \label{eq:power}
\end{equation}
which is obtained by taking the inner product of (\ref{eq:disturbance_v2}), and introduces the strain rate tensor ${\mathcal H}$, 
\begin{equation}
\mathcal{H} = \frac{1}{2}\left[\nabla {\boldsymbol u}_0 + (\nabla {\boldsymbol u}_0)^{\dagger}\right].
\end{equation}
Here we have introduced the notation $\langle {\boldsymbol f}, {\boldsymbol g} \rangle = \int_{V} {\boldsymbol f} \cdot {\boldsymbol g}^{\dagger} \,\mathrm{d}V$.
This teaches us that amplification of ${\boldsymbol v}$, and thus the presence of instability, requires that the right-hand side of (\ref{eq:power}) does not vanish. Using (\ref{eq:u(q)}),(\ref{eq:sol_qx})-(\ref{eq:sol_qy}), $\mathcal H$ reads
\begin{equation}
{\mathcal H} = \left(
\begin{array}{ccc}
0 & 0 & \chi_1 \sin(\omega_L t) \\
0 & 0 & \chi_2 \cos(\omega_L t) \\
\chi_1 \sin(\omega_L t) &  \chi_2 \cos(\omega_L t) & 0 
\end{array} 
\right), \label{eq:strainratetensor}
\end{equation}
with
\begin{eqnarray}
\chi_1 & =  & \frac{\varepsilon \omega_L \beta_{ac} \beta_{bc}}{\omega_L^2-f^2} = \frac{\varepsilon \omega_L \beta_{ac} \beta_{bc}}{\omega_L^2-|1+\beta_{ac}||1+\beta_{bc}|} ,  \label{eq:chi1}\\
\chi_2 &  = & \frac{\varepsilon (\omega_L^2-f_0^2)\beta_{bc}}{\omega_L^2-f^2} = \frac{\varepsilon (\omega_L^2-1-\beta_{ac})\beta_{bc}}{\omega_L^2-|1+\beta_{ac}||1+\beta_{bc}|} \label{eq:chi2}.
\end{eqnarray}
By virtue of the Cauchy-Schwarz inequality, we infer that
\begin{equation}
 \frac{1}{2} \frac{\mathrm{d}}{\mathrm{d}t} \langle {\boldsymbol v}, {\boldsymbol v} \rangle \le |\chi_1  \sin(\omega_L t) + \chi_2 \cos(\omega_L t)|  \langle{\boldsymbol v}, {\boldsymbol v} \rangle.
\end{equation}
This allows us to bound the growth rate $\sigma$ of a perturbation ${\boldsymbol v} \propto \exp(\sigma t)$, since
\begin{equation}
\sigma = \overline{ \frac{1}{2}\left( \frac{\mathrm{d}}{\mathrm{d}t} \langle {\boldsymbol v}, {\boldsymbol v} \rangle\right) \langle {\boldsymbol v}, {\boldsymbol v} \rangle^{-1}} \le | h_1(\omega_L) \chi_1| + |h_2(\omega_L) \chi_2| \le|\chi_1| + |\chi_2|, \label{eq:bound_sigma}
\end{equation}
where  $|h_1(\omega_L)|,|h_2(\omega_L)| \le 1$, and where the bar notation denotes an adequate time average that allows us to account for an additional harmonic time-dependence of ${\boldsymbol v}$. Given that $h_1$ and $h_2$ only depend on $\omega_L$, any dependence of the instability on the geometry and $\varepsilon $ is captured by the parameters $\chi_1$ and $\chi_2$. We immediately observe stability for $\beta_{bc}=0$. Note that for $\omega_L=f_0^2=1+\beta_{ac}$, $\chi_1 = \chi_2$, and for $\omega_L=1$ and arbitrary $\beta_{ac}$, $\chi_1 = - \chi_2$. In these cases, there is thus only a single control parameter. In particular, $\chi_1=\chi_2=0$ for $\omega_L=1$ and ${\beta}_{ac}=0$. No instability can grow for such a configuration regardless of the value of the topographic coupling ${\beta}_{bc}$. Intuitively, this can be understood from inspection of the flow (\ref{eq:u(q)}), (\ref{eq:sol_qx})-(\ref{eq:sol_qy}) which has only unstrained, circular streamlines in planes $y=cst$. 

\subsection{Local method: Short--wavelength Lagrangian stability analysis} \label{subsec:lifschitz}

The approach we follow here is based on the short--wavelength Lagrangian theory, used by \cite{bayly1986three} and \cite{craik1986evolution}, and then generalized in \cite{friedlander1991instability}, \cite{lifschitz1991local,lifschitz1993localized} and \cite{lifschitz1994instability} where the whole theory is thoroughly explained. This theory is now rather classical in stability studies of flows \cite[e.g.][]{bayly1996three,lebovitz1996short,leblanc1997three}, and we thus only recall below some basic elements of the stability analysis in following the approach of \cite{le1999short}. We found it simplest to work in the inertial frame of reference. The perturbation velocity $\boldsymbol{v}$ is written in the geometrical optics, or WKB (Wentzel-Kramers-Brillouin) form:
\begin{eqnarray}
\boldsymbol{v}(\boldsymbol{r},t)=\boldsymbol{a}(\boldsymbol{r},t)\, \mathrm{e}^{\textrm{i} \psi(\boldsymbol{r},t)/ \vartheta} \, . \label{eq:pert}
\end{eqnarray}
Here, the amplitude $\boldsymbol{a}(\boldsymbol{r},t)$ and phase $\psi(\boldsymbol{r},t)$ are real functions dependent on space $\boldsymbol{r}$ and time $t$. The characteristic wavelength $\vartheta \ll 1$ is the small parameter used for the asymptotic (WKB) expansion. In the inviscid limit, the evolution of (\ref{eq:pert}) is governed by the linearized Euler equations. Along the pathlines of the flow $\boldsymbol{U}$ in the inertial frame of reference, the leading-order problem can then be written in Lagrangian form as a system of ordinary differential equations \cite[][]{lifschitz1994instability}:
\begin{eqnarray}
\frac{\textrm{d} \boldsymbol{X}}{\textrm{d} t}&=&\boldsymbol{U}(\boldsymbol{X},t) \, , \label{eq:lifshitz1} \\
\frac{\textrm{d} \boldsymbol{\mathcal{K}}}{\textrm{d} t}&=&-(\boldsymbol{\nabla} \boldsymbol{U})^{\textrm{T}}(\boldsymbol{X},t)\, \boldsymbol{\mathcal{K}} \, , \label{eq:lifshitz2}\\
\frac{\textrm{d} \boldsymbol{a}}{\textrm{d} t}&=& \left( \frac{2 \boldsymbol{\mathcal{K}} \boldsymbol{\mathcal{K}}^{\textrm{T}}}{|\boldsymbol{\mathcal{K}}|^2}-\boldsymbol{I} \right) \boldsymbol{\nabla} \boldsymbol{U}(\boldsymbol{X},t)\, \boldsymbol{a} \, , \label{eq:lifshitz3}
\end{eqnarray}
with constraint
\begin{eqnarray}
\boldsymbol{\mathcal{K}} \cdot \boldsymbol{a} =0 \, . \label{eq:lifshitz4}
\end{eqnarray}
Here $\textrm{d}/\textrm{d}t=\partial_t + \boldsymbol{U} \cdot \boldsymbol{\nabla}$ are Lagrangian derivatives, $\boldsymbol{I}$ is the identity matrix, $\boldsymbol{\mathcal{K}}=\boldsymbol{\nabla} \chi$ is the (local) wavevector along the Lagrangian trajectory $\boldsymbol{X}$. The incompressibility condition (\ref{eq:lifshitz4}) is always fulfilled if the initial condition $(\boldsymbol{X}_0,\boldsymbol{\mathcal{K}_0},\boldsymbol{a}_0)$ satisfies $\boldsymbol{\mathcal{K}_0} \cdot \boldsymbol{a}_0=0 $ \cite[][]{le2000three}. As shown by \cite{lifschitz1991local}, the existence of an unbounded solution for $\boldsymbol{a}$ provides a sufficient condition of instability. Assuming closed pathlines, stability is naturally analysed over one turnover period $T$ along the pathline. Note that this system of equations can be seen as an extension of rapid distortion theory (RDT) to non-homogeneous flows \cite[][]{cambon1985etude,cambon1994stability,sipp1998elliptic}.

In practice, the equation (\ref{eq:lifshitz1}) has to be solved as a first step to know the trajectory $\boldsymbol{X}$ emerging out of initial position $\boldsymbol{X}_0$. Knowing $\boldsymbol{X}$, one can solve the wavevector equation (\ref{eq:lifshitz2}) for an initial vector $\boldsymbol{\mathcal{K}}_0$. As the magnitude of $\boldsymbol{\mathcal{K}}_0$ does not influence the growth of $\boldsymbol{a}$, we only have to consider the spherical surface $||\boldsymbol{\mathcal{K}_0}||=1$ in wave space to find the maximum growth rate.  
Knowledge of $\boldsymbol{X}$ and $\boldsymbol{\mathcal{K}}$ finally allows us to solve equation (\ref{eq:lifshitz3}) for the amplitudes $\boldsymbol{a}$ and to look for growing solutions.

\begin{figure}
\begin{center}
\begin{tabular}{ccc}
\setlength{\epsfysize}{5.0cm}
\subfigure[]{\epsfbox{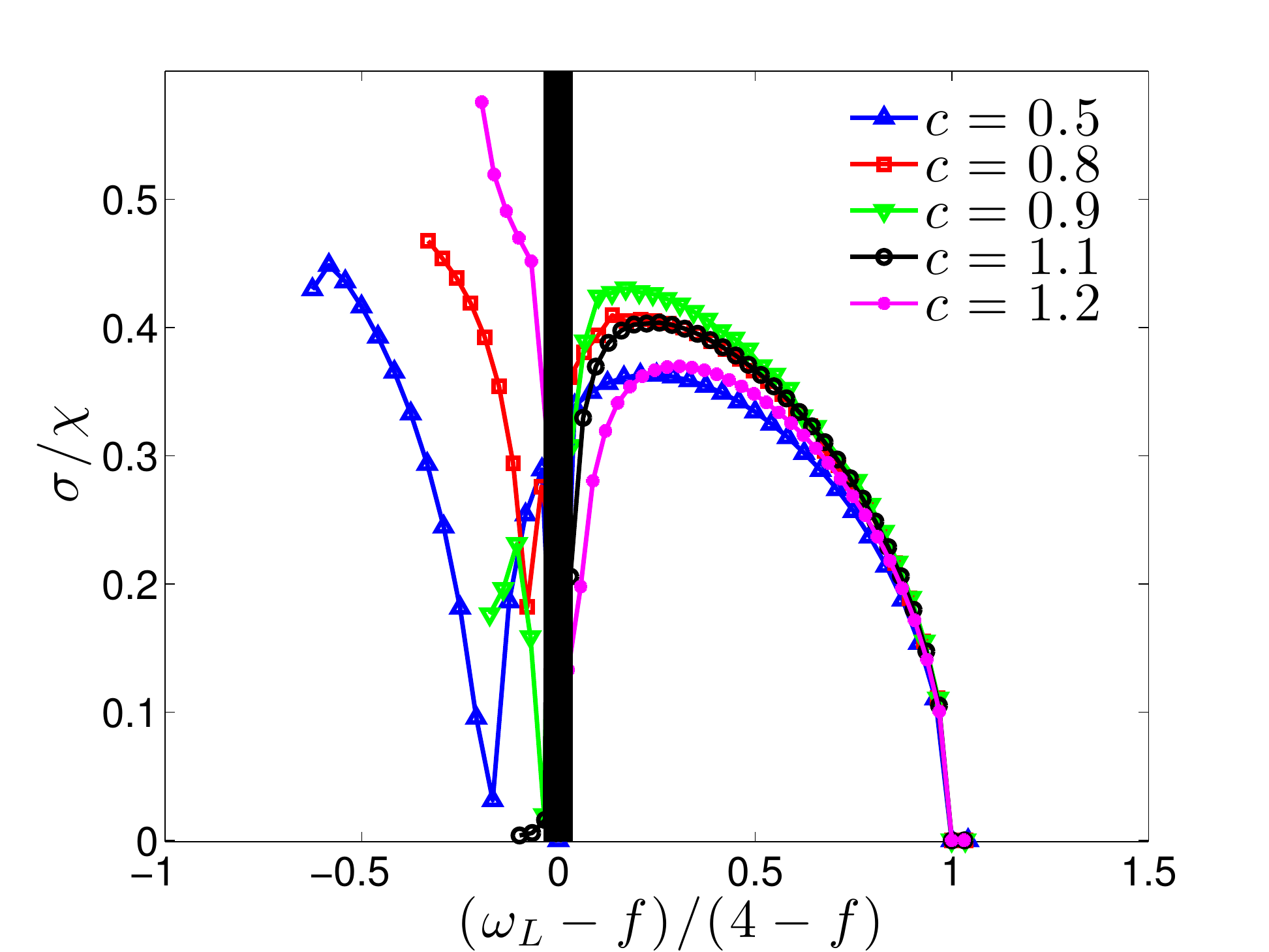}} &
\setlength{\epsfysize}{5.0cm}
\subfigure[]{\epsfbox{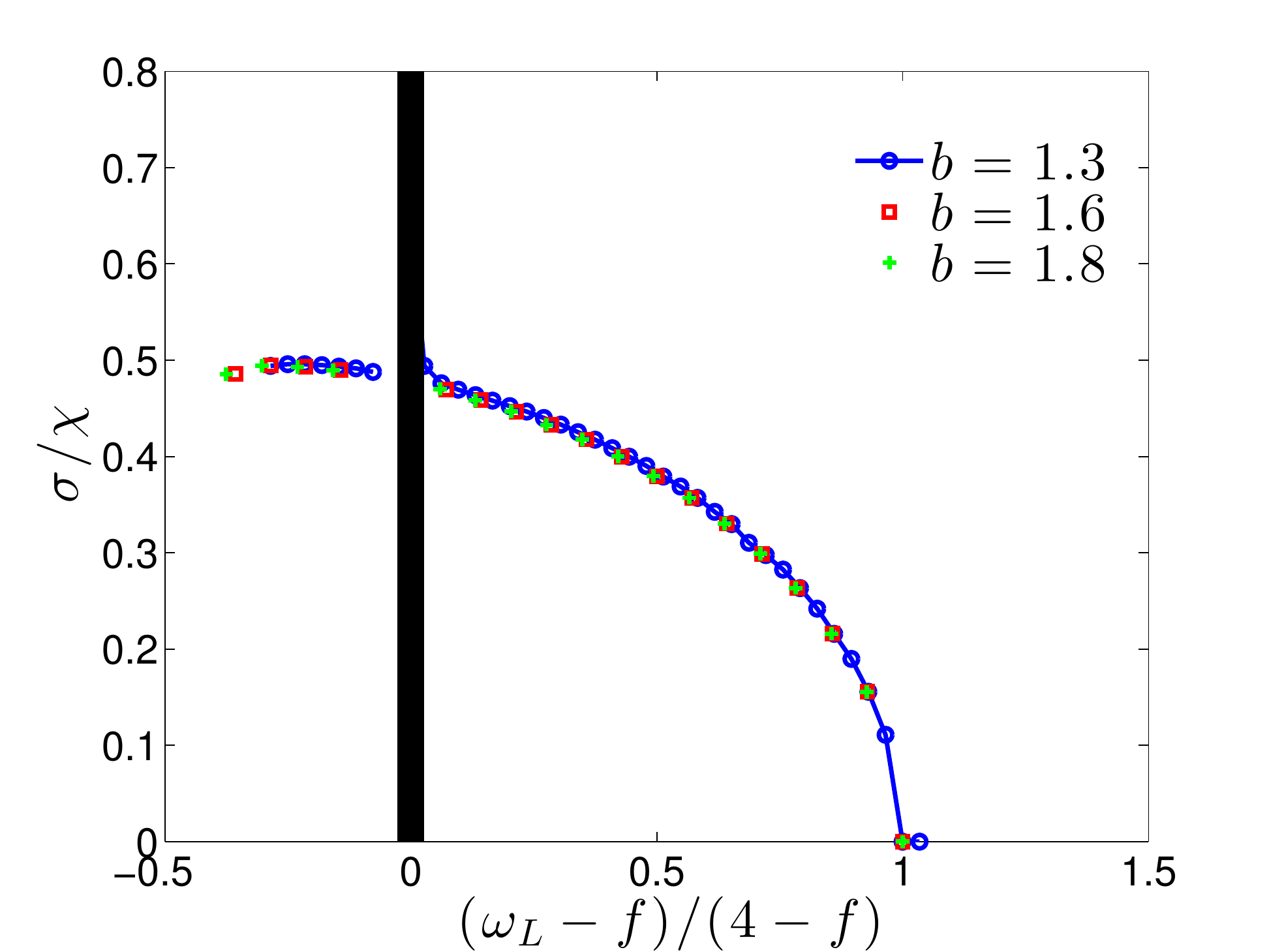}}
\end{tabular}
\caption{Results of the inviscid local stability analysis, carried out at fixed $\varepsilon=0.04$. The thick vertical black line indicate the resonance area. (a) $\sigma / \chi $ for a spheroid ($a=b=1$). (b) $\sigma / \chi $ for a spheroid ($a=c=1$). In this case, there is only one control parameter, i.e $\chi = |\chi_2|$.}
\label{fig:compWKB}
\end{center}
\end{figure}

In figure \ref{fig:compWKB}, we present typical results of the local stability analysis of the basic flow (\ref{eq:sol_qx})-(\ref{eq:sol_qy}) for different geometries. We vary the libration frequency $\omega_L$ between 0 and 4, but exclude a small frequency band around the spin-over frequency $f$ where the inviscid base flow diverges. We choose to present our results as a function of $(\omega_L-f)/(4-f)$ so that this frequency band collapses for all geometries. More precisely, we display $\sigma/\chi$, where $\chi=|\chi_1|+|\chi_2|$. Both figurea \ref{fig:compWKB}(a) and \ref{fig:compWKB}(b) clearly illustrate that $\sigma/\chi < 1$, which is consistent with the upper bound (\ref{eq:bound_sigma}).     
We consider the simple case of a rotating spheroid $a=b$ in figure \ref{fig:compWKB}(a). Since two control parameters ($\chi_1$, $\chi_2$) are available in this case, it is not possible to collapse all the curves on a single one. 
For the case $a=c$ however, there is only a single control parameter $\chi_2$, and thus $\chi=|\chi_2|$. Figure \ref{fig:compWKB}(b) clearly shows that all curves then collapse onto a single master curve, which actually depends only on the function $h_2(\omega_L)$ introduced in subsection \ref{subsec:energy}. One can notice that the local stability results indicate that, for $a=c$, the function $h_2(\omega_L)$ is bounded by $1/2$ when we are not on the resonance peak.

\subsection{Global method: Gledzer-Ponomarev (GP) polynomial perturbation analysis \label{subsec:GP}}
Anticipating the results of the GP method, we first put forward that the base flow ${\boldsymbol u}_0$ is prone to inertial instabilities, similar to those that can grow upon uniform vorticity flows driven by precession, tides and longitudinal libration \cite[]{kerswell1993instability,kerswell1994tidal,lacaze2004elliptical,le2010tidal,Cebron_Pof,WR2012}. Underlying these instabilities is a parametric resonance mechanism, whereby the strain imposed by the geometry couples two inertial modes of the unforced system, provided certain resonance conditions are met.  

We follow roughly an approach set out by \cite{tilgner2007treatise}, and start our analysis by recasting (\ref{eq:disturbance_v2}) in a more condensed form,
\begin{equation}
\frac{\partial {\boldsymbol v}}{\partial t} + 2\hat{\boldsymbol z} \times {\boldsymbol v} + \nabla p = -\varepsilon {\mathcal L} {\boldsymbol v}  \label{eq:nullspace}, 
\end{equation}
where ${\mathcal L}$ is a time-dependent linear operator, which acts on ${\boldsymbol v}$, and is characterized by one single frequency $\omega_L$.  It will now be instructive to consider perturbations ${\boldsymbol v}$ that are superpositions of two inertial modes of the cavity, i.e.
\begin{equation}
 {\boldsymbol v}({\boldsymbol r},t) = A(t){\boldsymbol V}_A({\boldsymbol r}) \exp(\mathrm{i}\mu_{A}t) + B(t){\boldsymbol V}_B({\boldsymbol r}) \exp(\mathrm{i}\mu_{B}t). \label{eq:two_modes}
\end{equation}
Here $(\mu_{A},{\boldsymbol V}_A)$ and $(\mu_{B},{\boldsymbol V}_B)$ are eigenvalue-eigenmode pairs of the inviscid  inertial mode problem 
 \begin{eqnarray}
\nabla \cdot {\boldsymbol V}_{A,B} & = & 0, \\
2 \hat{\boldsymbol z} \times {\boldsymbol V}_{A,B} - \nabla p_{A,B} & = &  \mathrm{i}\mu_{A,B} {\boldsymbol V}_{A,B} \label{eq:inertial_v},
\end{eqnarray}
previously encountered in its `vorticity' form  (\ref{eq:inertialmode}). The modes ${\boldsymbol V}_{A,B}$ satisfy (\ref{eq:no-penetration}), and are mutually orthogonal, i.e. $\langle {\boldsymbol V}_B, {\boldsymbol V}_A \rangle =\delta_{AB}$, with $\delta_{AB}$ the Kronecker delta. Substituting (\ref{eq:two_modes}) into (\ref{eq:nullspace}) and projection on ${\boldsymbol V_A}$, ${\boldsymbol V}_B$ respectively, leads to 
 \begin{eqnarray}
 \dot A & = & \varepsilon \langle {\boldsymbol V}_A, {\mathcal L}{\boldsymbol V}_A\rangle A + \varepsilon \langle {\boldsymbol V}_A, {\mathcal L}{\boldsymbol V}_B \rangle B, \label{eq:amplitude_A}\\
 \dot B & = & \varepsilon \langle {\boldsymbol V}_B, {\mathcal L}{\boldsymbol V}_A \rangle A + \varepsilon \langle {\boldsymbol V}_B, {\mathcal L}{\boldsymbol V}_B \rangle B.  \label{eq:amplitude_B}  \label{eq:amplitude_B}
 \end{eqnarray}
We note that all terms on the right-hand side of (\ref{eq:amplitude_A}) and (\ref{eq:amplitude_B}) are of order of magnitude $\varepsilon$. This suggests that its solution will be characterized by the time scale $\varepsilon^{-1}$. Thus, the rationale for the ansatz (\ref{eq:two_modes}) is that it permits a time scale separation between the amplitudes $A$ and $B$, characterized by the slow time scale $\varepsilon^{-1}$, and the fast rotation scale of order of magnitude $1$, captured by the inertial mode factors $\exp(\mathrm{i}\mu_{A,B}t)$.
The fact that $A$ and $B$ only depend on the slow time scale $\varepsilon^{-1}$ suggests the following asymptotic expansion:
\begin{equation}
A(t) = \sum_{k=0}^{\infty }\varepsilon^{k} A_k(t), B(t) = \sum_{k=0}^{\infty }\varepsilon^{k} B_k(t) \label{eq:scale_expansion}
\end{equation}
Substituting this into (\ref{eq:amplitude_A})-(\ref{eq:amplitude_B}) yields, at order 0 in $\varepsilon$, to $\dot A_0 = \dot B_0 = 0$. At the next order in $\varepsilon$, we obtain, after projection on ${\boldsymbol V_A}$, ${\boldsymbol V}_B$ respectively:
 \begin{eqnarray}
 \dot A_k & = & \varepsilon \langle {\boldsymbol V}_A, {\mathcal L}{\boldsymbol V}_A\rangle A_{k-1} + \varepsilon \langle {\boldsymbol V}_A, {\mathcal L}{\boldsymbol V}_B \rangle B_{0}, \\
 \dot B_k & = & \varepsilon\langle {\boldsymbol V}_B, {\mathcal L}{\boldsymbol V}_A\rangle A_{k-1}  + \varepsilon \langle {\boldsymbol V}_B, {\mathcal L}{\boldsymbol V}_B \rangle B_{k-1}. 
 \end{eqnarray}
The time dependence  of the diagonal terms in the above system is $\exp(\pm \mathrm{i}\omega_L t)$; for the off-diagonal terms, it is $\exp(\pm \mathrm{i}(\mu_A - \mu_B \pm \omega_L) t)$. Resonant coupling between the two modes, tantamount to the occurrence of instability, requires that $A_{k},B_{k}$ can grow unbounded as $t \rightarrow \infty$. This demands that
\begin{equation}
 \mu_A \pm \omega_L = \mu_B.  \label{eq:parametric_resonance}
\end{equation}
Since $-2 \le \mu_{A,B} \le 2$, couplings of the form (\ref{eq:parametric_resonance}) are limited to the libration frequency range $|\omega_L| \le 4$.  
\par
A further resonance condition can be found in the case of spheroidal geometries, for which the inertial modes ${\boldsymbol V}_{A,B}({\boldsymbol r})$ are of the form ${\boldsymbol W}_{A,B}(r,z)\exp(\mathrm{i}m_{A,B}\varphi)$, with $m_{A,B} \in \mathbb Z$, and $(r,\varphi,z)$ conventional cylindrical coordinates \cite[]{zhang2004inertial}. After substituting this into (\ref{eq:two_modes}), we find that the requirement that the right-hand side of (\ref{eq:power}) does not cancel, leads to the resonance condition
\begin{equation}
m_A \pm 1 = m_B. \label{eq:parametric_resonance_m}
\end{equation} 
\par
The above theoretical considerations are closely related to the GP method, which solves (\ref{eq:disturbance_v2}) in ellipsoidal enclosures. Devised originally by \cite{gledzer1978finite} and \cite{lebovitz1989}, it has since been employed by, amongst others, \cite{kerswell1993instability}, \cite{wu2011precession,WR2012} and \cite{roberts2011}. More precisely, this method restricts ${\boldsymbol v}$ to the vector space of perturbations that (1) are divergence-free, (2) satisfy the impermeability condition, and (3) have a vorticity field whose cartesian components are polynomials that have degree less than $n$ in the cartesian coordinates.
For a given polynomial degree $n$, this defines a vector space of finite dimension $N$, and thus one can expand the perturbation in a finite series
\begin{equation}
{\boldsymbol v} = \sum_{k=1}^{N} a_k(t){\boldsymbol V}^{\star}_j. \label{eq:GP_expansion}
\end{equation}
In this expression, the vectors ${\boldsymbol V}^{\star}_j$ denote the basis vectors of the GP subspace of degree $n$, and are cartesian polynomials of degree $n$ or less.
As a consequence of the particular nature of the Coriolis operator and of the uniform vorticity character of the base flow ${\boldsymbol u}_0$, the expansion (\ref{eq:GP_expansion}) reduces (\ref{eq:disturbance_v2})  to a closed system of $N$ linear ODEs for the functions $a_1(t),a_2(t),...,a_N(t)$. Since the coefficients of this system are time-periodic with period $T=2\pi/\omega_L$, one can solve it by means of Floquet analysis, or directly by brute numerical force; the last method was applied in this work. First, we integrate the ODEs in time from $t=0$ to $t=3000$. Then, for $k=1, 2, ..., N$, we fit $\log |a_k(t)|$ with $\log |a_k(t)|=A_k + \sigma_k t$, which gives the fit parameters $A_k$ and $\sigma_k$, from which we determine the growth rate $\sigma=\max_k \sigma_k $. Since this procedure is the source of small uncertainties in the measurement of $\sigma$, we filter away values of $\sigma < 0.02\varepsilon$.
\begin{figure}  
\begin{center}
\includegraphics[width=\textwidth]{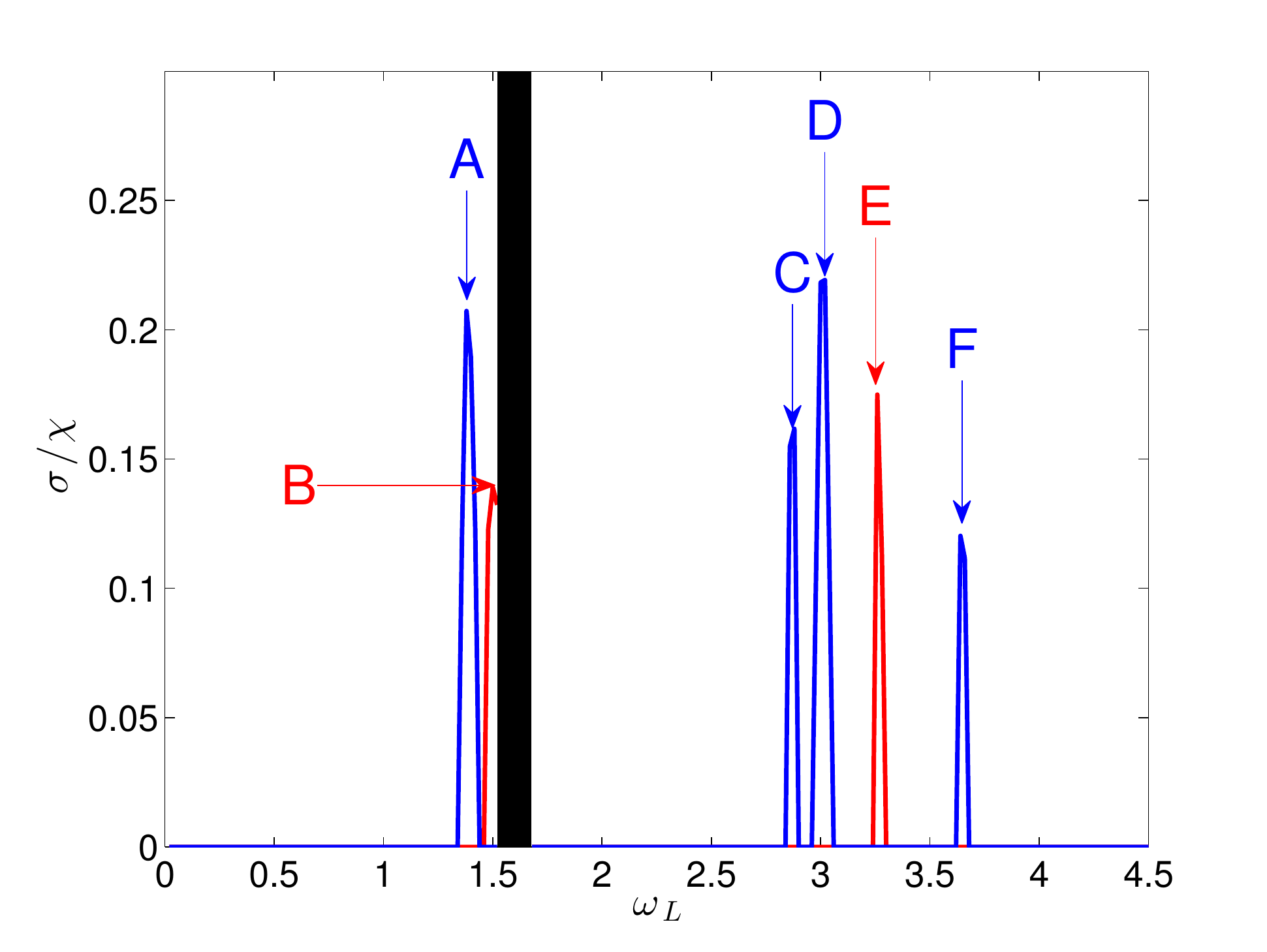} \\
\begin{tabular}{ccc}
    \begin{tabular}{  l  l l l }
    \hline
    \hline
    \multicolumn{2}{c}{Inertial modes} \\
    \hline
    Quadratic & Cubic  \\ 
    \hline
    $\mu_{1} = 0.18046$ ($m=1$) & $\mu_{5}=  0.24499$ ($m=2$)\\
    $\mu_{2} = 1.3333$   ($m=2$) & $\mu_{6} = 1.1399$ ($m=1$) \\
    $\mu_{3} = 1.4142$   ($m=0$) & $\mu_{7} = 1.1429$ ($m=3$) \\
    $\mu_{4} = 1.8471$   ($m=1$) & $\mu_{8} = 1.2921$ ($m=1$) \\
                                                             & $\mu_{9} = 1.7187$ ($m=2$) \\
                                                             & $\mu_{10} = 1.7321$ ($m=0$) \\
                                                             & $\mu_{11} = 1.9169$ ($m=1$) \\
                                                             & $\mu_{12} = 0$ ($m=2$)\\
    \hline
    \hline
\end{tabular} & &
    \begin{tabular}{  c  c c  }
     \hline
    \hline
    \multicolumn{3}{c}{Parametric resonances} \\
    \hline
    & $\omega_{resonance}$ & $(\mu_a,\mu_b)$  \\ 
    \hline  
        A & 1.40 & ($\mu_5,\mu_6$), ($\mu_5,\mu_7$)  \\
    B & 1.52 & ($\mu_1,\mu_2$) \\
C & 2.88 &($\mu_6,\mu_{10}$) \\ 
  D & 3.00 & ($\mu_8,\mu_9$)\\
    E & 3.26 & ($\mu_3,\mu_{4}$) \\
    F & 3.64 &  ($\mu_9,\mu_{11}$), ($\mu_{10},\mu_{11}$) \\
        \hline
        \hline
\end{tabular}
\end{tabular}
    \caption{GP analysis for the spheroidal geometry with $(a,b,c)=(1,1,0.5)$ using quadratic (red, dashed) and cubic (blue) polynomial perturbations. Inviscid normalized growth rate $\sigma/\chi$ as function of the libration frequency $\omega_L$. 
 Notations A-F indicate the main resonance peaks and are given in the table at the right-hand side. The black rectangle indicates the frequency band centered around the direct resonance at $\omega_L=1.6$ where the inviscid base flow ${\boldsymbol U}_0$ diverges.}               
  \label{fig:GPspheroid}
  \end{center}

\end{figure}

\begin{figure}  
\begin{center}
\includegraphics[width=\textwidth]{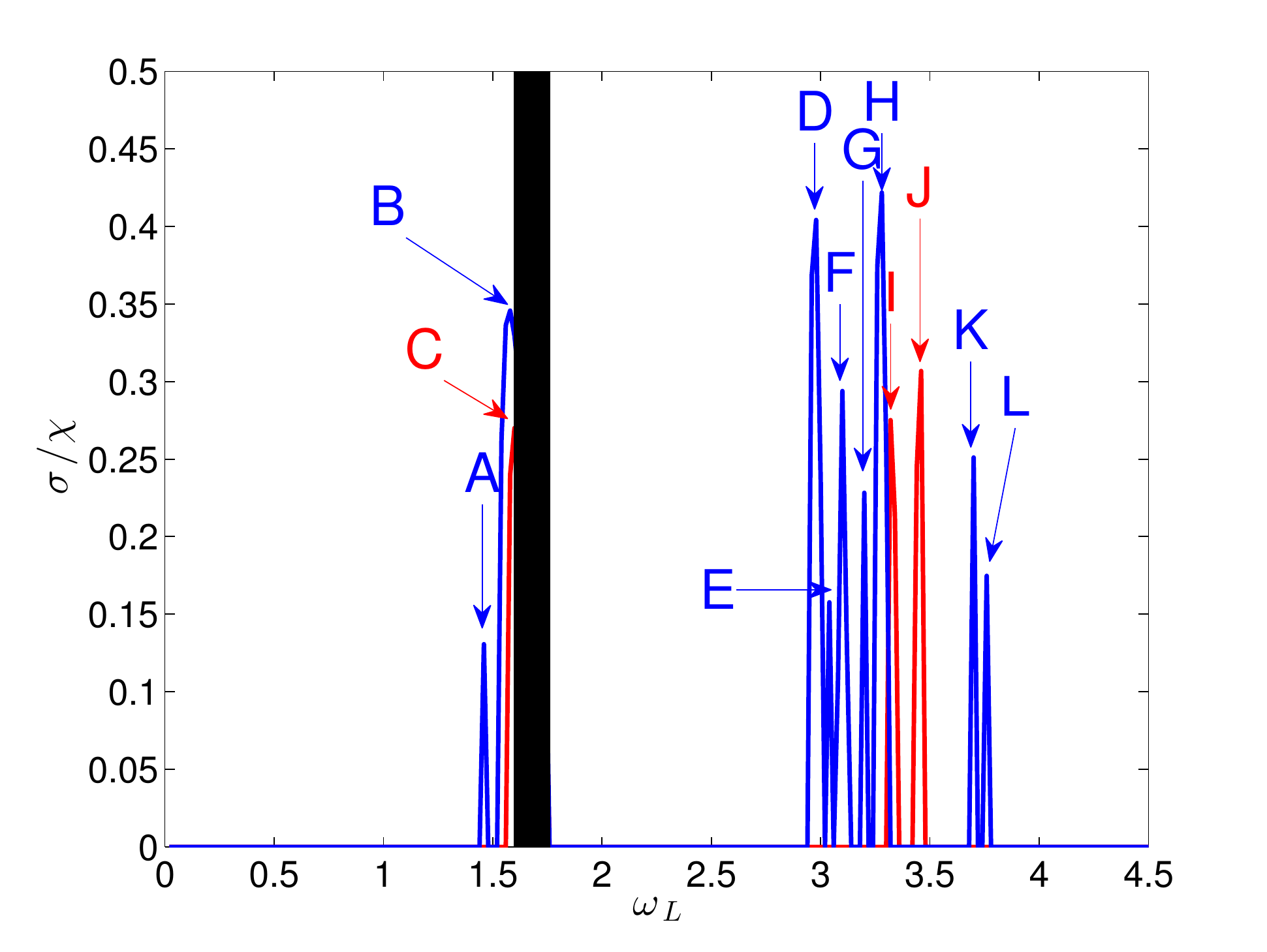} \\
\begin{tabular}{ccc}
    \begin{tabular}{  l  l l l }
    \hline
    \hline
    \multicolumn{2}{c}{Inertial modes} \\
    \hline
    Quadratic & Cubic  \\ 
    \hline
    $\mu_{1} = 0.1696 $& $\mu_{5}=  0.23553$ \\
    $\mu_{2} = 1.4406 $& $\mu_{6} = 1.2181$ \\
    $\mu_{3} = 1.5674 $  & $\mu_{7} = 1.3417$ \\
    $\mu_{4} = 1.8874 $  & $\mu_{8} = 1.4481$ \\
                                                   & $\mu_{9} = 1.7596$ \\
                                                   & $\mu_{10} = 1.8266$ \\
                                                   & $\mu_{11} = 1.9394$ \\
                                                   & $\mu_{12} = 0$ \\
    \hline
    \hline
\end{tabular} & &
    \begin{tabular}{  c  c c | c c c }
     \hline
    \hline
    \multicolumn{6}{c}{Parametric resonances} \\
    \hline
    & $\omega_{resonance}$ & $(\mu_a,\mu_b)$  & &$\omega_{resonance}$ & $(\mu_a,\mu_b)$ \\ 
    \hline
    A & 1.46 & ($\mu_{5},\mu_{6}$)&  G & 3.20 &  ($\mu_{8},\mu_{9}$) \\ 
    B & 1.58 &  ($\mu_{5},\mu_{7}$)&  H & 3.28 & ($\mu_8,\mu_{10}$) \\ 
    C& 1.60 &  ($\mu_{1},\mu_{2}$) &   I & 3.32 &  ($\mu_2,\mu_{4}$) \\ 
    D & 2.98 &($\mu_6,\mu_{9}$) &   J & 3.46 &  ($\mu_3,\mu_{4}$) \\
    E& 3.04 &  ($\mu_{6},\mu_{10}$) &   K & 3.70 &  ($\mu_9,\mu_{11}$) \\
    F& 3.10 &  ($\mu_7,\mu_{9}$) &   L & 3.76 &  ($\mu_{10},\mu_{11}$) \\

        \hline
        \hline
\end{tabular}
\end{tabular}
    \caption{GP analysis for the triaxial geometry with $(a,b,c)=(1,1.5,0.5)$ using quadratic (red,dashed) and cubic (blue) polynomial perturbations.  Inviscid normalized growth rate $\sigma/\chi$ as function of the libration frequency $\omega_L$.Notations A-K indicate the main resonance peaks and are given in the table at the right-hand side. The black rectangle indicates the frequency band centered around the direct resonance at $\omega_L = 1.68$ where the inviscid base flow ${\boldsymbol U}_0$ diverges.}      
         \label{fig:GPtriax}
            
  \end{center}

\end{figure}
\par
Recently, \cite{vantieghem2014inertial} established that one can always construct an orthogonal basis of inertial modes for the (finite-dimensional) GP subspaces of polynomial vector fields. Hence, the GP method describes the dynamics of the perturbation ${\boldsymbol v}_2$ in terms of a superposition of inertial modes, and thus allows the identification of possible resonant couplings. This sets out the basic principles for a global analysis that is valid for ellipsoids of arbitrary shape. 
\par
To illustrate this, we solve (\ref{eq:disturbance_v2}) using quadratic ($n=2$) and cubic ($n=3$) GP subspaces for two configuration: (1) a spheroid with $a=b=1,c=0.5$, (2) a fully triaxial geometry with $a=1, b=1.5, c=0.5$. We consider the libration frequency band between 0 and 4.5; the value of $\varepsilon$ is fixed at 0.1. Because the base flow ${\boldsymbol u}_0$ diverges as $\omega_L$ approaches the spin-over frequency $f$, we exclude a small frequency window of width $\Delta \omega = 0.15$ around $f$. In figures \ref{fig:GPspheroid} and \ref{fig:GPtriax}, we plot $\sigma/\chi$ for the spheroidal and triaxial geometry, respectively. These figures also provide the inertial mode spectrum of quadratic and cubic polynomial inertial modes. For the spheroidal geometry, we tabulate the azimuthal wave number $m$ of the mode as well.
\par
For both configurations, we observe a spectrum characterized by sharp peaks, which reflects the parametric-resonance-like nature of the instability. We see that the upper bound (\ref{eq:bound_sigma}), i.e. $\sigma \le \chi = |\chi_1|+|\chi_2|$, indeed holds. Further inspection of the main peaks (denoted by capital letters in figures \ref{fig:GPspheroid} and \ref{fig:GPtriax}) shows that these resonant frequencies are indeed associated with the criterion (\ref{eq:parametric_resonance}), where $\mu_{A}$ and $\mu_{B}$ correspond to values that are listed in the left-hand side of the respective figures.  For the spheroidal geometry (figure \ref{fig:GPspheroid}), we see that every resonance is associated with an inertial mode pair that also satisfies the condition (\ref{eq:parametric_resonance_m}). The inertial mode spectrum of the spheroid is such that one resonance peak can sometimes be related to multiple pairs of modes. The triaxial configuration discussed in figure \ref{fig:GPtriax} exhibits more resonance peaks. In terms of mode interactions, one can argue intuitively that the equatorial deformation induces additional strain components that give birth to an extended set of mode couplings. 
\par
The results of the GP analysis thus give evidence that libration in latitude indeed drives parametric resonances that arise from inertial mode couplings. Having focused on quadratic and cubic inertial modes, we have (only) found a handful of resonances. More resonances would be expected, however, if we were to include higher-order polynomial GP bases. Indeed, since the inertial mode spectrum is dense \cite[]{greenspanbook}, it is always possible to find two inertial mode frequencies that satisfy the criterion (\ref{eq:parametric_resonance}) for a given frequency $\omega_L$. As such, we expect that instability can take place for any libration frequency $|\omega_L| \le 4$.
\subsection{Direct numerical simulations} \label{subsec:instab_simu}
In this subsection, we compare the results of the LH and GP analysis against direct numerical simulations. In contrast to the preceding theoretical approaches, it is not feasible to carry out numerical simulations in the inviscid regime. Therefore, we have to reintroduce a viscous term into the perturbation equation (\ref{eq:disturbance_v2}). However, we replace the no-slip boundary condition by the stress-free condition
\begin{eqnarray}
\hat{\boldsymbol n} \cdot {\boldsymbol u}= 0, & \hspace{1em} & \hat{\boldsymbol n} \times \left[\hat{\boldsymbol n} \cdot \left(\nabla {\boldsymbol u} + (\nabla {\boldsymbol u})^T \right) \right] = {\boldsymbol 0}. \label{eq:stress_free}
\end{eqnarray}
The rationale for this choice is that the no-slip condition may give rise to centrifugal-type instabilities if $\varepsilon$ is too high and/or $E$ too low \cite[]{noir2009,sauret2012fluid}. Stress-free boundary conditions on the other hand, allow isolation of inertial instabilities, such as the parametric resonances discussed above. A further advantage of this choice is that we avoid the computationally expensive task of resolving thin Ekman boundary layers. We keep the value of the Ekman number fixed at $E=10^{-4}$, and we will focus on one particular geometry, characterized by the length of the semi-axes $(a,b,c)=(1,1.5,0.5)$, i.e. the same triaxial geometry as studied in subsection \ref{subsec:GP}. The deformation chosen is thus rather large. The motivation for this choice resides in the fact that our theoretical methods are not restricted to asymptotically small deformations, in contrast to other theoretical analyses \cite[e.g.]{kerswell1994tidal,le2010tidal}. 
\par
\begin{figure}                 
  \begin{center}
  \setlength{\epsfysize}{5.0cm}
    \begin{tabular}{ccc}
          \setlength{\epsfysize}{5.0cm}
      \subfigure[]{\epsfbox{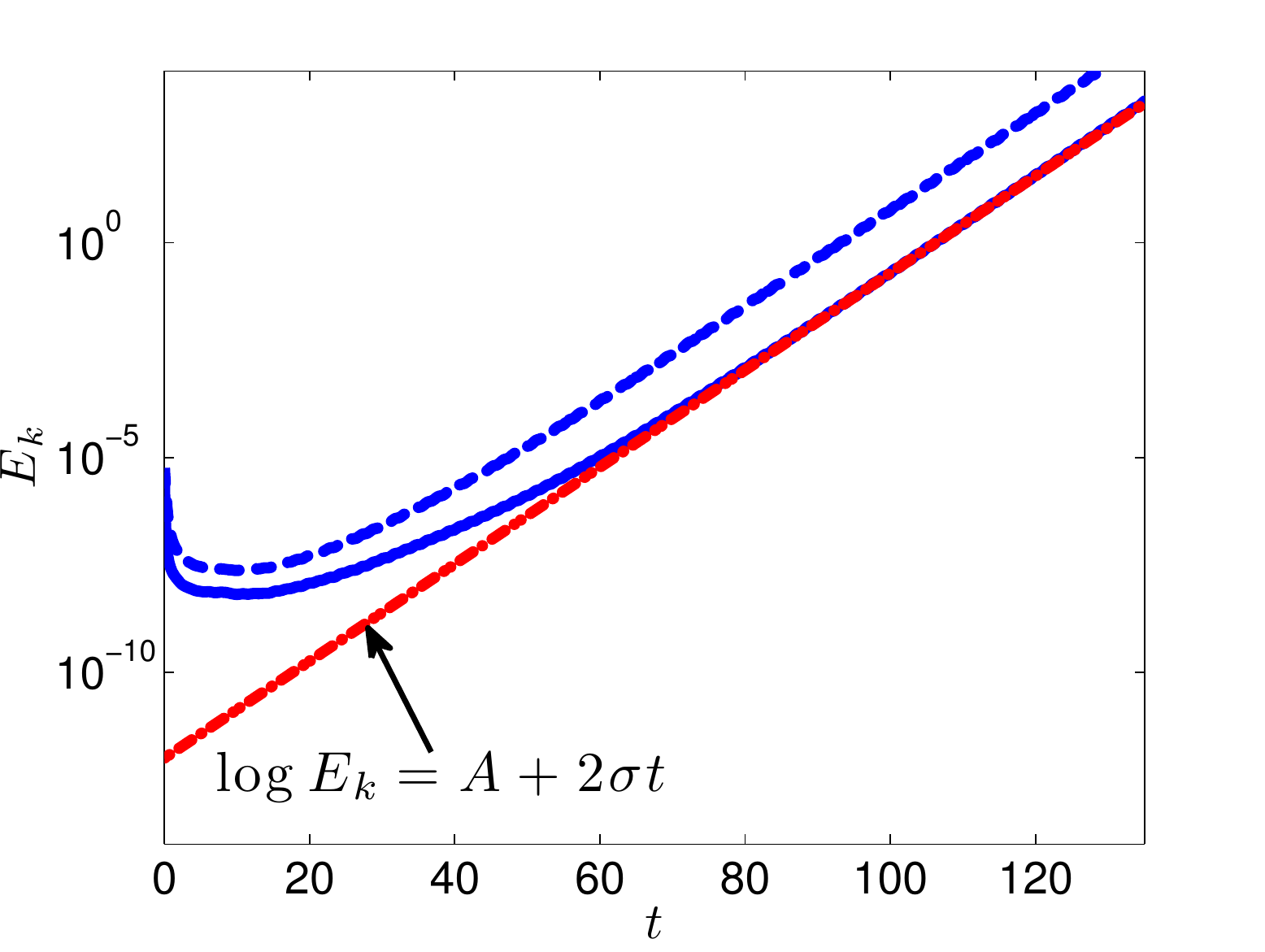}} &
      \setlength{\epsfysize}{5.0cm}
        \subfigure[]{\epsfbox{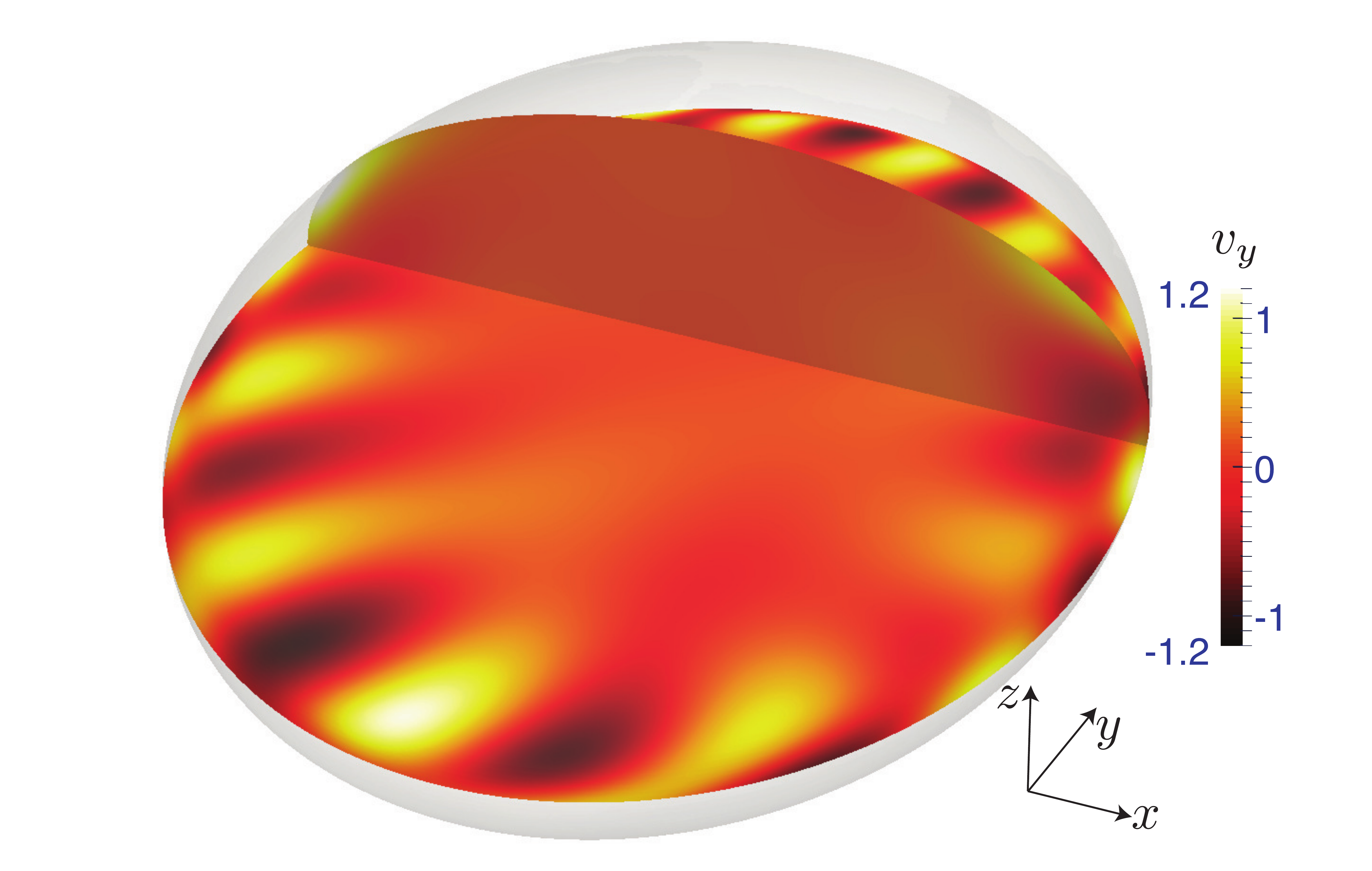}}
      \end{tabular}
    \caption{Direct numerical solution of the (viscous) perturbation equation for $\varepsilon=0.3$ and $\omega_L=2.1$. (a) Time series of the kinetic energy $E_k$ for a simulation using 1.3 million (dashed, blue) and 10.5 million (solid, blue) mesh elements. Also shown is an exponential fit to determine the growth rate $\sigma$ (dot-dashed, red). (b) Snapshot of the velocity field $v_y$ in the planes $y=0$ and $z=0$, illustrating the spatial structure of the growing instability. }
         \label{fig:timeseries_snapshot}                
  \end{center}
\end{figure}

In a first series of numerical simulations, we seek solutions of the linear perturbation equation (\ref{eq:disturbance_v2}). As initial condition we impose a random velocity field  with a root-mean-square amplitude of $5\cdot 10^{-3}$. We use the same numerical method as described in subsection \ref{subsec:numerics}. The computational mesh consists of uniformly distributed tetrahedral elements. In order to verify that our spatial resolution is adequate, two different mesh sizes have been considered, containing 1.3 to 10.5 million CVs respectively. The results reported hereafter are those obtained with the highest resolution; the quantities of interest do not differ more than $\sim 1\%$ from those at lower resolution. This is shown in figure \ref{fig:timeseries_snapshot}a, where we compare the evolution of perturbation kinetic energy $E_k$ for $\varepsilon=0.3,\omega_L=2.1$ for the two different resolutions. Figure \ref{fig:timeseries_snapshot}b illustrates the spatial structure of the growing instability for this parameter set.
\par

\begin{figure}  
\begin{center}
\includegraphics[width=0.8\textwidth]{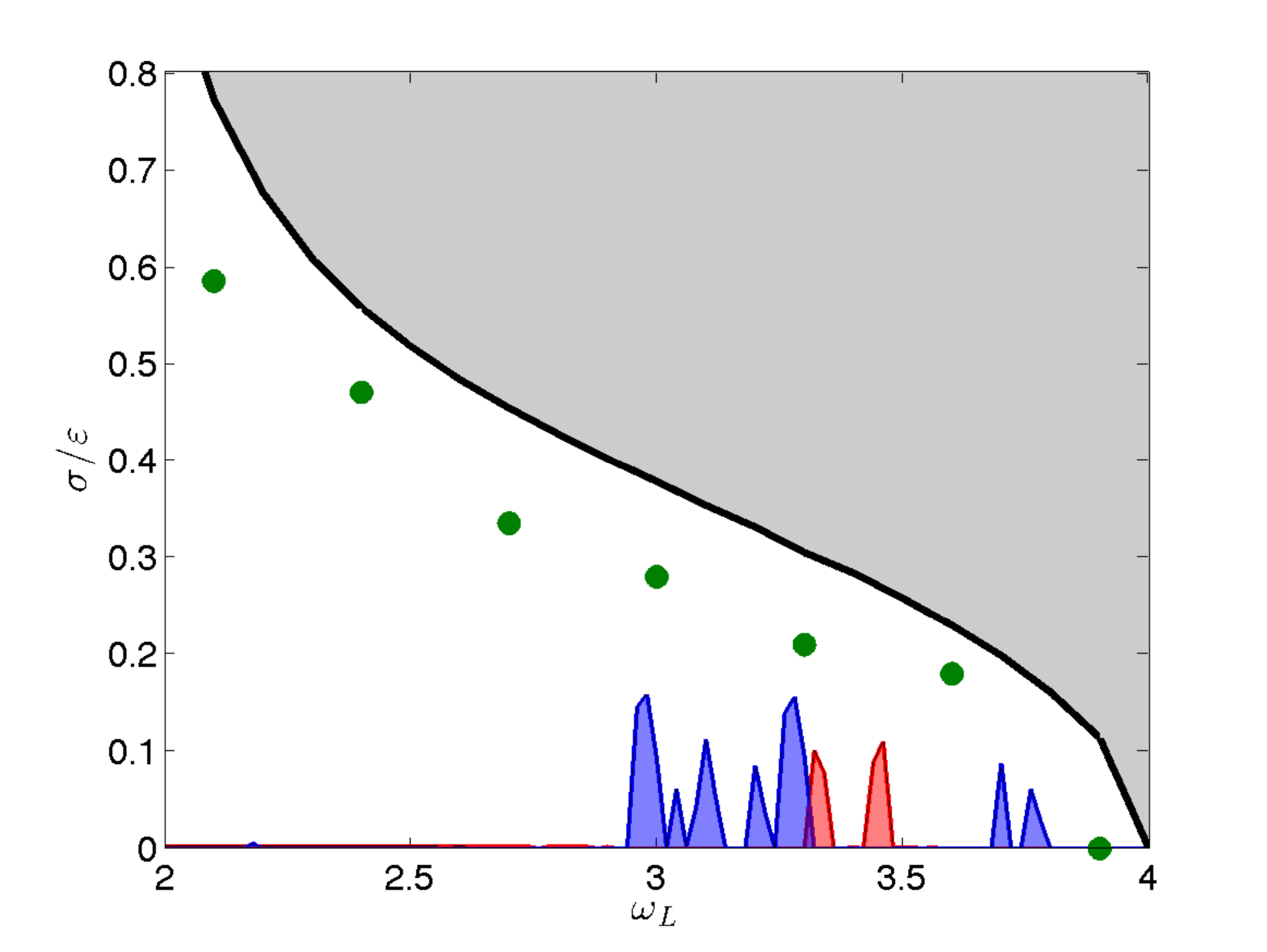}
    
    \caption{Stability analysis in a triaxial configuration with $(a,b,c)=(1,1.5.0.5)$. Comparison of growth rates resulting from the local LH analysis (solid black line), the global GP analysis for quadratic (red dashed line) and cubic (blue dash-dotted line) polynomial perturbations and direct-numerical simulation (green circles).}             
    \label{fig:comparison_LH_GP_num}   
   
  \end{center}
\end{figure}

We are now concerned with comparing the growth rates between the LH, GP and numerical approaches. To extract growth rates from our simulations, we proceed as follows. For a given value of $\varepsilon$, we fit a curve of the form $\log E_k = A + 2 \sigma t$ to the numerical time series, as can be seen from figure \ref{fig:timeseries_snapshot}a. The fitting window is chosen large with respect to $\omega_L$. The slope of this line gives a certain value for $\sigma(\varepsilon,\omega_L,E=10^{-4})$. Because we use stress-free boundary conditions and $\varepsilon$ is small compared to one, we can argue that (1) $\sigma$ is linear in $\varepsilon$, and (2) the viscously modified growth rate $\sigma_{visc}$ is related to the inviscid growth rate $\sigma$ by
\begin{equation}
\sigma_{visc} = \sigma - \kappa E =  \alpha(\beta_{ac},\beta_{bc},\omega_L)\varepsilon - \kappa E,
\end{equation}
with $\alpha$ independent of $\varepsilon$. To eliminate the unknown coefficient $\kappa$, we perform simulations for different values of $\varepsilon$ and report $\displaystyle{\alpha =  \sigma / \varepsilon = \Delta \sigma_{visc}/\Delta \varepsilon}$ in figure \ref{fig:comparison_LH_GP_num}.
\par
The values of the growth rate retrieved from the numerical simulations are located in the non-shaded area in figure \ref{fig:comparison_LH_GP_num}, i.e they are contained between the GP lower  and LH upper bound. The numerics and the local LH analysis exhibit the same trend for the dependence of $\sigma$ on $\omega_L$. However, the simulations yield growth rates that are about 35\% smaller than the ones found in the local LH analysis. This is consistent with the interpretation of the LH growth rates as upper bounds. Indeed, we recall that the LH approach does not constrain the perturbations to be bounded, and therefore provides a supremum rather than a maximum for the growth rate. It is worth noting that discrepancies of the same order of magnitude have been observed between the local and global stability analysis for flows driven by precession, libration and tides \cite[]{kerswell1993instability,kerswell1994tidal,cebronAA}.  
\par
\begin{figure}                 
  \begin{center}
  \setlength{\epsfysize}{7.0cm}
    \begin{tabular}{ccc}
          \setlength{\epsfysize}{7.0cm}
      \subfigure[]{\epsfbox{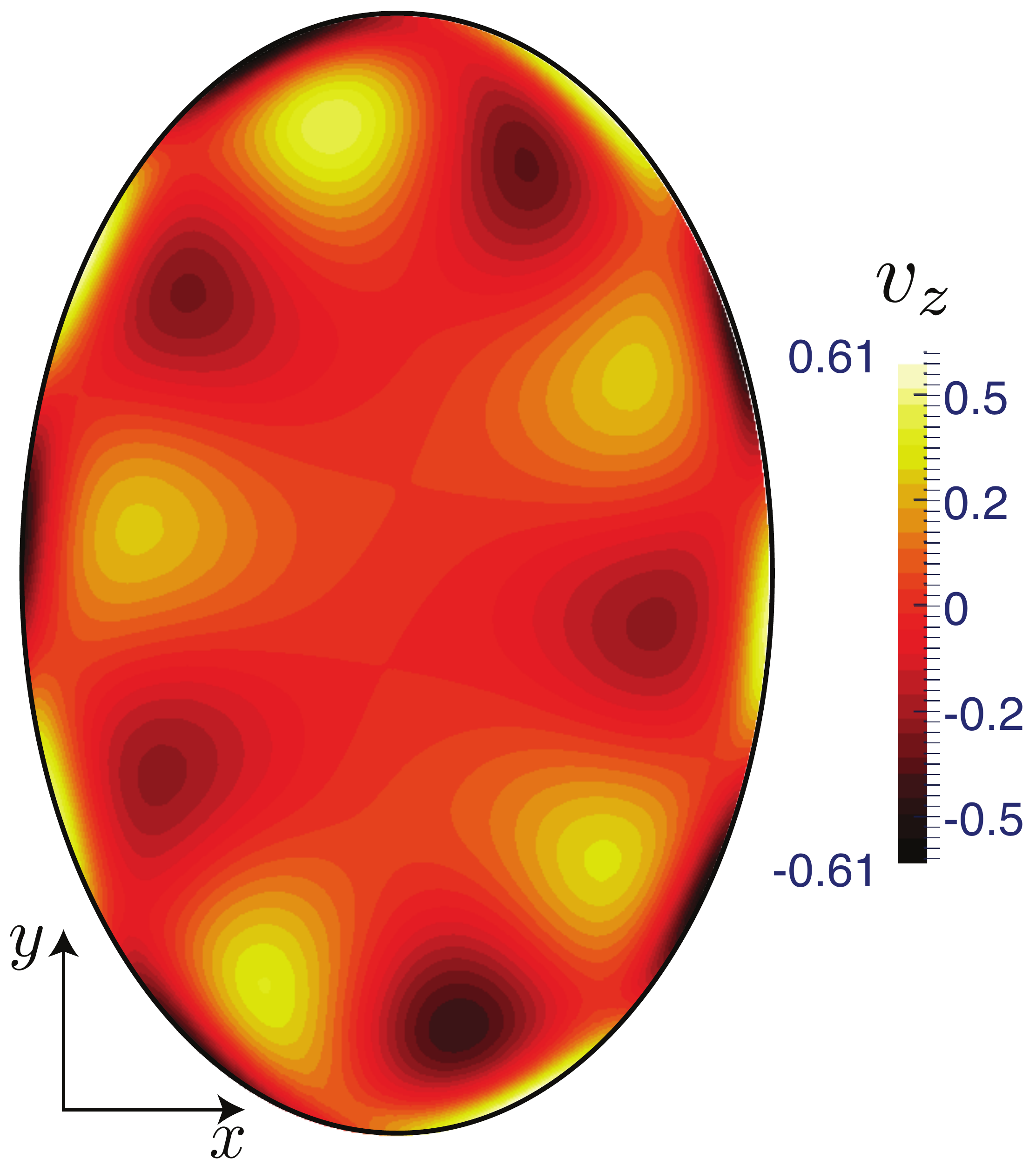}} &
      \setlength{\epsfysize}{7.0cm}
        \subfigure[]{\epsfbox{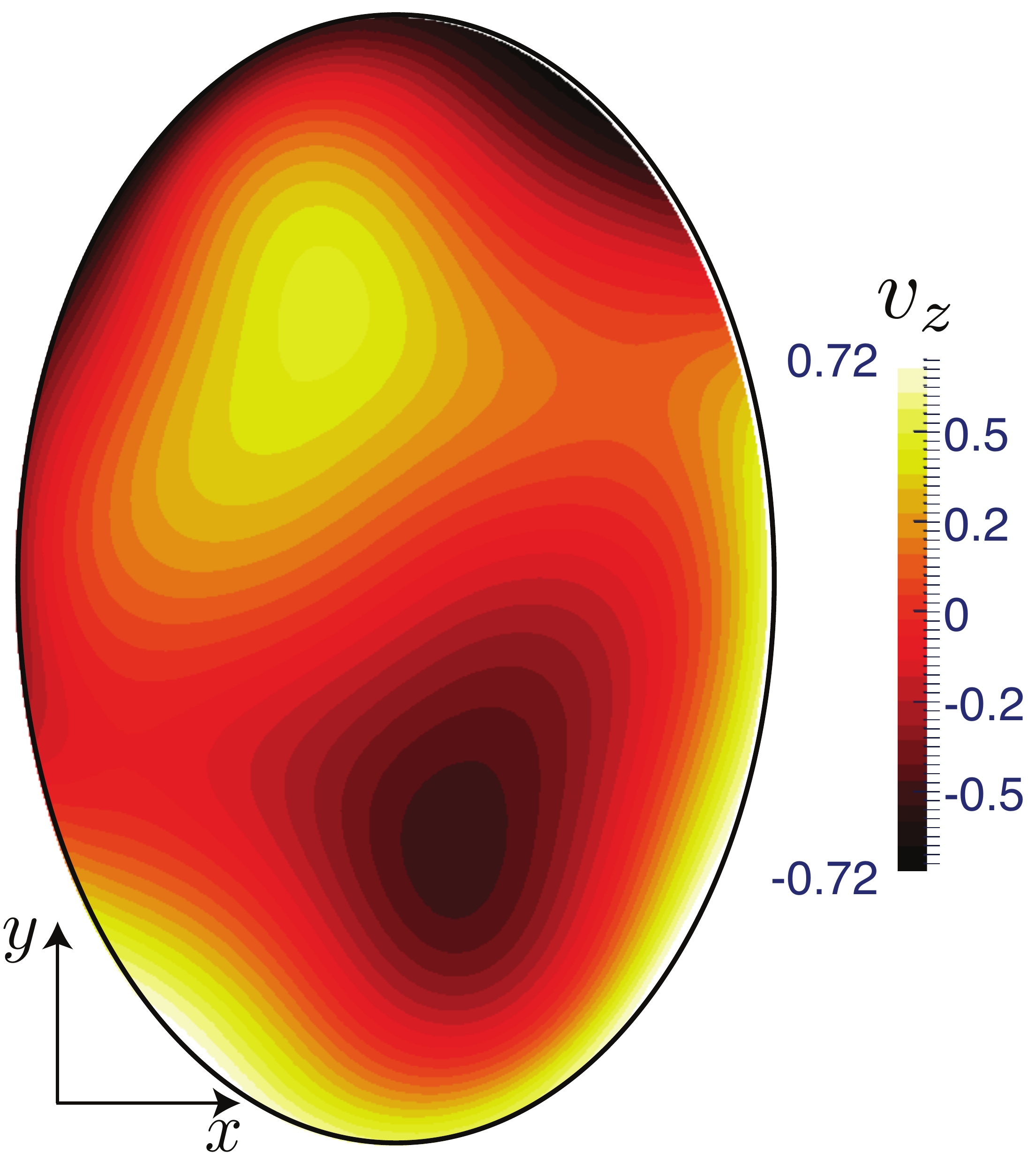}}
      
    \end{tabular}
    \caption{Instantaneous profile of $u_z$ in the $z=0$-plane for $\varepsilon=0.3, \omega_L=2.7$ (a) and $\varepsilon=0.3, \omega_L=3.3$ (b).}
         \label{fig:snapshot_instab}                
  \end{center}
\end{figure}

According to figure \ref{fig:comparison_LH_GP_num} (or equivalently figure \ref{fig:GPtriax}), the GP analysis does not predict instability for $\omega_L$ between 2 and roughly 2.9. Our simulations on the other hand show that instability occurs for almost the whole frequency window $\omega_L \in [2,4]$. This implies that the numerically observed instabilities for $\omega_L=2.1,2.4,2.7$ should be associated with inertial modes of polynomial degree $n>3$. This is indeed the case, as is illustrated in figure \ref{fig:snapshot_instab}, where we show a snapshot of $u_z$ in the equatorial plane for $\omega_L=2.7$ and $3.3$.  The velocity profile for $\omega_L=2.7$ (figure \ref{fig:snapshot_instab}a) is  too  complex to be well captured by a quadratic or cubic polynomial. The corresponding profile for  $\omega_L=2.7$ on the other hand has a spatially simpler structure.
\par
\begin{figure}                 
  \begin{center}
  \setlength{\epsfysize}{5.0cm}
    \begin{tabular}{ccc}
          \setlength{\epsfysize}{5.0cm}
      \subfigure[]{\epsfbox{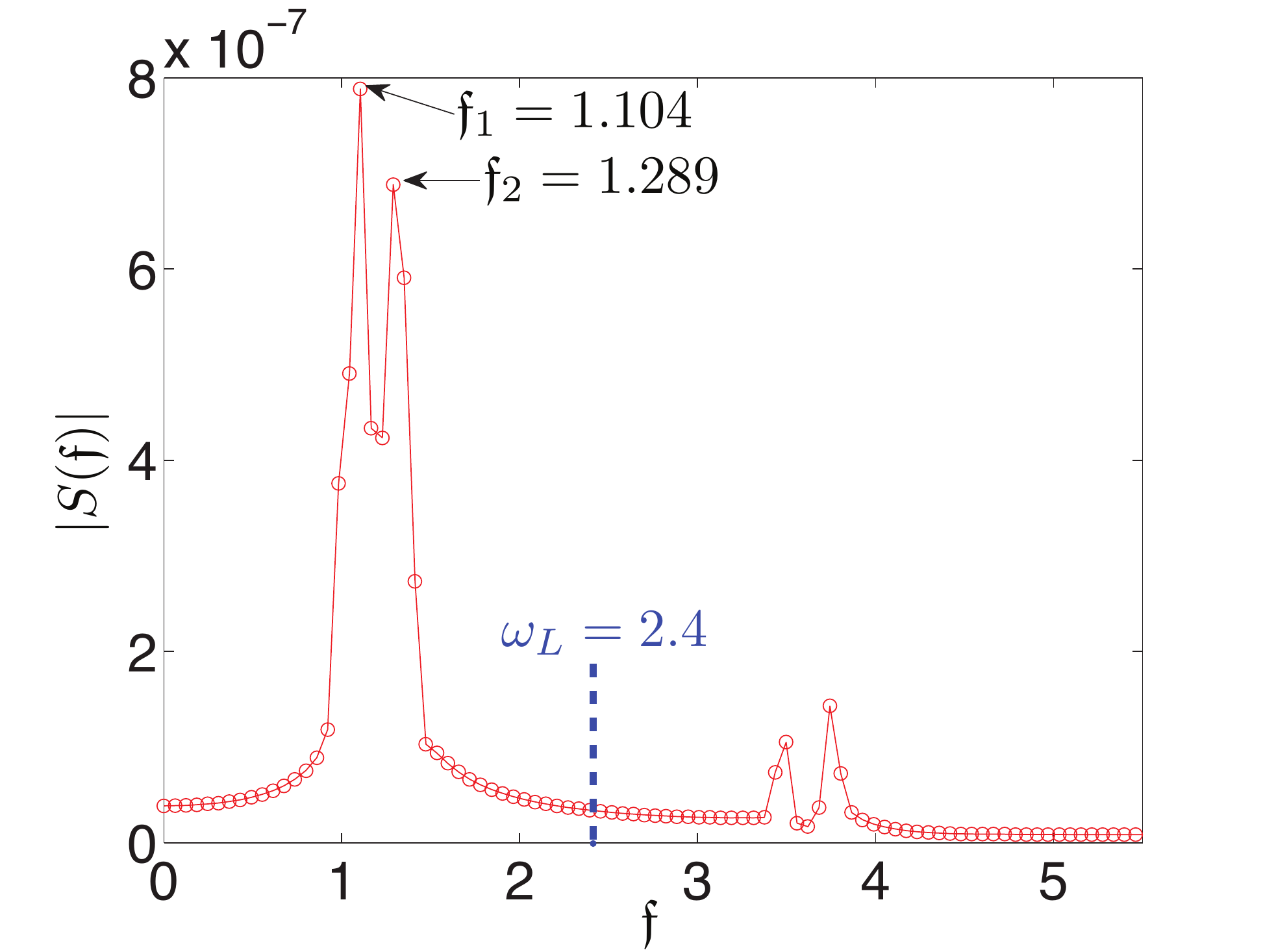}} &
      \setlength{\epsfysize}{5.0cm}
        \subfigure[]{\epsfbox{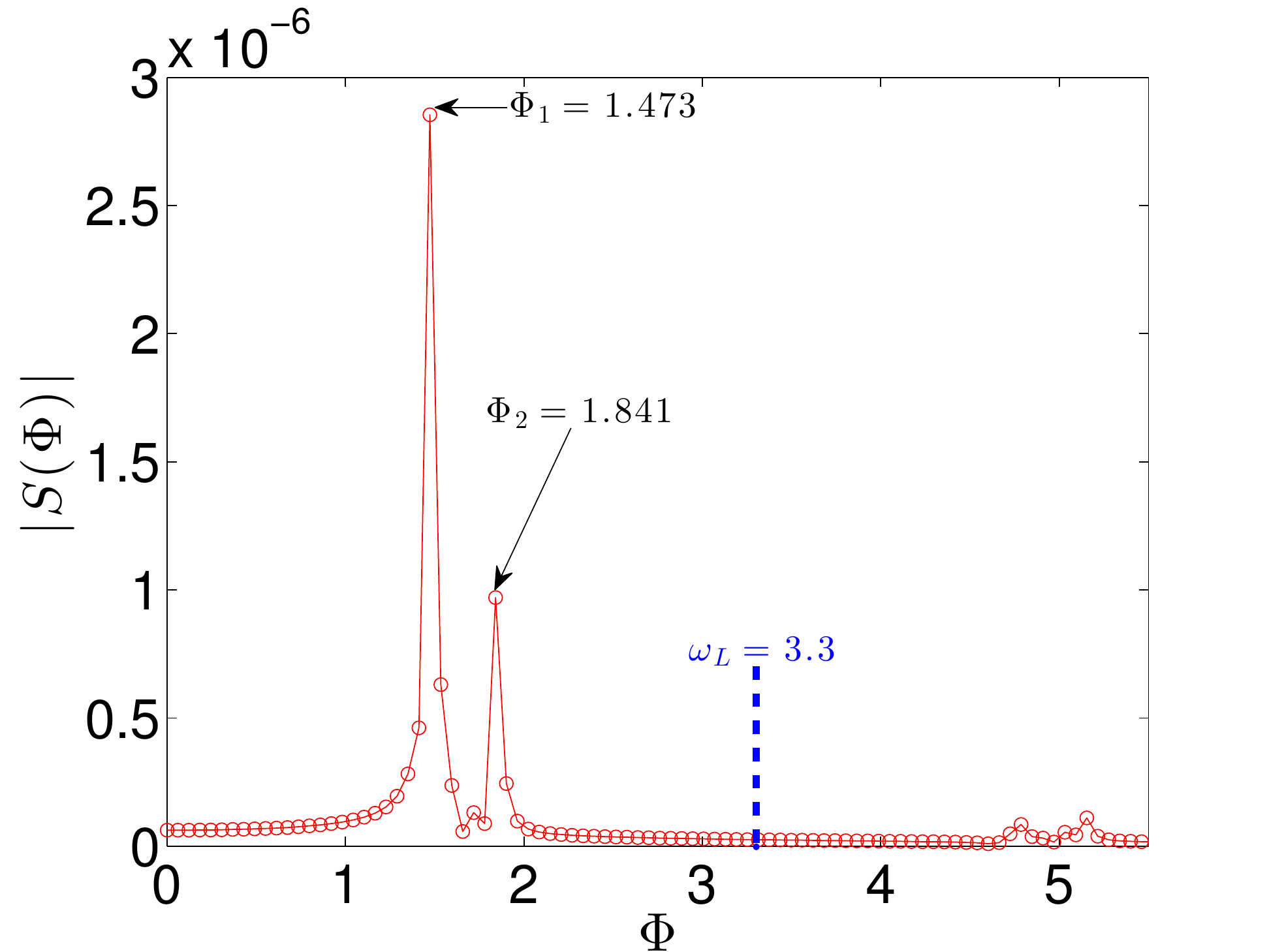}}
      \end{tabular}
    \caption{Discrete Fourier Transform of ${\boldsymbol s}(t)={\boldsymbol v}({\boldsymbol r}_0,t)\exp(-\sigma t)$ at location ${\boldsymbol r}_0 = (0.25, 0.6,0.15)$ for $\varepsilon=0.2, \omega_L=2.4$ (a) and $\varepsilon=0.2, \omega_L=3.3$ (b). Dashed blue lines indicate the driving frequency $\omega_L$.}
         \label{fig:FFT}                
  \end{center}
\end{figure}

In the earlier discussion we have shown that the instability arises by virtue of a parametric resonance mechanism coupling two inertial modes. We now investigate whether this can also be achieved in our numerical simulations. To this end, we consider time series of the velocity field at one single point ${\boldsymbol r}_0=(0.25,0.6,0.15)$. In figure \ref{fig:FFT}, we trace the frequency content $|S(\mathfrak{f})|$ of ${\boldsymbol s}(t) = {\boldsymbol v}({\boldsymbol r}_0,t)\exp(-\sigma t)$. The factor $\exp(-\sigma t)$ is a compensation for the exponentially growing amplitude of ${\boldsymbol v}$. $|S(\mathfrak{f})|$ is defined by $|S(\mathfrak{f})|^2=|S_x(\mathfrak{f})|^2+|S_y(\mathfrak{f})|^2+|S_z(\mathfrak{f})|^2$, where $S_{x,y,z}(\mathfrak{f})$ denote the discrete Fourier transforms of the respective components of ${\boldsymbol s}(t)$.  We find that the spectrum is clearly dominated by two peaks, that moreover satisfy the resonance condition $\mathfrak{f}_1 + \mathfrak{f}_2 \approx \omega_L$. This confirms that a parametric resonance mechanism is operating. 
In figure \ref{fig:FFT}a, we also observe smaller peaks around $\omega \approx 3.5, 3.7$. These are subharmonic effects that are due to the term ${\boldsymbol v} \cdot \nabla {\boldsymbol U}_0+{\boldsymbol U}_0 \cdot \nabla {\boldsymbol v}$ in (\ref{eq:disturbance_v2}) that couples the perturbation ${\boldsymbol v}$ with frequency content $f=1.104, 1.289$ and the base flow oscillating at $\omega_L=2.4$. A similar phenomenon is also present in figure \ref{fig:FFT}b.

\par
\begin{figure}                 
  \begin{center}
  \setlength{\epsfysize}{5.5cm}
    \begin{tabular}{ccc}
          \setlength{\epsfysize}{5.5cm}
      \subfigure[]{\epsfbox{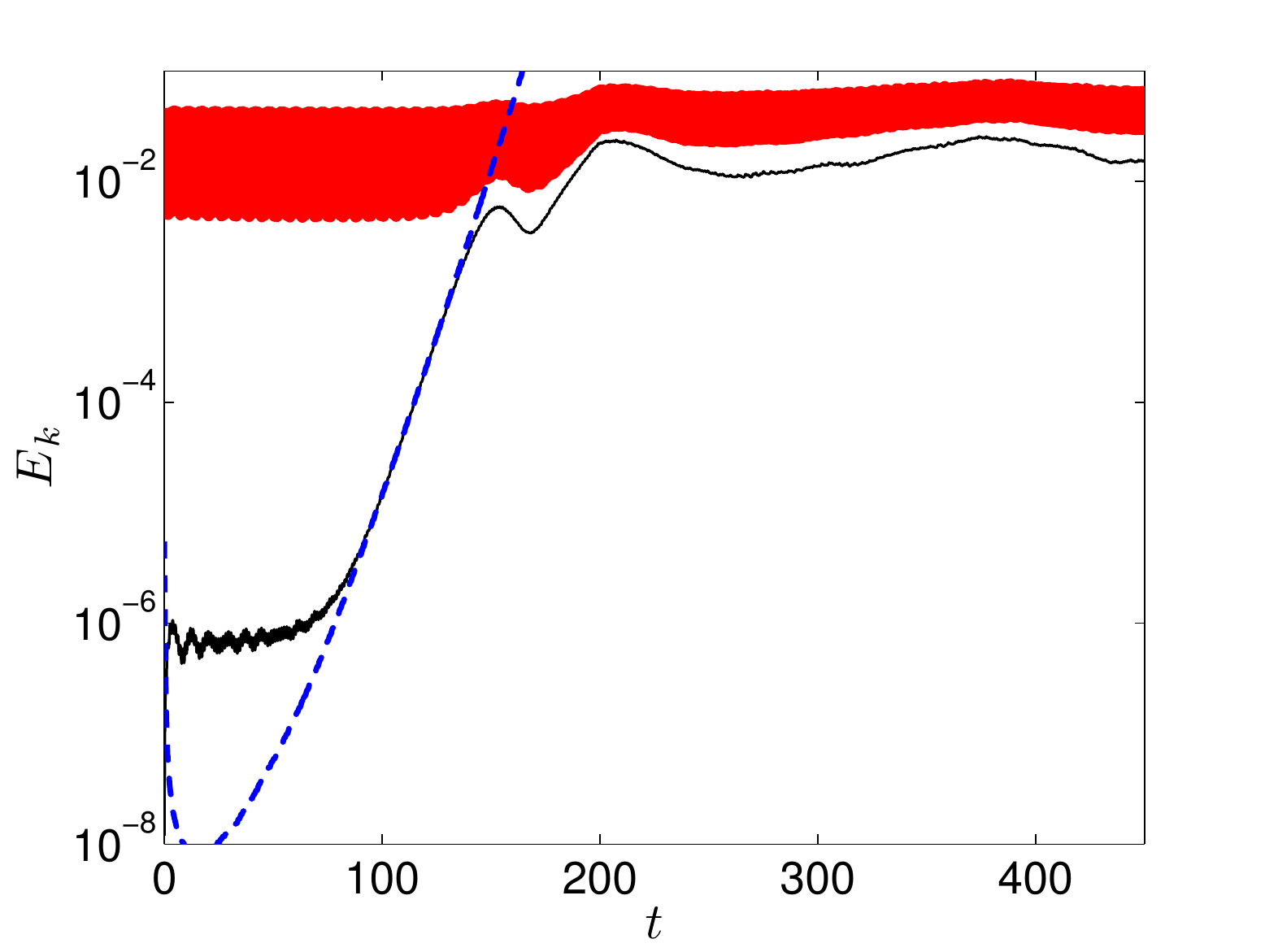}} &
      \setlength{\epsfysize}{5.5cm}
        \subfigure[]{\epsfbox{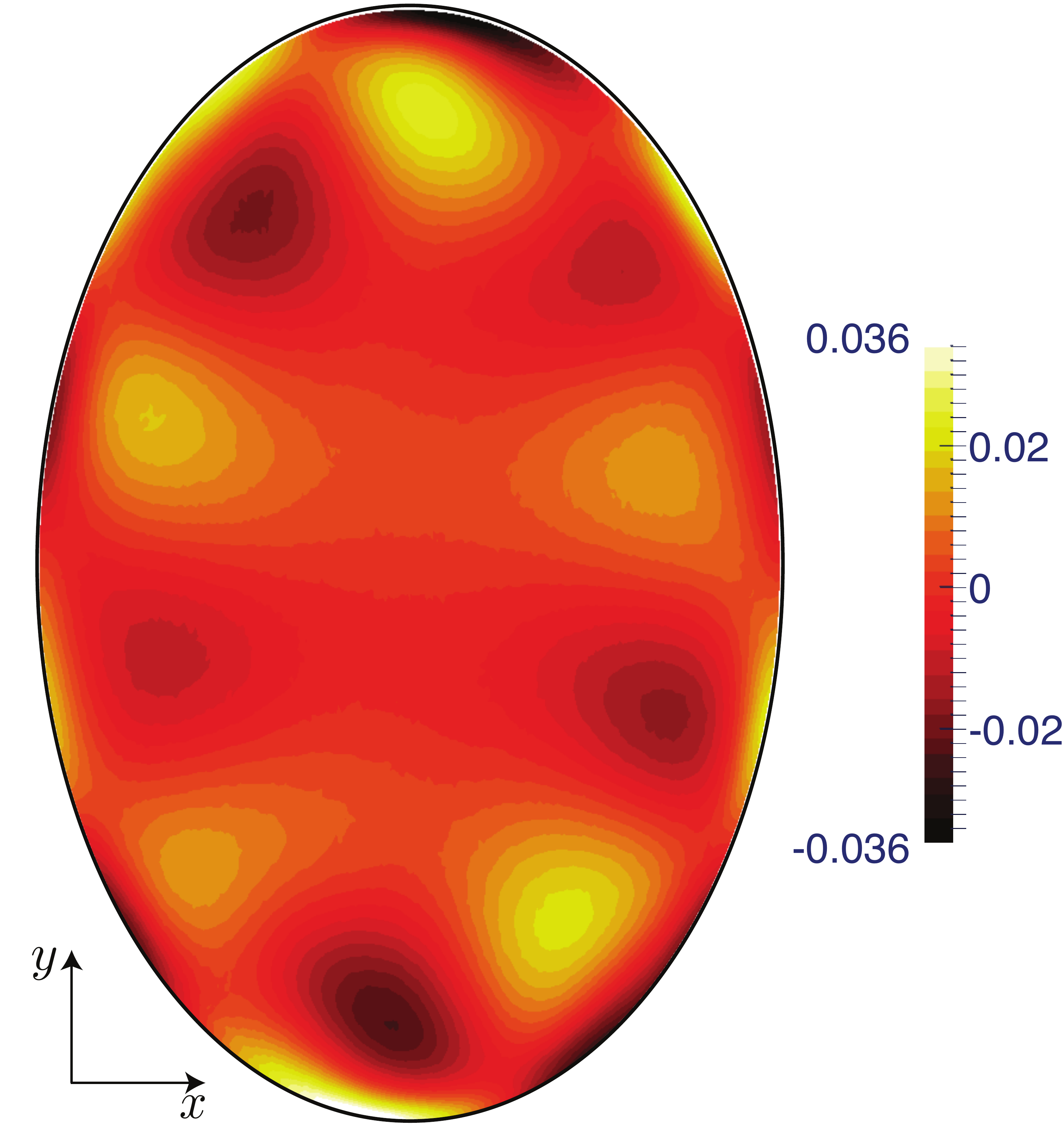}}
      \end{tabular}
    \caption{(a) Time series of non-linear simulations for $\varepsilon=0.3$ and $\omega_L=2.7$, showing the total kinetic energy (solid red thick line) and the kinetic energy of the non-uniform-vorticity flow (solid black thin line). The blue dashed line the evolution of perturbation kinetic energy in a corresponding linear simulation. (b) Snapshot of the $z$-component of the non-uniform vorticity flow in the plane $z=0$ at $t=125$.}
         \label{fig:timeseries_nonlinear}                
  \end{center}
\end{figure}
\begin{figure}                
  \begin{center}
  \setlength{\epsfysize}{5.0cm}

    \begin{tabular}{ccc}
          \setlength{\epsfysize}{4.0cm}
      \subfigure[]{\epsfbox{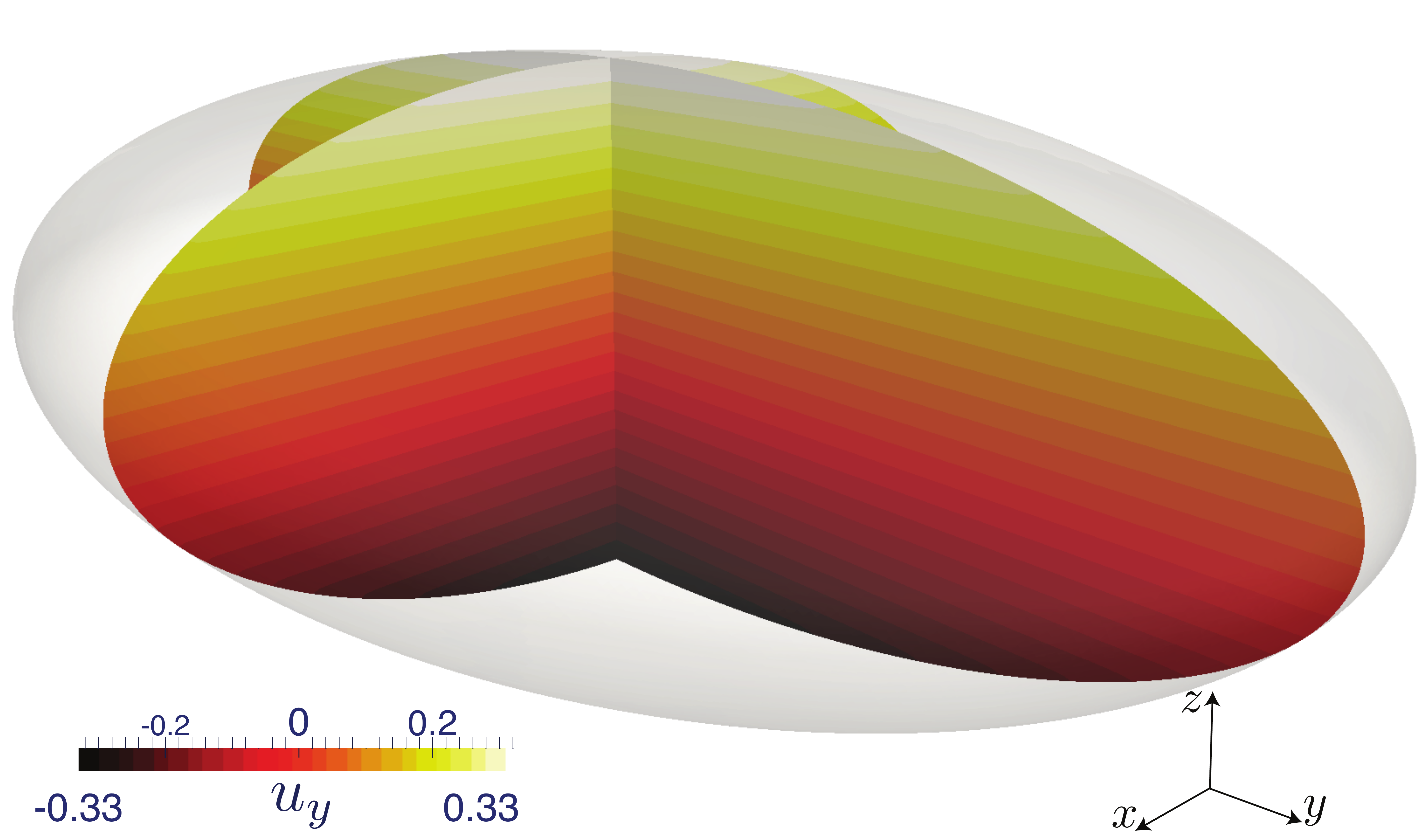}}  &
      \setlength{\epsfysize}{4.0cm}
      \subfigure[]{\epsfbox{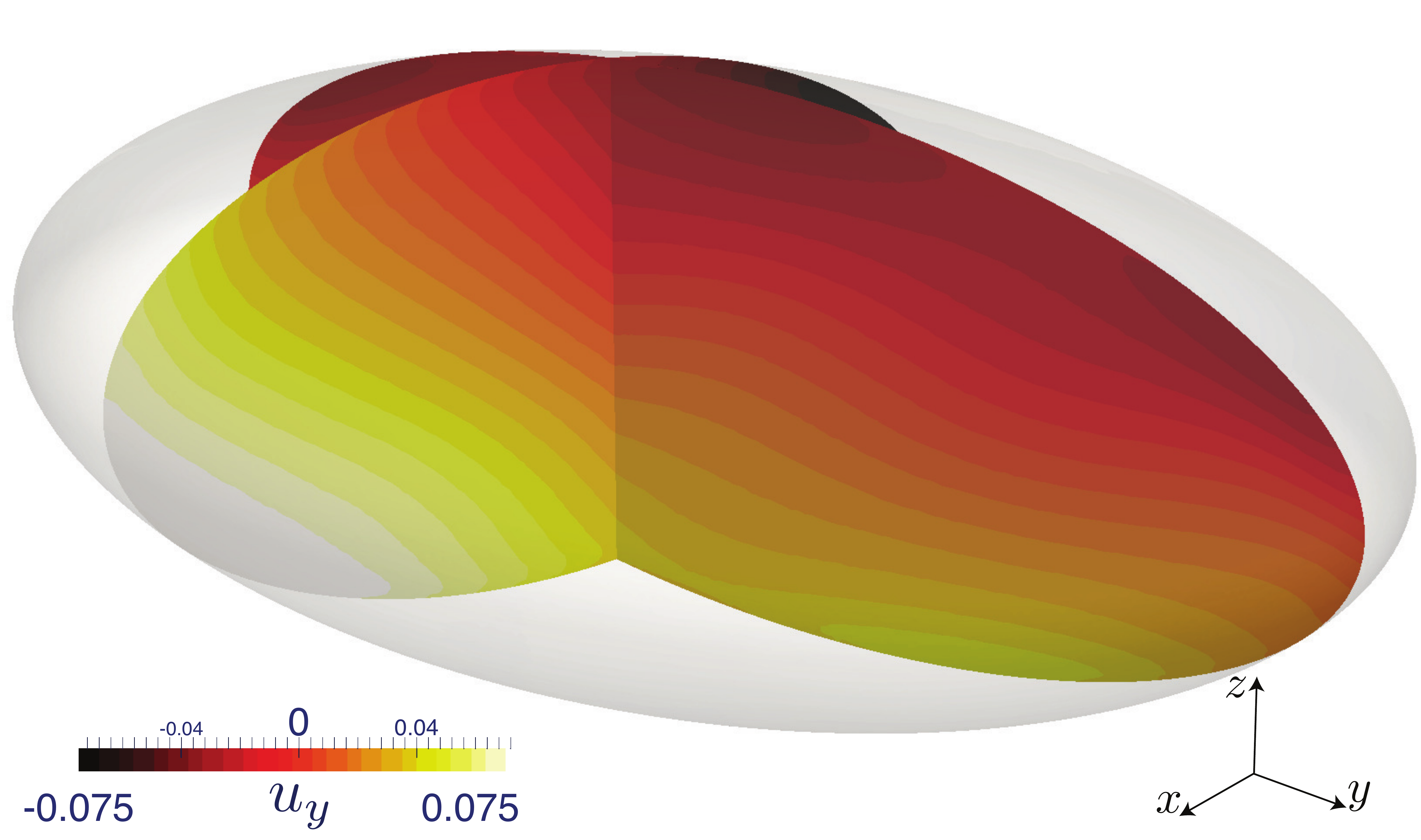}}  \\
      \setlength{\epsfysize}{4.0cm}
      \subfigure[]{\epsfbox{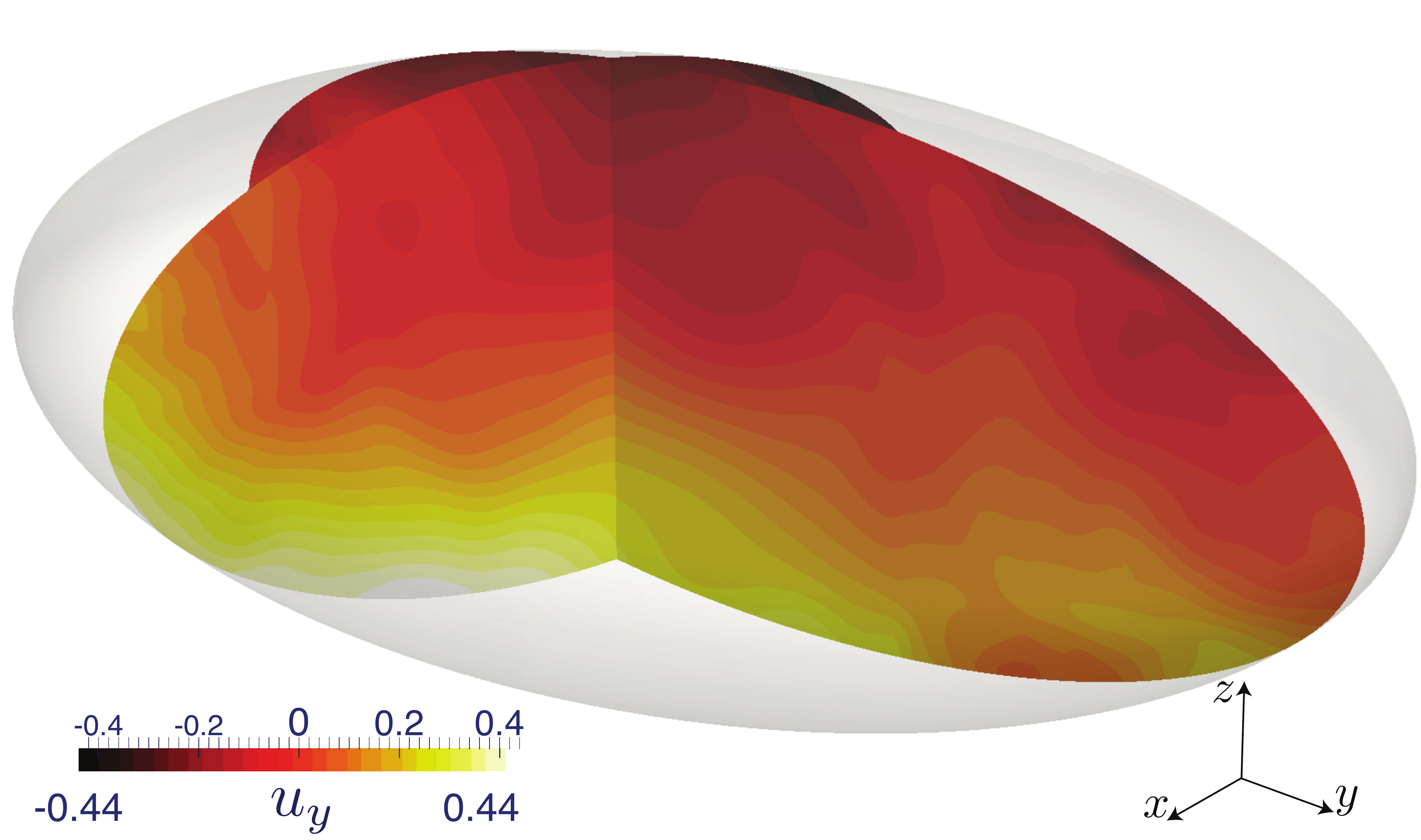}}  &
      \setlength{\epsfysize}{4.0cm}
       \subfigure[]{\epsfbox{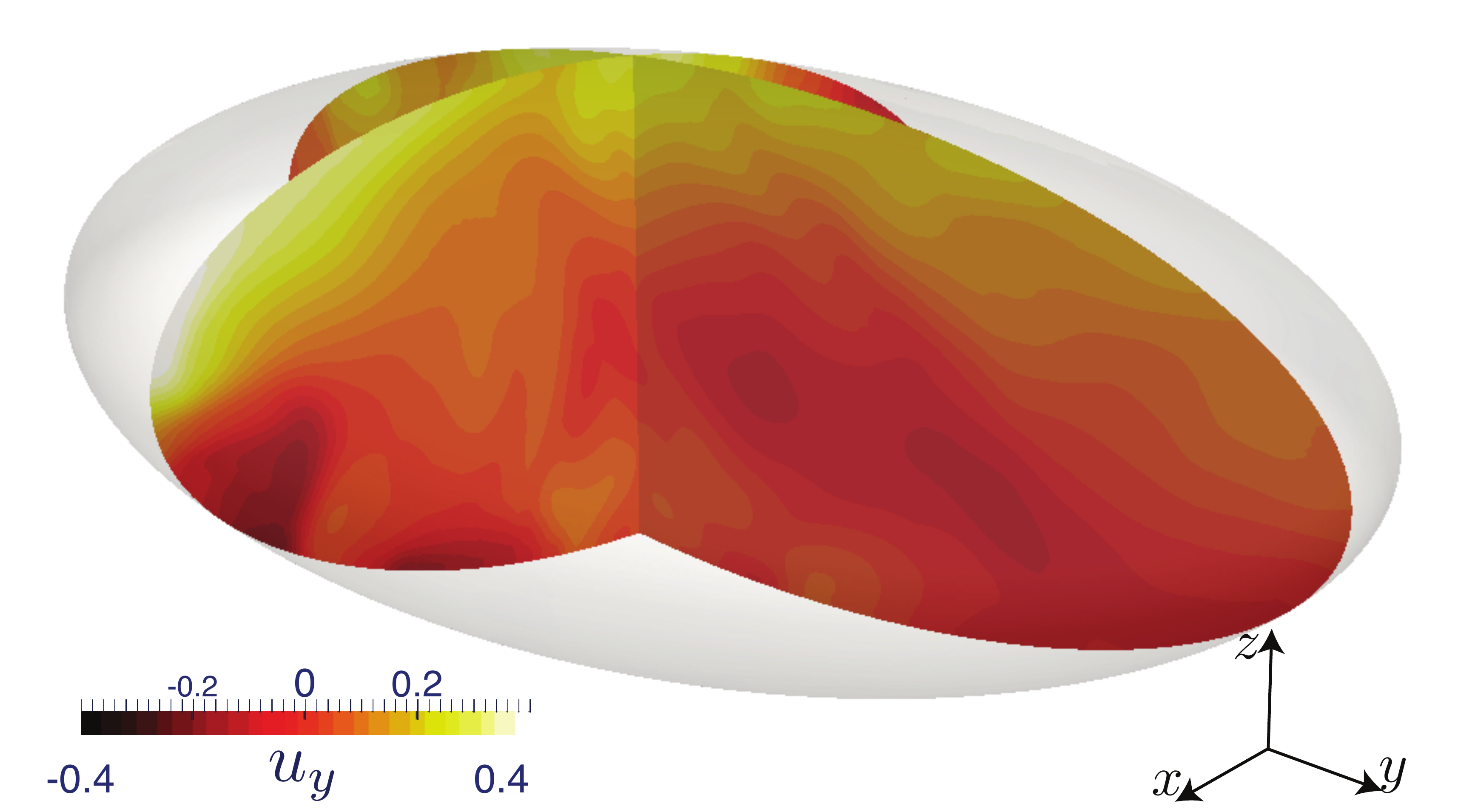}}  &
      \setlength{\epsfysize}{4.0cm}
    \end{tabular}
    \caption{Non-linear simulations of latitudinally libration driven flow for $\varepsilon=0.2$ and $\omega_L=2.7$. Instantaneous profiles of $u_y$ in the planes $x=0$ and $y=0$ at $t=35$ (a), $t=125$ (b), $t=215$ (c) and $t=275$ (d).  }
         \label{fig:snapshot_nonlinear}                
  \end{center}
\end{figure}

Finally, we also consider direct numerical simulations of the non-linear equations (\ref{eq:divu}) and (\ref{eq:navierstokes}) with $\varepsilon=0.3$ and $\omega_L=2.7$, supplemented with the stress-free boundary condition ({\ref{eq:stress_free}}). In figure \ref{fig:timeseries_nonlinear}a, we display the evolution in time of the kinetic energy, as well as the energy related to the non-uniform vorticity component of the flow. This quantity is obtained as the difference between ${\boldsymbol u}$ and the projection of ${\boldsymbol u}$ on the subspace of uniform vorticity flows. Starting from the solution (\ref{eq:u(q)}),(\ref{eq:sol_qx})-(\ref{eq:sol_qy}), we see that a weak non-uniform vorticity flow immediately emerges, which is related to a weak boundary layer flow required to match the stress-free boundary condition. However, until $t \approx 50$, the flow remains essentially of uniform vorticity, as can be seen from figure \ref{fig:snapshot_nonlinear}a, which shows an instantaneous velocity profile of $u_y$ in the planes $x=0$ and $y=0$ at $t=35$. 
For $t \gtrsim 50$, the non-uniform vorticity component undergoes exponential growth over several orders of magnitude. This gives evidence of the presence of instability. Furthermore, we see that the growth rate of the the non-uniform- vorticity component in the non-linear simulation is in good agreement with the results from the corresponding simulation of the linear equation (\ref{eq:disturbance_v2}).  In figure \ref{fig:timeseries_nonlinear}b, we depict an instantaneous profile of the non-uniform vorticity component of $u_z$ in the plane $z=0$ at $t=125$; we find that its spatial structure is similar to its counterpart from the linear stability analysis, displayed in figure \ref{fig:snapshot_instab}a. The growth of the non-uniform vorticity component also affects the structure of the total velocity field. This can be seen from figure \ref{fig:snapshot_nonlinear}b, where we see slightly rippled isolines of $u_y$ at $t=125$. Eventually, the instability saturates when the non-uniform vorticity kinetic energy is approximately of the same order of magnitude as the total kinetic energy. The flow has now completely lost its uniform vorticity character as can be seen from figure \ref{fig:timeseries_nonlinear}c-d. 

\section{Planetary core flow estimates \label{sec:planeto}}
Having adopted a fluid mechanist's perspective in the preceding sections, we now return to the initial problem of the latitudinal libration in a planetary context. With the exception of Mercury, satellites that undergo libration in latitude and possess a liquid core are in a 1:1 spin-orbit resonance, i.e. $\omega_L \approx 1$.  Following \cite{murraybook}, the semi-axes $a$, $b$ and $c$ for such synchronized, homogeneous satellites in hydrostatic equilibrium are
\begin{equation}
\begin{array}{ccc}
a = R\left(1-\frac{H}{3}\right), & b = R\left(1+\frac{7H}{6}\right) & c= R\left(1-\frac{5H}{6}\right). \label{eq:abc_satellite}
\end{array}
\end{equation} 
Here, $R$ is the estimated mean core radius, and $H$ denotes the static tidal bulge, which is the product
\begin{equation}
H= h_{2t}  \frac{m}{M} \left(\frac{R_p}{R_{orb}}\right)^{3}.
\end{equation}  
In this expression, $h_{2t}$ is the tidal Love number, which characterizes the satellite's rigidity and is between 1 and 2.5. $R_{p}$ and $R_{orb}$ stand for the planet's radius and mean orbital radius, respectively, and $m$ and $M$ denote the mass of the satellite and its host. 
In Table \ref{tab:planets}, we compute ${\beta}_{ac}$ and ${\beta}_{bc}$ for the Moon and Io, using values for $h_{2t}$ that are reported in the literature. Provided that ${\beta}_{ac},\beta_{bc} \ll 1$, it follows from (\ref{eq:abc_satellite}) that
\begin{equation}
4\beta_{ac} \approx \beta_{bc}.
\end{equation}
Using (\ref{freq:spin_over}), we then obtain a linearized expression for the spin-over frequency as a function of $\beta_{bc}$,
\begin{equation}
f = 1 + \frac{5}{8}\beta_{bc} + \mathcal{O}(\beta_{bc}^2).
\end{equation}
In the case of Mercury, we cannot apply the above formulas, and therefore adopt estimates of ${\beta}_{ac}$ and ${\beta}_{bc}$ used in a previous study \cite[]{rambaux2007}. 
\par For the satellites of interest (see Table \ref{tab:planets}), we conclude that $f- \omega_L$ is much larger than $E^{1/2}$. This statement is a fortiori true for Mercury, since its principal libration frequency is close to $2/3$ so that $f - \omega_L = \mathcal{O}(1/3)$.
Hence, we do not expect a resonance of the spin-over mode forced by libration in latitude for the satellites under consideration. Therefore, we can use the solution (\ref{eq:sol_qx})-(\ref{eq:sol_qz}) to estimate the typical amplitude of the forced uniform vorticity flow, which we compute as $\displaystyle \sqrt{2 \overline{e_k}}\Omega_0 R$. For the cores of the Moon and Io, this yields $0.13$ mm/s, $0.019$ mm/s, respectively. In the case of Mercury, we find an upper bound of approximately $0.00053$ mm/s. All this is rather small, when compared to the typical convective velocities in the core of the Earth, which are about $0.4$ mm/s. 
\begin{table}
\def~{\hphantom{0}}
\begin{minipage}{\textwidth}
\begin{center}
\begin{tabular}{cccc}
\hline \hline
Satellites & Moon & Io  & Mercury \\
\hline
$\sim R$ [km] & $^a$350 & $^b$900 & $^d$1900  \\
$\sim h_{2t}$ & $^a$1.026 & $^b$2.292 & \---  \\
$\sim \beta_{ac} \cdot 10^{-5}$ & 0.387 & 197 & $^d$$10$ \\
$\sim \beta_{bc} \cdot 10^{-5}$ & 1.55 & 785  & $^d$$10$\\
$\Omega_0$  [$\mu$rad/s] & $^a$2.66 & $^c$41.1 & 1.24 \\
$\Delta \theta $ [rad] & $^a$$3.8 \cdot 10^{-4}$ & $^c$$3.8 \cdot 10^{-7}$ & $^e$$< 5 \cdot 10^{-8}$ \\
$\omega_L$ & $^a$1.0044 & $^c$1.0036 & $\sim 2/3$ \\
$\sim E$ & $3.0 \cdot 10^{-12}$ & $3.0 \cdot 10^{-14}$  & $2.5 \cdot 10^{-13}$ \\
\hline
$ \left|\frac{f-\omega_L}{\sqrt{E}}\right|$ & $2.5 \cdot 10^3$ & $7.5 \cdot 10^3$ & $7.1 \cdot 10^5$ \\
$|\chi_1| + |\chi_2|$ & $5.9 \cdot 10^{-9}$ & $8.1 \cdot 10^{-9} $ & $4.8 \cdot 10^{-10}$ \\
\hline \hline
\end{tabular}

\caption{Orbital dynamical parameters of the Moon, Io and Mercury. Estimated liquid core size, ellipticites, spin rate, principal libration amplitude $\Delta \theta$, libration frequency $\omega_L$ and Ekman number $E$ based on a value of the kinematic viscosity of $\nu = 10^{-6} \mathrm{m^2/s}$ corresponding to an iron-rich molten core. References: $^a$\cite{williams2002proc}, $^b$\cite{anderson2001}, $^c$\cite{noyelles2012},$^d$\cite{rambaux2007}, $^e$\cite{dufey2009}.}
\label{tab:planets}
\end{center}
\end{minipage}
\end{table}

\par
\begin{figure}                 
  \begin{center}
\includegraphics[width=0.7\textwidth]{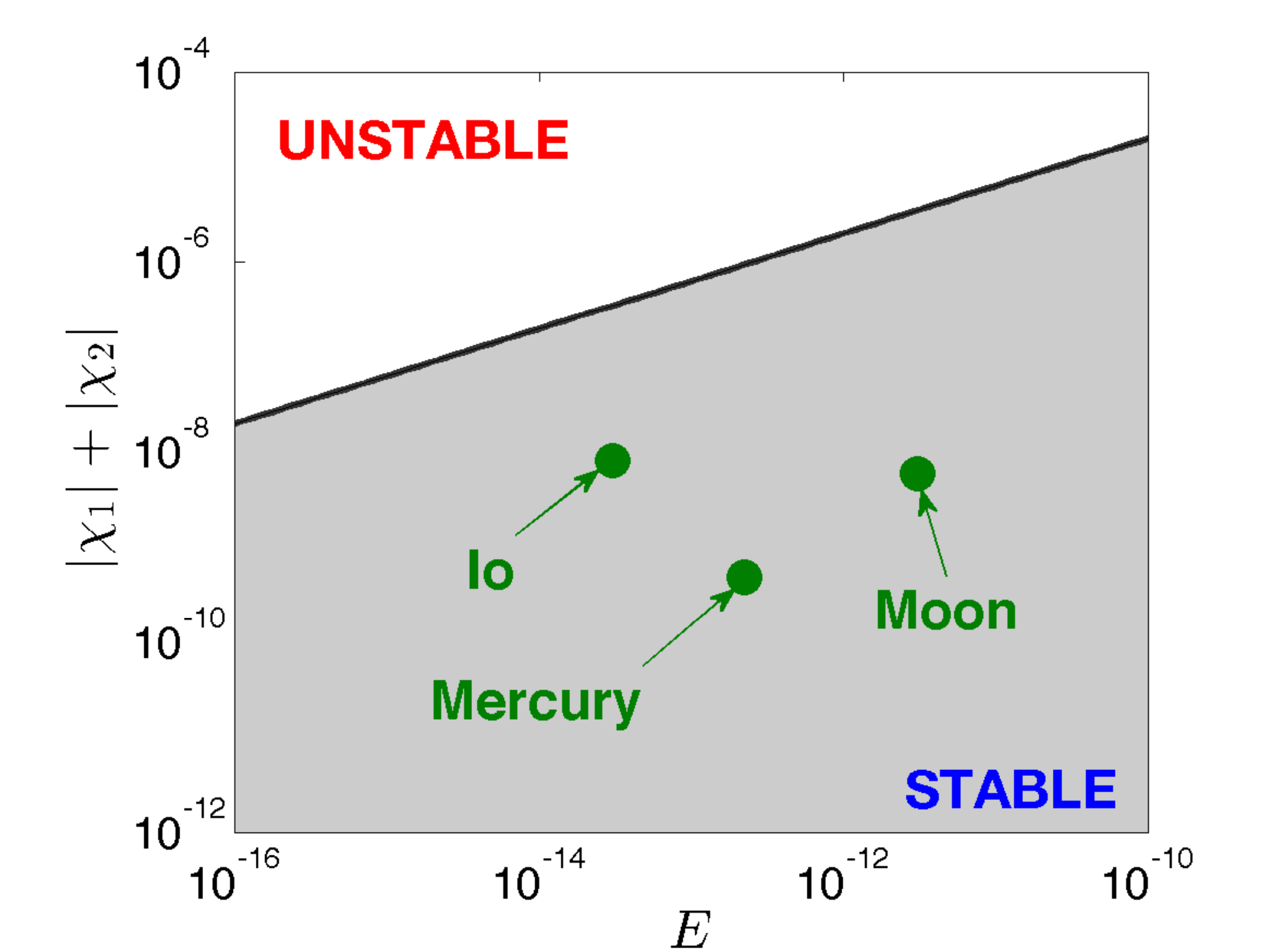}
    \caption{Criterion for marginal stability (\ref{eq:criterion_planets}) (solid line) and location of the Moon, Io and Mercury (symbols) in the $(E,|\chi_1|+|\chi_2|)$ plane.}
         \label{fig:planets}                
  \end{center}
\end{figure}

Given that the Moon, Io and Mercury operate in the non-resonant regime, we can also apply the linear stability analysis discussed in section \ref{sec:stability}. Using the data in Table \ref{tab:planets}, we can compute the quantity $|\chi_1| + |\chi_2|$ that bounds the growth rate of an inviscid instability according to (\ref{eq:bound_sigma}). In the preceding sections, we have considered inviscid instabilities. In the presence of viscosity and rigid walls, the onset of the inertial instability requires that this inviscid growth rate is larger than the viscous dissipation rate in the Ekman layer $K\sqrt{E}$, where $K\gtrsim 1$ is a constant that depends on the inertial modes participating in the parametric coupling \cite[]{le2000three}. Therefore, a minimum condition on $|\chi_1| + |\chi_2|$ is obtained with $K=1$,
\begin{equation}
|\chi_1| + |\chi_2| >  \sqrt{E} \label{eq:criterion_planets}
\end{equation}
Typical values of $|\chi_1| + |\chi_2|$ are given in figure \ref{fig:planets} and Table \ref{tab:planets}. We see that these are at least one order of magnitude smaller than $\sqrt{E}$. According to our analysis, none of these celestial objects are thus subject to latitudinal libration driven instabilities. 

\section{Conclusion and perspectives} \label{sec:discussion}
Motivated by modelling the core dynamics of planets and moons, we have presented a theoretical-numerical study of the flow driven by latitudinal libration within a triaxial ellipsoid. We have found that the inviscid equations of motion allow simple uniform vorticity solutions, which may enter in resonance with the spin-over inertial mode. This result is an extension of the theory of \cite{chan2011simulations-2} and \cite{zhang2012} to triaxial ellipsoids. Numerical simulations have shown that the uniform vorticity formalism, in combination with a reduced model of viscosity, is a powerful approach to the study of integral quantities of the flow, such as the total angular momentum and kinetic energy. Nevertheless, it does not encompass all effects of viscosity, like the boundary layer driven circulation in the bulk.
\par
Once the uniform vorticity solution was derived, we investigated its dynamical stability. We have revealed that a parametric resonance mechanism underlies  the onset of instability. Its growth rate is governed by two control parameters $\chi_1$ and $\chi_2$, which capture the geometry of the cavity and the libration amplitude and frequency.
\par
Application of our results to planetary settings has enabled us to conclude that the liquid cores of the Moon, Io and Mercury are not presently unstable or in a state of direct resonance. However, it cannot be excluded that such phenomena exist for other satellites, or might have taken place in the early Solar System. Deciding this would require accurate estimates of the libration frequencies and amplitudes, which have yet to be established.
\par       
In conclusion, we identify some aspects of latitudinally libration driven flow that, in our opinion, deserve further exploration. Of primary importance is the experimental validation of the results set out in this work, and their extension to the strongly non-linear regime, otherwise inaccessible numerically. Furthermore, it has been argued that inertial instabilities can be driven by the internal conical shear produced by the boundary layer \cite[]{Kida2013,Lin2014conical}. Taylor-G\"ortler instabilities, associated with viscous Ekman layers, may also exist, as it is the case for longitudinal libration \cite[]{noir2009,calkins2010axisymmetric}. Finally, we recall that we have only considered the case of a rigid shell. However, it \cite{Goldreich2010} suggested that, for a visco-elastic ice shell overlying subsurface oceans, the gravitational torque can cause dynamical tides in the liquid-solid boundary in addition to the rigid response. One may argue that this problem is very similar to that presented here \cite[]{cebronAA}. Nevertheless, the thin shell geometry of a subsurface ocean leads to the well-known ill-posed Cauchy problem, which raises considerable mathematical difficulties \cite[]{rieutord1997inertial}. 
 \begin{acknowledgements}
S. V. and J. N. are supported at ETH Z\"urich by ERC grant 247303 (MFECE).  S.V. also acknowledges financial support from the Swiss National Science Foundation (SNF) through the project $200020\_143596$. D. C. is partially supported by the ETH Z\"urich Postdoctoral Fellowship Progam as well as by the Marie Curie Actions for People COFUND Program. This work was supported by a grant from the Swiss National Supercomputing Centre (CSCS) under project ID s369. We thank four anonymous reviewers for their constructive comments.
\end{acknowledgements}
\appendix
\section{Equations of motion in different frames of reference.} \label{app:reference_frame}
The mantle frame, i.e. the reference frame attached to the walls of the ellipsoid, is a non-inertial frame. Therefore, the equations of motion written in this frame involve two fictitious forces: the Coriolis force, which requires an expression for the total rotation vector ${\boldsymbol \Omega}$, and the Poincar\'e force, which depends on $\dot {\boldsymbol \Omega}$
In this appendix, we compute the projection of these forces onto the cartesian (unit) axes of the mantle frame. To this end, we will consider three different frames of reference. The first one is an inertial frame, and will be denoted by a subscript $I$. The `frame of mean rotation' (subscript $R$) rotates at constant angular speed $\Omega_0$ around the $\hat{\boldsymbol z}$ axis with respect to the inertial frame. The cartesian unit vectors between the frame of mean rotation and inertial frame are related via the transformation
\begin{equation}
\left(
\begin{array}{c}
\hat {\boldsymbol x}_R \\ \hat {\boldsymbol y}_R\\ \hat {\boldsymbol z}_R
\end{array}
\right)
=
\left(
\begin{array}{ccc}
\cos(\Omega_0 t) & \sin(\Omega_0 t) & 0 \\
-\sin(\Omega_0 t) & \cos(\Omega_0 t) & 0 \\
0 & 0 & 1
\end{array}
\right)
\left(
\begin{array}{c}
\hat {\boldsymbol x}_I \\ \hat {\boldsymbol y}_I\\ \hat {\boldsymbol z}_I
\end{array}
\right).  \label{eq:rotation_inertial}
\end{equation}
On the other hand, the mantle frame (subscript $M$) undergoes the librating motion of the ellipsoidal container. The following transformation holds between the unit axes of the mean rotation and mantle frame,
\begin{equation}
\left(
\begin{array}{c}
\hat {\boldsymbol x}_R \\ \hat {\boldsymbol y}_R\\ \hat {\boldsymbol z}_R
\end{array}
\right)
=
\left(
\begin{array}{ccc}
1 & 0 & 0 \\
0 & \cos\theta  & -\sin \theta \\
0 & \sin \theta & \cos \theta
\end{array}
\right)
\left(
\begin{array}{c}
\hat {\boldsymbol x}_M \\ \hat {\boldsymbol y}_M\\ \hat {\boldsymbol z}_M
\end{array}
\right), \label{eq:rotation_mantle}
\end{equation}
where the angle $\theta(t)$ is defined by (\ref{eq:def_theta}), and denotes the instantaneous tilt between the axes of the mean rotation and mantle frame.
\par
For latitudinal libration, the rotation vector ${\boldsymbol \Omega}$ is most easily expressed in the frame of mean rotation:
\begin{equation}
{\boldsymbol \Omega} = \dot{\theta}\hat{\boldsymbol x}_R + \Omega_0 \hat{\boldsymbol z}_R.
\end{equation}
Using(\ref{eq:rotation_mantle}), we can derive an expression for the rotation vector ${\boldsymbol \Omega}$ in the mantle frame, 
\begin{equation}
{\boldsymbol \Omega} = \dot{\theta}\hat{\boldsymbol x}_M + \Omega_0 \sin \theta \hat{\boldsymbol y}_M +  \Omega_0 \cos \theta \hat{\boldsymbol z}_M.
\end{equation}
The computation of $\dot{\boldsymbol \Omega}$ is the most easily performed in the inertial frame, because the unit vectors are stationary in this frame. We obtain
\begin{equation}
{\boldsymbol \Omega} = \dot{\theta}\left(\cos (\Omega_0 t) \hat{\boldsymbol x}_I + \sin (\Omega_0 t) \hat{\boldsymbol y}_I \right) + \Omega_0 \hat{\boldsymbol z}_M,
\end{equation}
and hence,
\begin{equation}
\dot {\boldsymbol \Omega} = \ddot{\theta}\left(\cos (\Omega_0 t) \hat{\boldsymbol x}_I + \sin (\Omega_0 t) \hat{\boldsymbol y}_I \right) + \dot{\theta} \Omega_0 \left(-\sin (\Omega_0 t) \hat{\boldsymbol x}_I + \cos (\Omega_0 t) \hat{\boldsymbol y}_I \right).
\end{equation}
Using (\ref{eq:rotation_inertial}) and (\ref{eq:rotation_mantle}), this can be recast in the mantle frame as
\begin{equation}
\dot {\boldsymbol \Omega} = \ddot{\theta} \hat{\boldsymbol x}_M + \dot{\theta} \Omega_0 \left(\cos \theta \hat{\boldsymbol y}_M - \sin \theta \hat{\boldsymbol z}_M \right),
\end{equation}
which gives expression (\ref{eq:omegadot}) for $\Omega_0 = 1$.
\bibliographystyle{jfm}
\bibliography{convection}
\end{document}